\documentclass[a4paper,11pt]{article}

\pdfoutput=1 

\usepackage{jcappub} 

\usepackage[T1]{fontenc} 

\usepackage{subfig}
\usepackage{graphicx}
\usepackage{amsmath}
\usepackage{mathtools}
\usepackage{xcolor}
\usepackage{booktabs}
\usepackage[normalem]{ulem}

\newcommand{\textcode}[1]{{\large\scshape#1}}

\newcommand{\Msun}{\, h^{-1} \,  M_{\odot}}
\newcommand{\hompc}{\,h\,{\rm Mpc}^{-1}}
\newcommand{\mpcoh}{\,h^{-1}\,{\rm Mpc}}

\title{Fast generation of mock galaxy catalogues in modified gravity models with COLA}

\author[a,1]{Bartolomeo Fiorini,\note{Corresponding author.}}
\author[a]{Kazuya Koyama,}
\author[a]{Albert Izard,}
\author[b]{Hans A. Winther,}
\author[c]{Bill S. Wright,}
\author[d]{Baojiu Li}

\affiliation[a]{Institute of Cosmology \& Gravitation, University of Portsmouth, Dennis Sciama Building, Burnaby Road, Portsmouth, PO1 3FX, United Kingdom}
\affiliation[b]{Institute of Theoretical Astrophysics, University of Oslo, PO Box 1029, Blindern 0315, Oslo, Norway}
\affiliation[c]{Queen Mary University of London, Mile End Road, London, E1 4NS, United Kingdom}
\affiliation[d]{Institute for Computational Cosmology, Department of Physics, Durham University, South Road, Durham, DH1 3LE, United Kingdom}

\emailAdd{bartolomeo.fiorini@port.ac.uk}

\abstract{We investigate the viability of producing galaxy mock catalogues with COmoving Lagrangian Acceleration (COLA) simulations in Modified Gravity (MG) models employing the Halo Occupation Distribution (HOD) formalism. In this work, we focus on two theories of MG: $f(R)$ gravity with the chameleon mechanism, and a braneworld model (nDGP) that incorporates the Vainshtein mechanism.  We use a suite of full {\it N}-body simulations in MG as a benchmark to test the accuracy of COLA simulations. At the level of Dark Matter (DM), we show that COLA accurately reproduces the matter power spectrum up to $k \sim 1 \hompc$, while it is less accurate in reproducing the velocity field. To produce halo catalogues, we find that the \textcode{rockstar} halo-finder does not perform well with COLA simulations. On the other hand, using a simple Friends-of-Friends (FoF) finder and an empirical mass conversion from FoF to spherical over-density masses, we are able to produce halo catalogues in COLA that are in good agreement with those in {\it N}-body simulations. To consider the effects of the MG fifth force on the halo profile, we derive simple fitting formulae for the concentration-mass and the velocity dispersion-mass relations that we calibrate using \textcode{rockstar} halo catalogues in {\it N}-body simulations. We then use these results to extend the HOD formalism to modified gravity simulations in COLA. We use an HOD model with five parameters that we tune to obtain galaxy catalogues in redshift space. We find that despite the great freedom of the HOD model, MG leaves characteristic imprints in the redshift space power spectrum multipoles and these features are well captured by the COLA galaxy catalogues.} 

\begin{document}
\maketitle
\flushbottom

\section{Introduction}
\label{sec:intro}

Stage IV dark energy surveys like Euclid\footnote{\url{https://www.euclid-ec.org/}} and the Vera C. Rubin Observatory's LSST\footnote{\url{https://www.lsst.org/}} are devoted to improving our understanding of the cosmological model: the investigation of the Large-Scale Structure (LSS) of the Universe can give insight on the nature of the dark sector components. Because structure formation is driven by gravitational forces, possible deviations of gravity from General Relativity (GR) on cosmological scales may affect our interpretation of Dark Matter (DM) and dark energy. 

On the other hand, a unified description of gravity with the other fundamental forces of nature is an active area of intense research, with many possible theoretical candidates but with a lack of experimental probes. Despite that this is a concern mainly for high energy physics, some of the alternative theories to GR known as modified gravity (MG) theories manifest a non-trivial behaviour on cosmological scales, where the theory of gravity has not been directly tested yet (for a review see \cite{Koyama:2015vza}).

It is crucial to make accurate theoretical predictions of the properties of the LSS on non-linear scales, both in terms of comparison to measurements and to construct realistic covariance matrices, to successfully constrain the gravity model (see e.g. \cite{Alam:2020jdv}). This can be achieved by means of cosmological simulations, but in excess of ${\cal O}(10^3)$ realisations must be produced to match the volume of Stage IV surveys and compute an accurate estimate of the covariance matrices. This poses a serious challenge for full {\it N}-body simulations in MG models due to their high computational cost. Such models usually have screening mechanisms that hide modified gravity effects on small scales. To describe the screening mechanism, an additional non-linear equation needs to be solved in {\it N}-body simulations, which significantly slows down the MG simulations \cite{Winther:2015wla}. 

An alternative to full {\it N}-body simulations is to exploit approximate methods (see \cite{Monaco16} for a comprehensive review and \cite{Chuang:2014toa} for a comparison project in GR), which lower the computational cost at the expenses of accuracy on non-linear scales. Amongst these, the COLA method \cite{Tassev:2013pn, Koda:2015mca, Izard:2015dja, Howlett:2015hfa} with its extension to MG \cite{Valogiannis:2016ane, Winther:2017jof, Wright:2017dkw} offers an interesting compromise between speed-up and accuracy without introducing any additional free parameter and is therefore ideal to access the MG information on (mildly) non-linear scales without losing predictability (see \cite{Moretti:2019bob} for an alternative approach based on \textcode{pinocchio}).

Having a bridge between observed galaxies and simulated dark matter distribution makes it possible to extract the LSS information encoded in the datasets from galaxy surveys.
Several prescriptions have been studied in the literature to do so. A popular and simple empirical prescription is the Halo Occupation Distribution (HOD), which starts from the DM halos identified in the density field by a halo-finder algorithm \cite{Knebe:2011rx} and populates them with galaxies with a probability distribution conditioned on some halo properties.
In particular, the HOD model proposed in \cite{Zheng:2007zg} relies on the halo mass and the density: using the analytical Navarro-Frenk-White (NFW) model \cite{Navarro:1995iw} for the halo profile, it is possible to apply this formalism to COLA simulations that do not resolve the internal halo properties \cite{Koda:2015mca}.
Due to the non-trivial dynamics of the screening mechanisms, the internal halo structure in MG theories can differ from the one in GR and this needs to be considered to produce realistic galaxy mocks \cite{Mitchell:2018qrg, Mitchell:2019qke}.

In this context, we investigate the feasibility of producing mock galaxy catalogues from COLA simulations in MG producing a simulation suite with \textcode{mg-picola}
\cite{Winther:2015wla, Wright:2017dkw} and using a suite of full {\it N}-body simulations performed by the \textcode{ecosmog} code \cite{Li:2011vk} to validate the COLA results and to estimate the accuracy in reproducing galaxy clustering statistics.

This paper is organised as follows. 
After introducing the MG theories of interest to this work in Section~\ref{sec:MG}, we discuss the simulation techniques and suites in Section~\ref{sec:Sims}. In Section~\ref{sec:Halos} we investigate the production of halo catalogues. We then apply the HOD formalism in Section~\ref{sec:Galaxies} to create mock galaxy catalogues and study the multipole moments of the galaxy spectrum in redshift space. We summarise our results in Section~\ref{sec:Conclusions}.

\section{Modified Gravity}
\label{sec:MG}

Einstein's theory of General Relativity (GR) has revolutionised our understanding of the Universe over the course of the last century and the $\Lambda$CDM model based on GR has been very successful in reproducing various cosmological observations. However, GR still lacks a high energy completion and the $\Lambda$CDM model requires the existence of a highly fine-tuned cosmological constant to explain the current accelerated expansion of the Universe. Furthermore, the so-called $H_0$ and $S_8$ tensions \cite{Planck:2018vyg,Riess:2020fzl,Troster:2019ean} may be hints of new physics beyond the $\Lambda$CDM model that may be due to the underlying gravity theory \cite{Raveri:2019mxg}.

These yet to be solved problems have motivated theorists to formulate alternatives to GR often referred to as modified gravity (MG) theories. The MG theories of interest to cosmology are the ones showing an infrared behavior, which is often associated with an additional force (referred as the fifth force), mediated by a scalar field. The fifth force dynamics can be interpreted as the result of two factors:
\begin{itemize}
  \item The effective mass of the scalar field, which limits the range of action of the fifth force.
  \item The screening mechanism, which may suppress the fifth force \cite{Joyce:2014kja, Koyama:2015vza}.
 \end{itemize}
Solar System tests put very stringent constraints on gravity \cite{Will:2014kxa}.
Because of this, theories that feature deviations from GR relevant to cosmology commonly incorporate a screening mechanism that let the theory evade these stringent constraints.
Amongst the possible screening mechanisms we focus on:
\begin{itemize}
    \item 
    Chameleon screening \cite{Khoury:2003aq,Khoury:2003rn}, occurring when the effective action for the scalar field accounts a potential term.
    \item 
    Vainshtein screening \cite{Vainshtein:1972sx}, present when the scalar field action contains a term with second derivatives of the scalar field.
\end{itemize} 

In this work we consider the Hu-Sawicki $f(R)$ and the Dvali-Gabadadze-Porrati gravity models, which implement the Chameleon and the Vainshtein screening mechanisms respectively, and study their impact on the large-scale structure compared to GR.

\subsection{Hu-Sawicki $f(R)$}
The class of theories where the modifications of gravity can be described by an additional term to the Einstein-Hilbert action in the form of a function of the Ricci scalar is known as $f(R)$ theories \cite{Starobinsky:1980te}:
\begin{equation}
    S=\int \mathrm{d}^{4} x \sqrt{-g}\left[\frac{R+f(R)}{16 \pi G}+\mathcal{L}_{\mathrm{m}}\right] \, .
\end{equation}

This action evades the Solar System constraints by means of the so-called Chameleon mechanism \cite{Khoury:2003aq,Khoury:2003rn}, where the density controls the shape of the scalar field potential, determining the background value of the scalar field and its effective mass $m$ (large masses in high density environments and vice-versa) and the fifth force is consequently screened.

In this paper we are interested in the Hu-Sawicki model with $n=1$ \cite{Hu:2007nk}
\begin{equation}
f(R)=-2 \Lambda+\left| f_{R 0} \right| \frac{\bar{R}_{0}^{2}}{R} \, ,
\label{Hu-Sawicki_fR}
\end{equation}
where $\bar{R}_{0}$ is the background curvature today and $f_{R0}$ is the value of $f_{R} \equiv d f / d R$ today.
For a small $|f_{R0}|$, the background cosmology can be approximated as the one given by the $\Lambda$CDM model. In the linear regime, it is described in Fourier space by the modified Poisson equation for the gravitational potential $\Psi$ \cite{Pogosian:2007sw}:
\begin{equation}
k^{2} \Psi=-4 \pi G_{\rm eff} a^{2} \delta \rho, \quad  
G_{\rm eff} = G\left(\frac{4+3 a^{2} m^{2} / k^{2}}{3+3 a^{2} m^{2} / k^{2}}\right),
\label{fR_Poisson}
\end{equation}
where $\delta \rho$ is the fluctuation around the mean energy density and $m^{2}=\frac{1}{6\left| f_{R 0} \right|} \frac{\bar{R}^{3}}{\bar{R}_{0}^{2}}$. Depending on scale and cosmic time, the effective gravitational constant $G_{\rm eff}$ in Eq.~\eqref{fR_Poisson} has two limits:
\begin{align}
& ma \gg k : G_{\rm eff} \rightarrow G  \, ,\\
& ma \ll k : G_{\rm eff} \rightarrow \frac{4}{3} G\, ,
\end{align}
where in the former case the deviation from GR are suppressed while in the latter gravity is enhanced by the fifth force.

In the non-linear regime, eq.~\eqref{fR_Poisson} for the Poisson equation does not hold and it is necessary to solve the non-linear scalar field equation with a potential. 
The chameleon screening is determined by the gravitational potential of the object as well as the environment. The effect of Chameleon screening can be approximated as 
\cite{winther15}
\begin{equation}
\label{Geff_fR}
G_{\rm eff} = \left(  1 + \frac{1}{3}  
\frac{k^2}{k^2 + a^2 m^2} 
 \text{Min}\left[1, f_s \, \left|\frac{\Phi_{\rm crit}(a)}{\Phi_N}\right|\right]   \right)
G,
\end{equation}
where $\Phi_N$ is the gravitational potential, $f_s$ is a parameter to tune the strength of the screening that we shall call screening efficiency, and
\begin{equation}
    \Phi_{\rm crit}(a) = \frac{3f_{R0}}{2}\left(\frac{\Omega_{m,0} a^{-3} + 4\Omega_{\Lambda,0}}{\Omega_{m,0} + 4\Omega_{\Lambda,0}} \right)^{2},
\end{equation}
with $\Omega_{m,0}$ and $\Omega_{\Lambda,0}$ the present-day density parameters for matter and dark energy.


\subsection{Dvali-Gabadadze-Porrati}
\label{dgp}
Another way to modify gravity is by defining a theory in a higher-dimensional space under the assumption that we are living in a 4D brane of the high dimensional space-time. This class of model is often known as braneworld gravity and the simplest of these is the Dvali-Gabadadze-Porrati (DGP) model \cite{Dvali:2000hr}, which is characterised by the action:
\begin{equation}
S=\frac{1}{16 \pi G_{(5)}} \int d^{5} x \sqrt{-\prescript{(5)}{}{g}}\prescript{(5)}{}{R}+\int d^{4} x \sqrt{-g}\left(\frac{1}{16 \pi G} R+\mathcal{L}_{m}\right) \, ,
\end{equation} 
where the index (5) indicates the quantity is the 5-D generalisation of a 4-D quantity.
The normal branch of this model (nDGP) still needs an additional dark energy component to explain the late-time acceleration and the tuning of the cross over radius $r_{c} \equiv G_{(5)} / 2 G$ to recover the standard cosmology at early times.

Although this model is not theoretically well motivated, it is useful as a toy model to study the effects of its screening mechanism. Indeed focusing on the normal branch, under quasi-static conditions, the theory incorporates the Vainsthein mechanism which screens the fifth force in high curvature environments, but allows for deviation from GR in low curvature environments. We assume that the background cosmology is the same as the $\Lambda$CDM model by introducing the appropriate dark energy in the model \cite{Schmidt:2009sg, Bag:2018jle}. Linearising the equation of motion, the resulting modified Poisson equation is \cite{Koyama:2005kd}:
\begin{equation}
\nabla^{2} \Psi=4 \pi G_{\rm eff} a^{2}\rho \delta, 
\quad 
G_{\rm eff}= G \left(1+\frac{1}{3 \beta}\right),
\label{DGP_Poisson}
\end{equation}
where $\beta=1-2 H r_{\mathrm{c}}\left(1+\frac{\dot{H}}{3 H^{2}}\right)$ and the overdot denotes the derivative with respect to the physical time. 
In the case of the Vainshtein mechanism, the scalar field equation satisfies a non-linear equation where non-linearity appears in the second derivative of the scalar field. The Vainshtein screening may be approximated using a density dependent effective Newton constant. For the DGP model, the approximation is given by \cite{winther15}
\begin{equation}
\label{Geff_DGP}
G_{\rm eff} = \left[ 1+ \frac{1}{3 \beta}\left( \frac{2(\sqrt{1 + x}-1)}{x}  \right)\right] G , \quad 
x = \frac{8(r_cH_0)^2\Omega_{m,0}}{9\beta^2}\frac{\rho}{\overline{\rho}}, 
\end{equation}
where $\rho$ is the average density within a given radius and $\bar{\rho}$ is the background density. Note that this approximation holds only for a spherical mass distribution and can violate the condition that the theory recovers the linear prediction on large scales \cite{Schmidt:2009yj}. Also the screening depends on the smoothing radius of the density field $\rho$, which needs to be tuned to match the exact result.
\section{Simulations}
\label{sec:Sims}

To produce mock catalogues for the modified gravity models previously described, we start by running dark matter simulations of the large-scale structure. In particular, we explore the following modified gravity models: the normal branch DGP model with $H_0 r_c=1$ (N1) and Hu-Sawicki $f(R)$ model with $|f_{R0}| = 10^{-5}$ (F5), plus the vanilla GR model.

Cosmological simulations can be accurate up to non-linear scales, but they are computationally very expensive, as forces between particles need to be calculated for thousands of time steps typically. The cost is even higher for MG models, which in addition also need to solve the non-linear equation for the fifth force. In recent years, the so-called approximate methods have become a popular alternative to full $N$-body simulations, and they are aimed at speeding up the creation of a density field at the expense of not resolving accurately sub-halo scales (see \cite{Chuang:2014toa,Lippich_2018,Blot_2019,Colavincenzo_2018} for studies on their accuracy, and  \cite{Monaco16} for a review). The simulations in this paper make use of the COmoving Lagrangian Acceleration (COLA) method \cite{Tassev:2013pn}, in particular, we use \textcode{mg-picola} \cite{Winther:2017jof}, that includes gravity models other than GR, and is an ideal tool to efficiently run simulations for MG.

\subsection{COLA method}
The COLA method \cite{Tassev:2013pn} uses the Particle-Mesh algorithm, a simple {\it N}-body technique, to evolve the displacement of particles with respect to their second-order Lagrangian Perturbation Theory (2LPT) positions (which are obtained analytically). Thanks to this, a small number of time-steps (typically of $\mathcal{O}(10)$) are enough to accurately recover the DM density field up to mildly non-linear scales, as well as the halo positions and masses. This results in a speed-up of a factor $100-1000$ with respect to full {\it N}-body simulations, at the expense of not resolving internal halo properties. For the production of accurate mock halo catalogues, \cite{Izard:2015dja} proposed an optimal configuration of the parameters in COLA that control the trade-off between accuracy and computational cost, such as the grid resolution used to compute forces and the distribution of time steps, which we adopt for this work.

\subsection{MG extension to COLA}
When MG is incorporated in cosmological simulation, the complexity increases because the motion of particles is governed by both the Newtonian force and the fifth force. This can lead to a slow-down of a factor of $\mathcal{O}(10)$ in theories where computing the fifth force requires the solution of non-linear equations. 
One way to avoid this significant slow-down is to use screening approximations \cite{winther15}: the fifth force dynamics can be described by means of an effective mass of the scalar field and a coupling determined by either the value of gravitational potential or its derivatives, depending on the type of screening mechanism. To extend the COLA method to MG, it is also necessary to reformulate the Lagrangian Perturbation Theory to take into account the effect of the fifth force on the dynamics \cite{Valogiannis:2016ane, Winther:2017jof, Aviles:2017aor}. 

\textcode{mg-picola} \cite{Winther:2017jof} is a publicly available code for cosmological simulations with the COLA method in modified gravity and it implements the scale-dependent 2LPT and some screening approximations. Its accuracy has already been tested against full {\it N}-body simulations at the level of the Dark Matter (DM) density field. In this work, we study the accuracy of \textcode{mg-picola} also for the DM velocity field, the halo abundance and halo clustering, and we show how it can be used in combination with a HOD algorithm to generate mock galaxy catalogues (with some tweaks relative to applications in GR).

\subsection{Simulation suites}
We use a full {\it N}-body simulation suite called \textcode{elephant}, which was introduced and validated in \cite{Cautun:2017tkc}, to benchmark the results of our runs based on COLA.
The \textcode{elephant} suite was produced using \textcode{ecosmog} \cite{Li:2011vk}, a MG extension of the adaptive mesh refinement code \textcode{RAMSES} \cite{Teyssier:2001cp} that solves the exact equations for the fifth force; these simulations were recently used to create mock galaxy catalogues in \cite{Hernandez-Aguayo:2018yrp,Hernandez-Aguayo:2018oxg, Alam:2020jdv}\footnote{see \cite{Barreira:2016ovx} and \cite{Devi:2019swk} for other attempts to create mock galaxy catalogues in MG theories}. Table~\ref{tab:elephant_details} describes the parameters of the suite such as the box size, the mass resolution and the gravity models implemented, and the cosmological parameters employed are:

\begin{equation}
\begin{array}{ccc}
\Omega_{m,0}=0.281 \, , & \Omega_{\Lambda,0}=0.719 \, , & \Omega_{b,0}=0.046 \, , \\
n_{s}=0.971 \, , & \sigma_{8}=0.842 \, , & h=0.697 \, .
\end{array}
\label{cosmology}
\end{equation}
The initial conditions were generated using the Zel'dovich approximation at $z=49$.
For a given realisation, the same initial seed was used for GR, F5 and N1\footnote{Due to an anomaly with one snapshot, we use only 4 realisations for N1 in this work.}.

We develop a new suite of simulations with \textcode{mg-picola} that matches the cosmology and parameters (such as the mass resolution, the box size and the number of realisations) of the fiducial \textcode{elephant} set. We call this new COLA suite PIpeline TEsting Run (\textcode{piter}). We stop the simulations at redshift $z=0.5057$ after 30 timesteps to match the redshift of the \textcode{elephant} snapshot closest to redshift $z=0.5$. As discussed in the previous section, with the screening approximation it is possible to tune the behaviour of the screening mechanism: in Eq.~\eqref{Geff_fR} we use a screening efficiency $f_s = 4.0$ for F5 and in Eq.~\eqref{Geff_DGP} we use a Gaussian smoothing with scale 1 $\mpcoh$ for N1. We have found these values to provide the best agreement for the dark matter power spectrum with {\it N}-body simulations amongst the values tested.
A summary of the \textcode{piter} simulation details is given in Table~\ref{tab:piter_details}.
Initial conditions are generated using the second-order Lagrangian perturbation theory (2LPT) at $z=49$. As in \textcode{elephant}, for a given realisation, the same initial seed is used for GR, F5 and N1. We also ran a GR COLA simulation using the same initial condition as one of the realisations in \textcode{elephant} and checked that our conclusions were not affected by the cosmic variance.
For simplicity of notation, in the following we will just use {\it N}-body and COLA to refer to \textcode{elephant} and \textcode{piter} respectively. 

\begin{table}
\centering 
\parbox{.45\linewidth}{
\caption{\textcode{elephant} simulations} \label{tab:elephant_details}
\begin{tabular}{cc}
\toprule
Models & GR, F5, N1 \\
Realisations & 5 \\
Box size & 1024 $\mpcoh$ \\
$N_{\mathrm{part}}$ & $1024^3$ \\
$M_{part}$ & $7.7985 \cdot 10^{10}\Msun$ \\
Domain grid & $1024^3$ \\
Refinement criterion & $8$ \\
Initial conditions & Zel'dovich, $z=49$ \\
\bottomrule
\end{tabular}
}
\hfill
\parbox{.45\linewidth}{
\caption{\textcode{piter} simulations} \label{tab:piter_details}
\begin{tabular}{ccccc}
\toprule
Models & GR, F5, N1 \\
Realisations & 5 \\
Box size & 1024 $\mpcoh$ \\
$N_{\mathrm{part}}$ & $1024^3$ \\
$M_{part}$ & $7.7985 \cdot 10^{10}\Msun$ \\
Force grid & $3072^3$ \\
Timesteps & $30$ \\
Initial conditions & 2LPT, $z=49$ \\
\bottomrule
\end{tabular}
}
\end{table}

\begin{figure}[tbp]
\centering 
\subfloat[][GR]{
\includegraphics[width=.48\textwidth,clip]{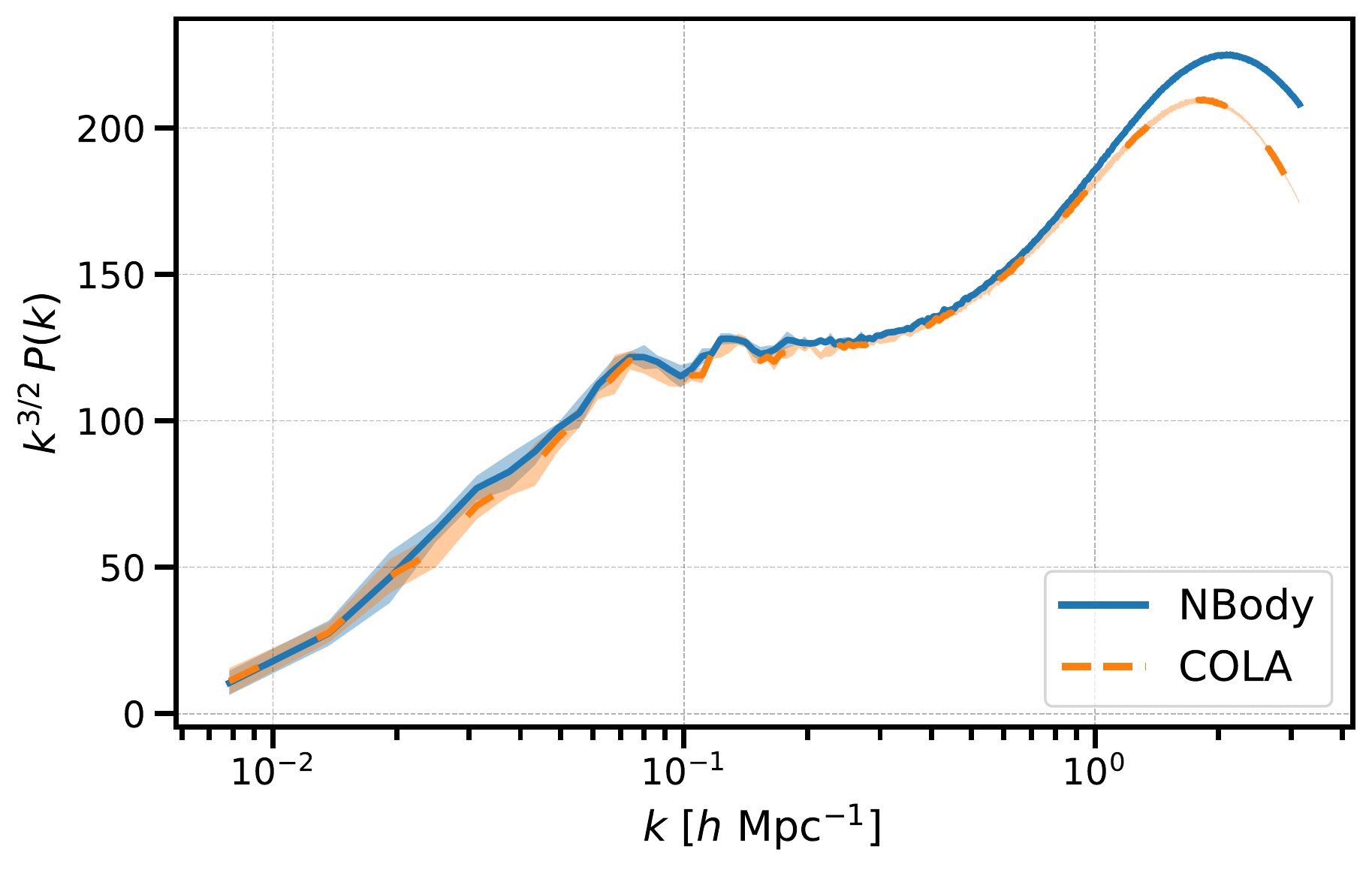}
}
\hfill
\subfloat[][GR ratio]{
\includegraphics[width=.48\textwidth]{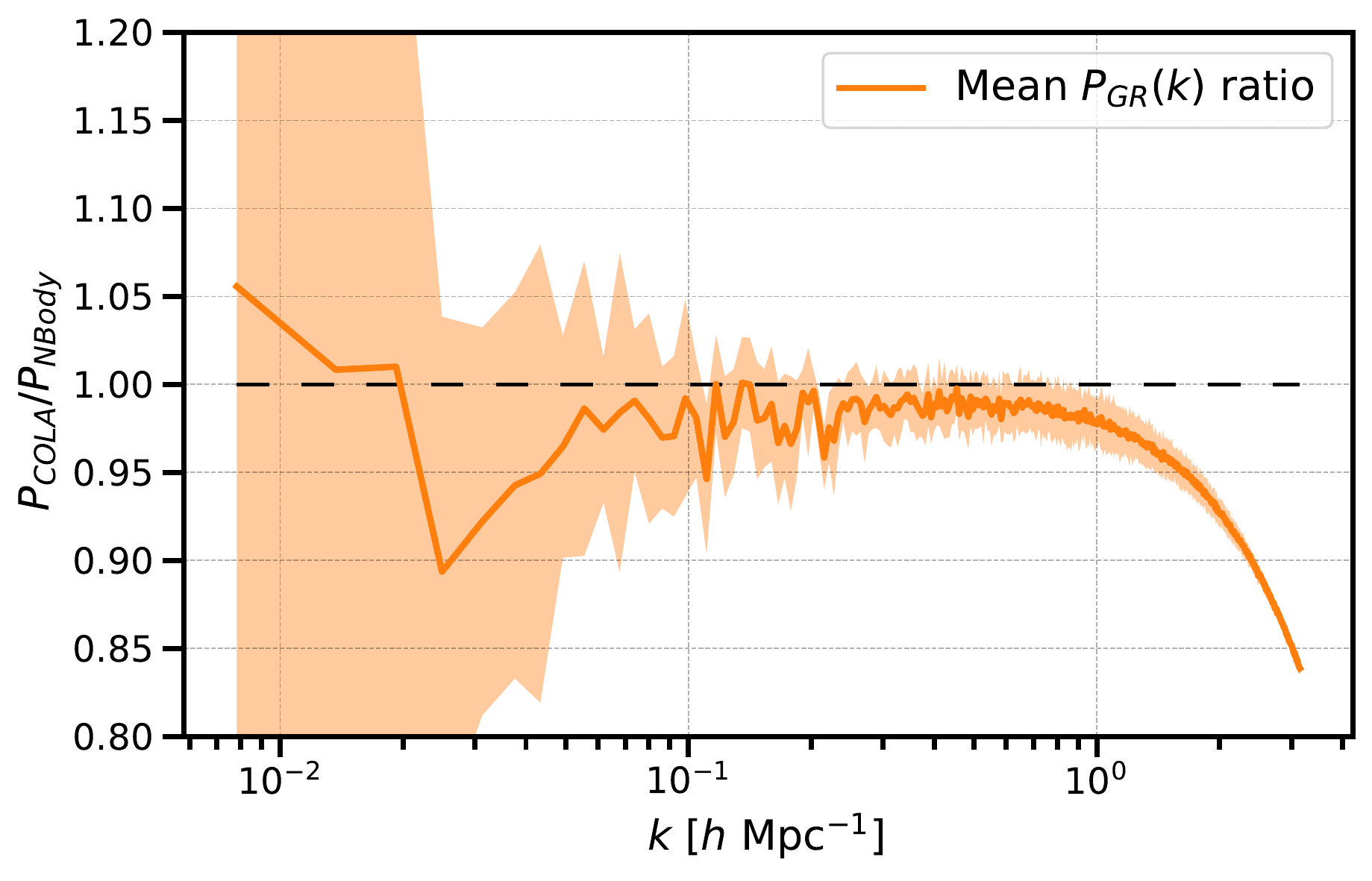}
}
\vfill
\subfloat[][F5 boost factor]{
\includegraphics[width=.48\textwidth]{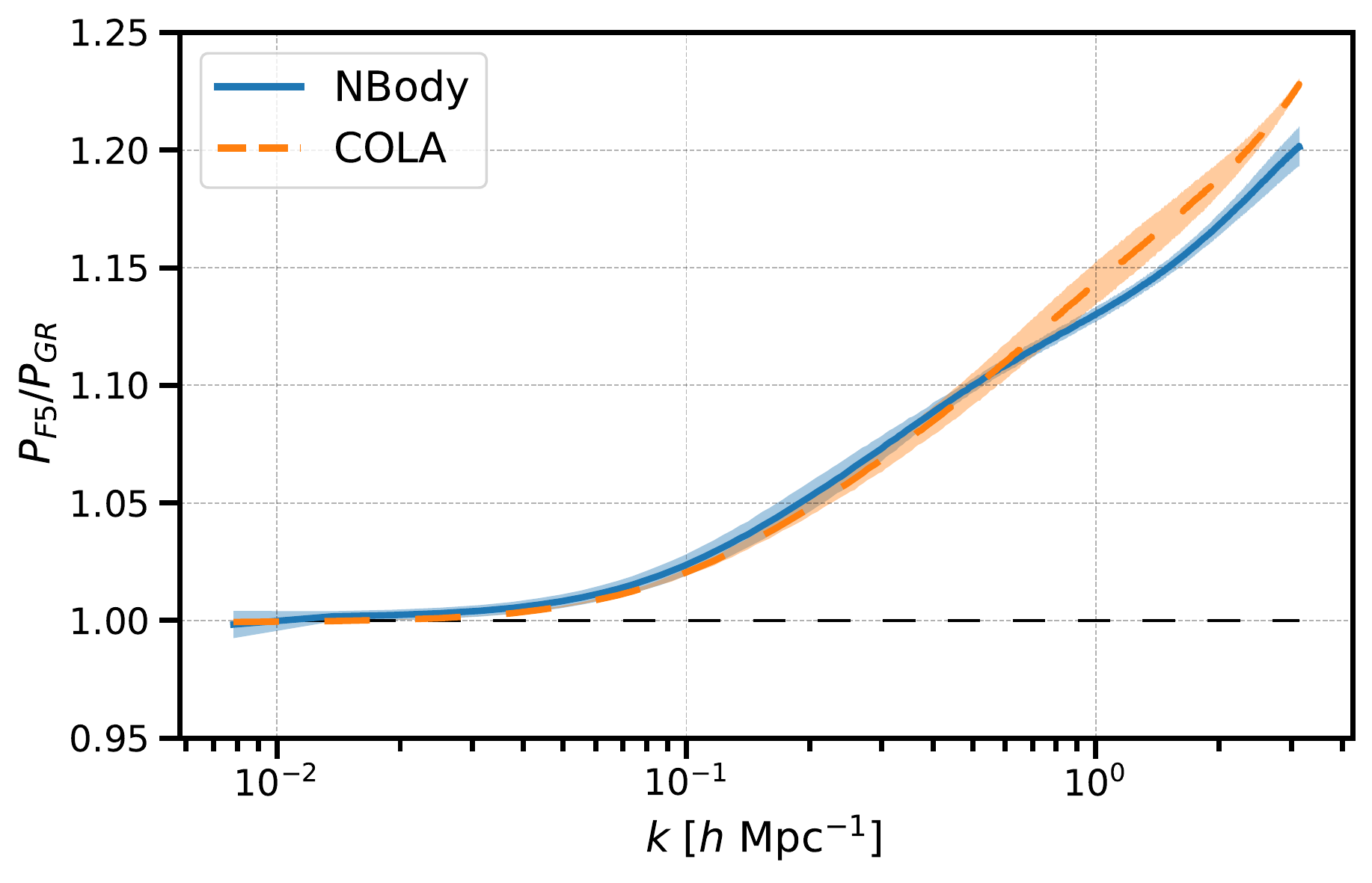}
}
\hfill
\subfloat[][N1 boost factor]{
\includegraphics[width=.48\textwidth]{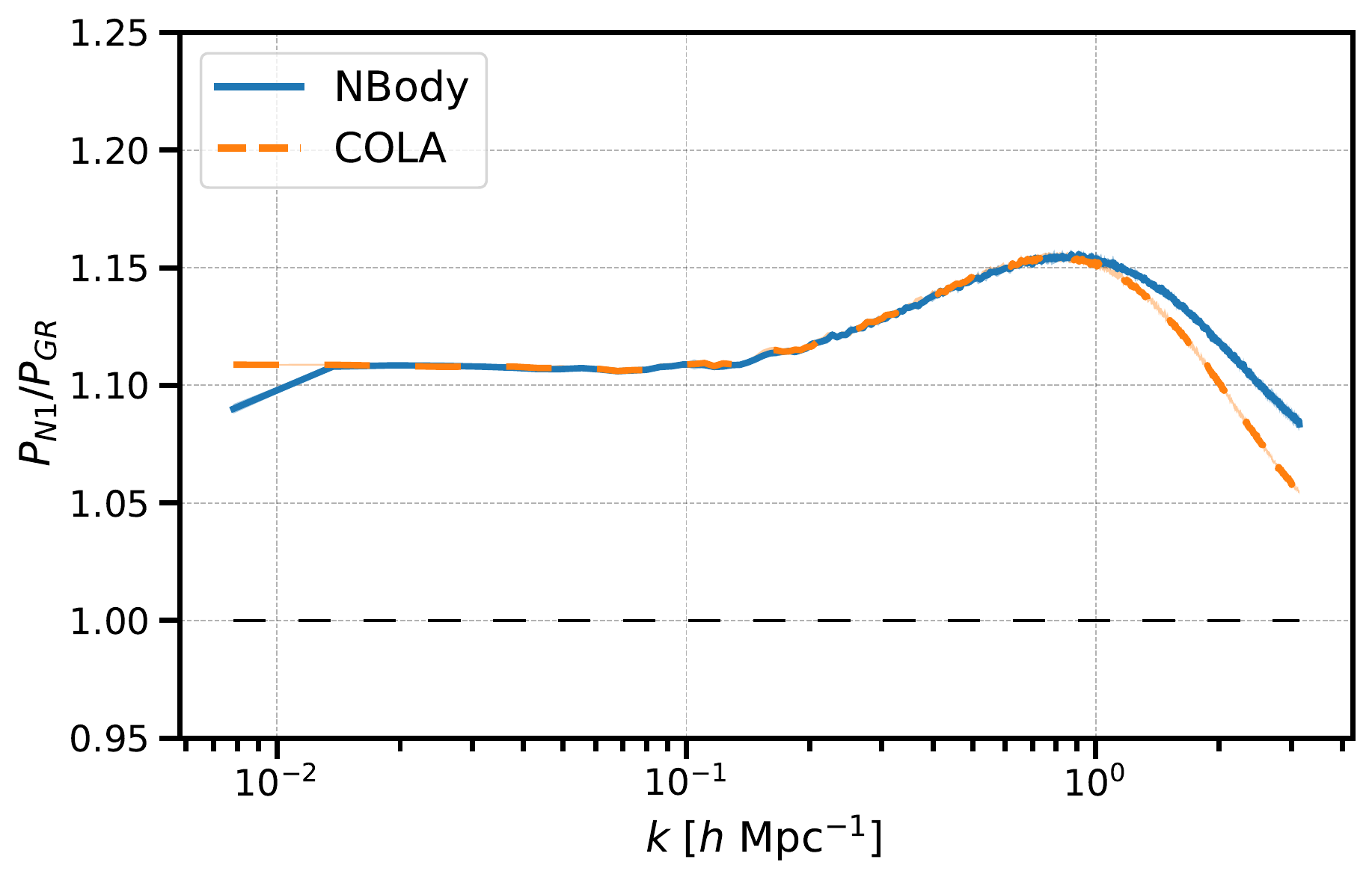}
}
\caption{\label{fig:DM_Pk} Dark Matter power spectrum comparison between COLA (orange dashed lines) and {\it N}-body (blue solid lines) obtained by taking the average over 5 realisations. The shaded regions represent the standard deviation over the 5 realisations. The power spectrum in GR (top panels) and the boost-factors in F5 (bottom left) and N1 (bottom right) show agreement within the variance up to $k \sim 1 \hompc$.}
\end{figure}

\begin{figure}[tbp]
\centering 
\subfloat[][GR]{
\includegraphics[width=.48\textwidth,clip]{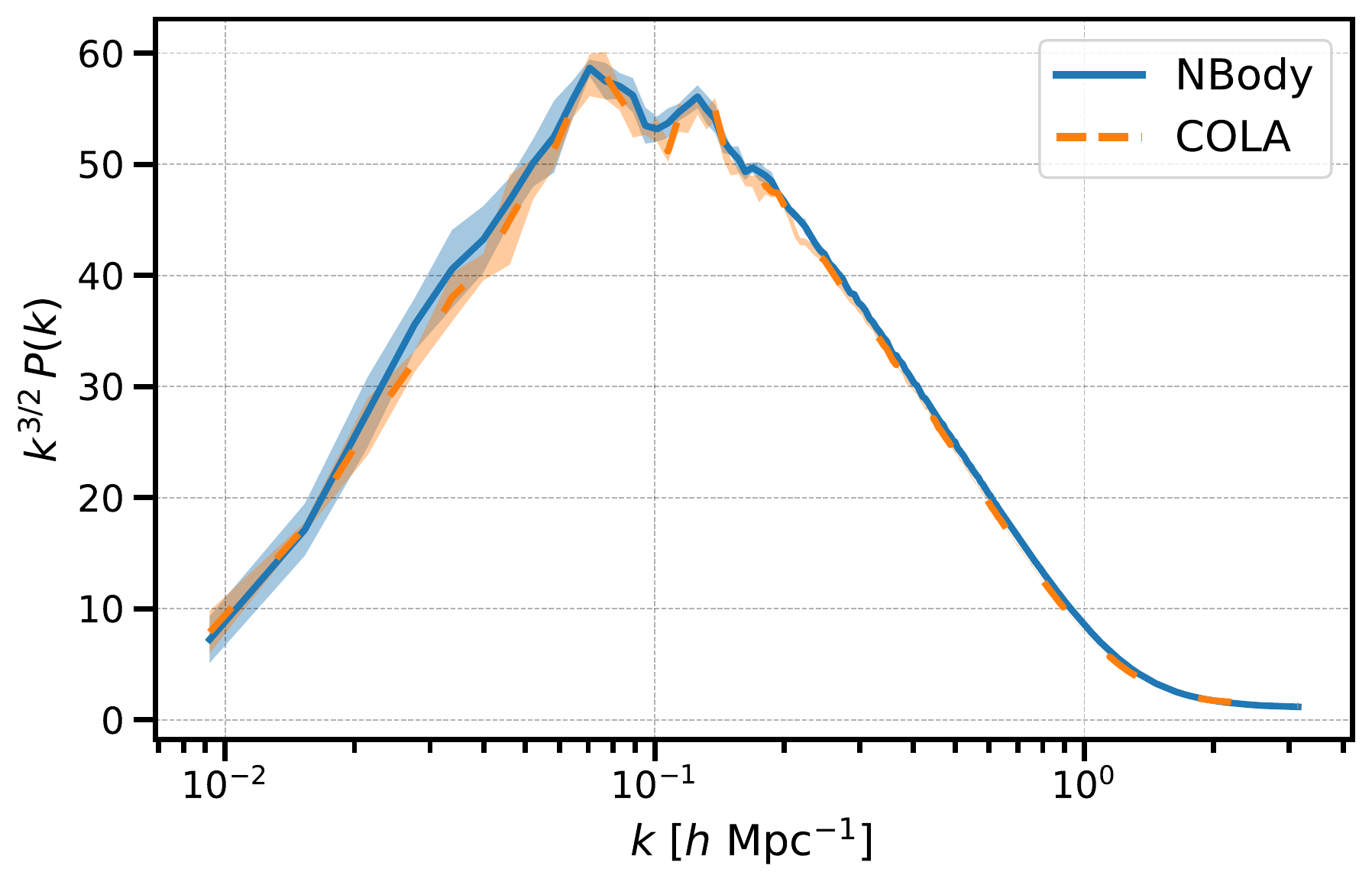}
}
\hfill
\subfloat[][GR ratio]{
\includegraphics[width=.48\textwidth]{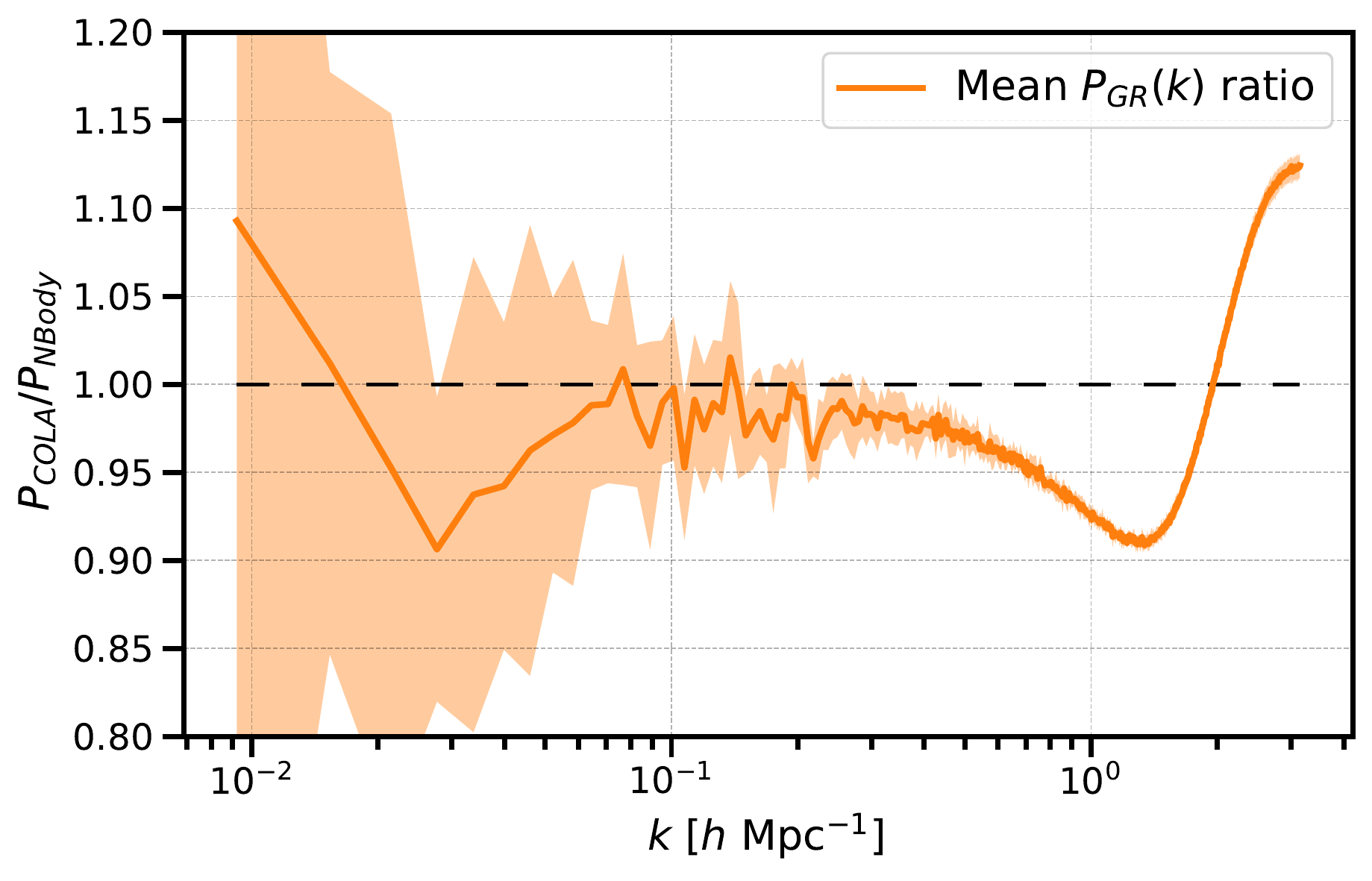}
}
\vfill
\subfloat[][F5 boost factor]{
\includegraphics[width=.48\textwidth]{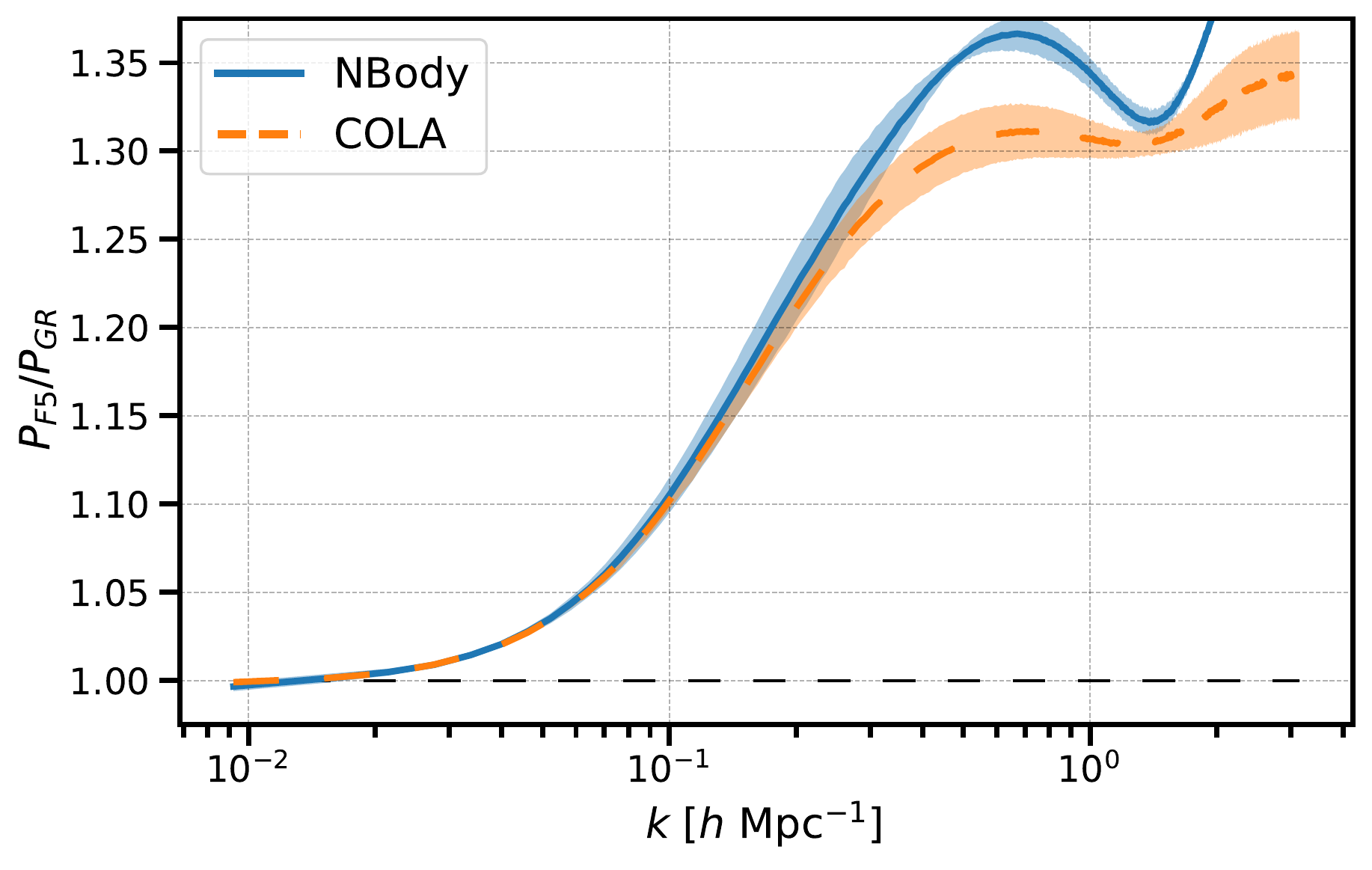}
}
\hfill
\subfloat[][N1 boost factor]{
\includegraphics[width=.48\textwidth]{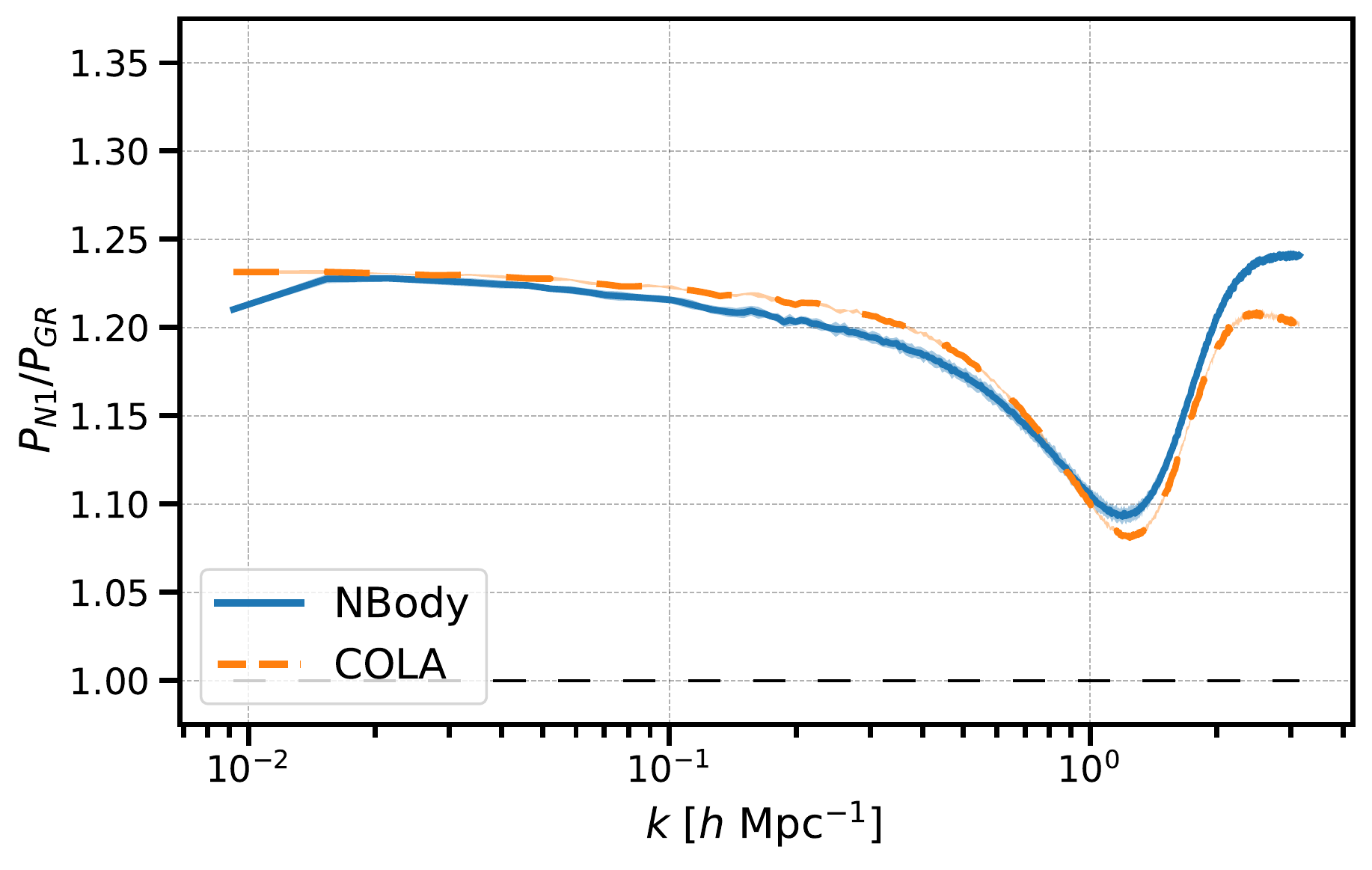}
}
\caption{\label{fig:DM_DTFE_Pk} Dark Matter velocity divergence power spectrum comparison between COLA (orange dashed lines) and {\it N}-body (blue solid lines) obtained by taking the average over 5 realisations. The shaded regions represent the standard deviation over the 5 realisations. The power spectrum in GR (top panels) is in agreement within $3\%$ up to $k \sim 0.4 \hompc$ where COLA starts to lose power. The boost-factor in F5 (bottom left) and N1 (bottom right) show better than $5\%$ accuracy even deeper in the non-linear regime, up to $k \sim 2 \hompc$.}
\end{figure}

We use the public codes PowerI4\footnote{\url{https://github.com/sefusatti/PowerI4}} and DTFE \cite{Cautun:2011gf} to compute and compare the density and velocity-divergence power spectra in the \textcode{elephant} and \textcode{piter} simulations. In Figures~\ref{fig:DM_Pk} and \ref{fig:DM_DTFE_Pk} we compare the clustering signals in GR (top left panel), the ratio between COLA and {\it N}-body (top right panel) and the enhancement in MG with respect to GR, which we refer as the boost-factor (bottom panels, F5 on the left and N1 on the right). The measurements correspond to the mean over 5 realisations and the shaded regions display the standard deviation (just to have a rough estimate of the errors to guide the eye). The errors are smaller in the boost-factor plots because the noise cancels out: the MG and the GR simulations were started from the same initial conditions and therefore have a very similar cosmic variance at late redshifts that vanishes in ratios. 
We find better than $3\%$ agreement between COLA and {\it N}-body for the density power spectra up to modes of $\sim 1 \hompc$ in GR while the boost-factors in F5 and N1 show agreement within the variance up to $k \sim 1 \hompc$. This shows that COLA simulations capture the MG effects very accurately even with the use of approximations for screening. The agreement in the velocity divergence power spectra is within $3\%$ up to $k \sim 0.4 \hompc$ in GR and for the boost factor in F5 while the boost factor in N1 agrees within $2\%$ up to $k \sim 2 \hompc$. In GR, after losing power at $k \sim 1 \hompc$, the velocity divergence power spectrum in COLA shows an enhancement compared with {\it N}-body at very high $k$. This is because COLA has less random motion at small scales that washes out the signal in {\it N}-body more than in COLA.
The better agreement in the density power spectra is expected because the velocity divergence power spectrum is more affected by non-linear physics.

From the boost-factors, we can already appreciate the intrinsic difference between F5 and N1: while F5 has a scale-dependent enhancement, N1 is almost scale-invariant across most scales. This behavior can be explained by comparing the respective Poisson equations: $G_{\rm eff}$ is a function of $k$ in Eq.~\eqref{fR_Poisson} for $f(R)$ gravity while it is scale-independent in  Eq.~\eqref{DGP_Poisson} for DGP.  

\section{Halos}
\label{sec:Halos}
Halos can be found by means of specific algorithms often referred to as halo-finders. Amongst the several halo-finders proposed in literature we focus on two popular codes: \textcode{rockstar} \cite{Behroozi13} and the Friends-of-Friends (FoF) finder \cite{Davis85} (as implemented in \texttt{nbodykit}\footnote{\url{https://nbodykit.readthedocs.io}}). \textcode{rockstar} uses a 6D phase-space FoF finder and is capable of measuring many halo properties like the spherical over-density mass, the concentration parameter and the velocity dispersion just to name a few. 
The FoF finder instead is purely based on particles positions.

\subsection{\textcode{rockstar} halos}
\begin{figure}
        \centering 
        \subfloat[][GR]{
        \includegraphics[width=.48\textwidth,clip]{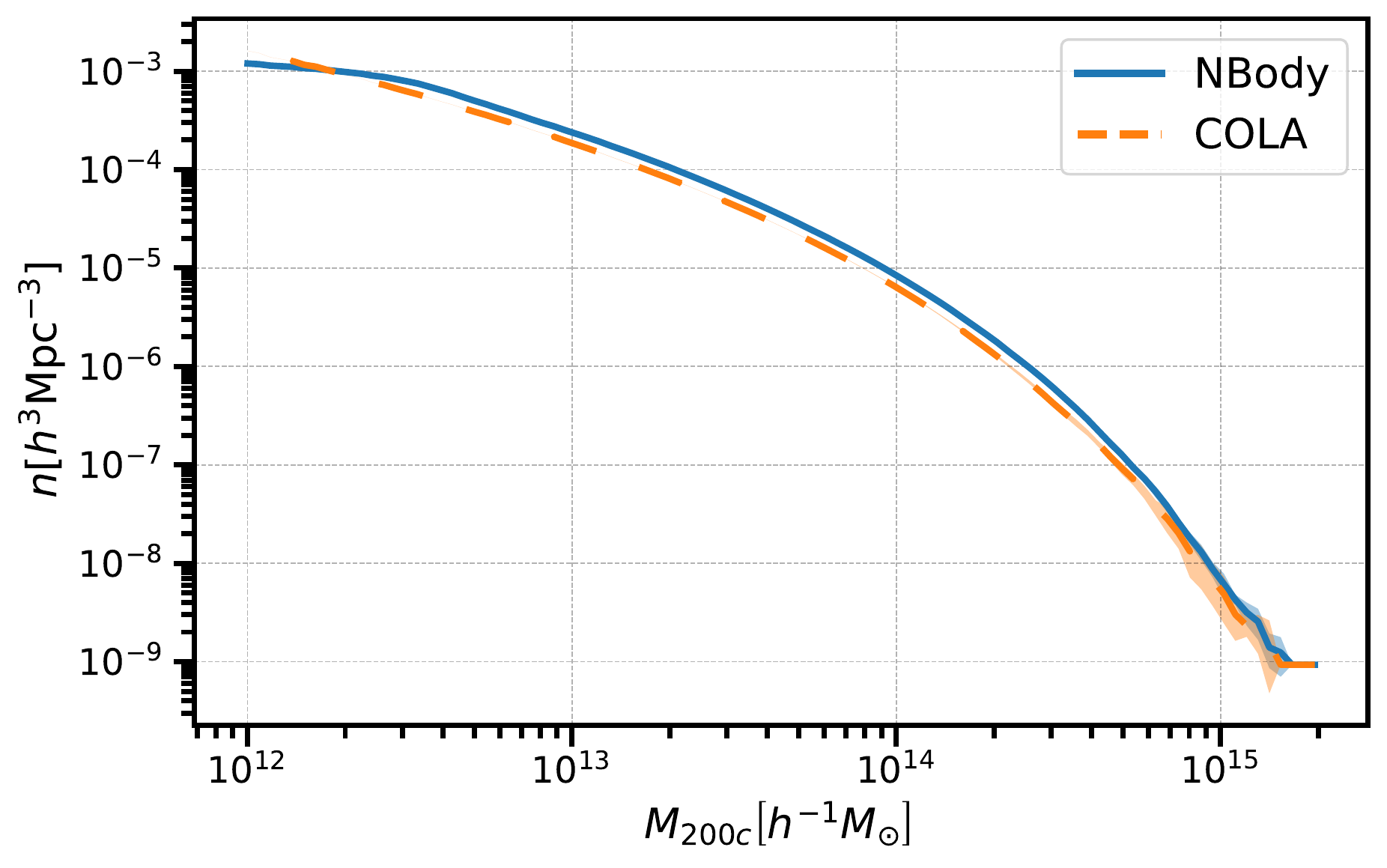}
        }
        \hfill
        \subfloat[][GR ratio]{
        \includegraphics[width=.48\textwidth]{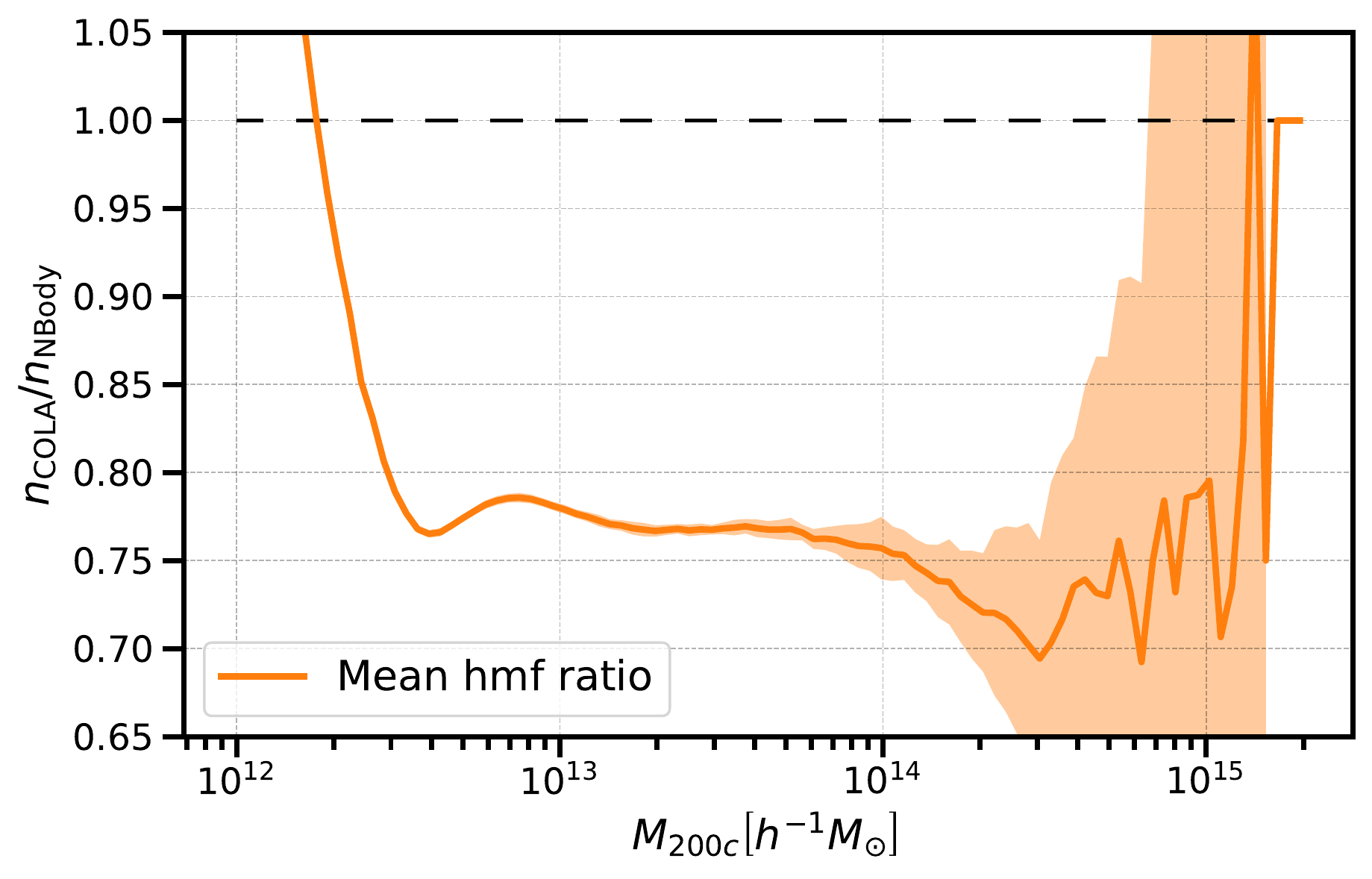}
        }
    \caption{\textcode{rockstar} halo mass function comparison in GR. The hmf in COLA and {\it N}-body (left panel) and their ratio (right panel) show $\sim 25 \%$ discrepancy for masses greater than $10^{13} \Msun$.}
    \label{fig:hmf_Rock}
\end{figure}

\begin{figure}
        \centering 
        \subfloat[][GR]{
        \includegraphics[width=.48\textwidth,clip]{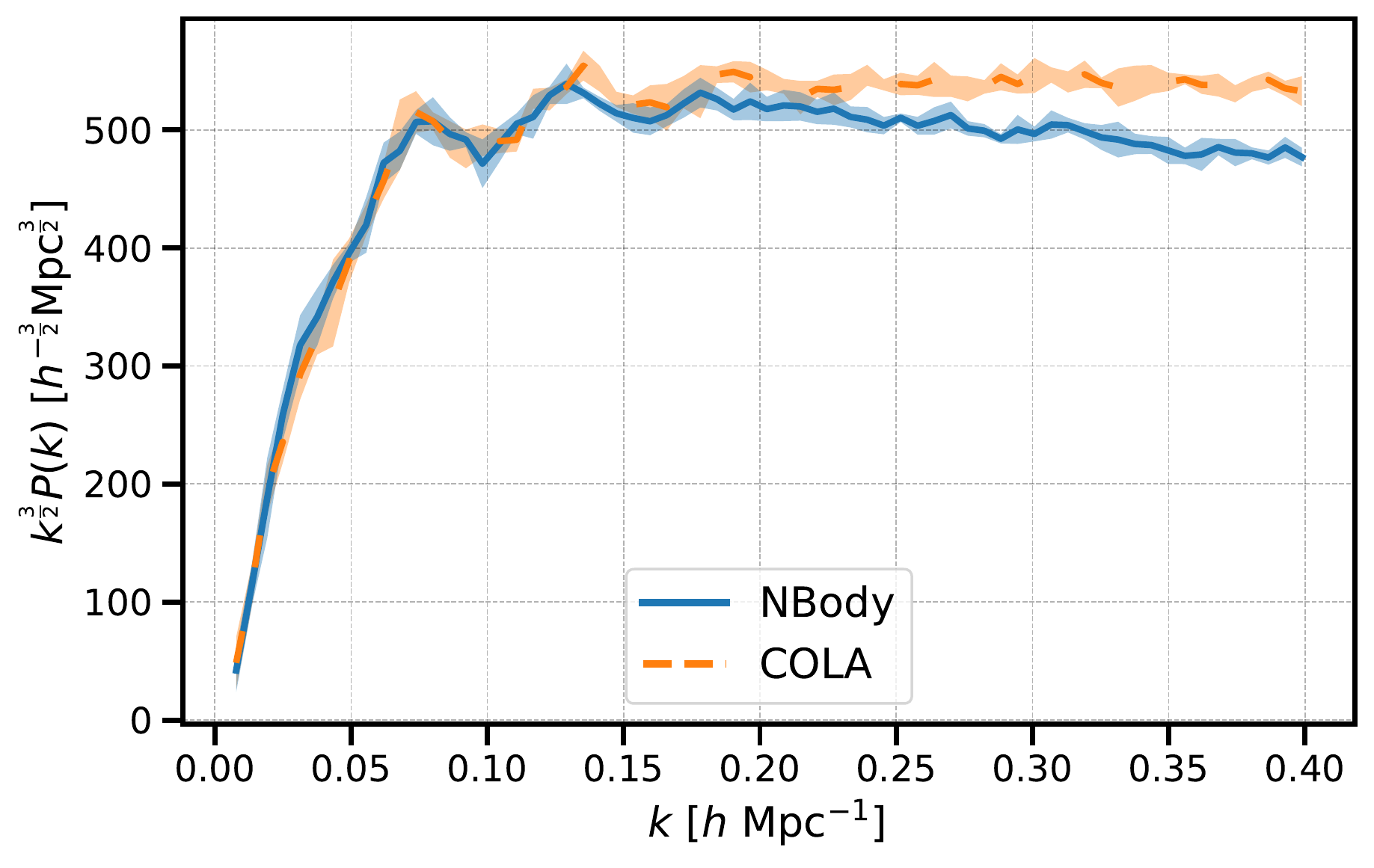}
        }
        \hfill
        \subfloat[][GR ratio]{
        \includegraphics[width=.48\textwidth]{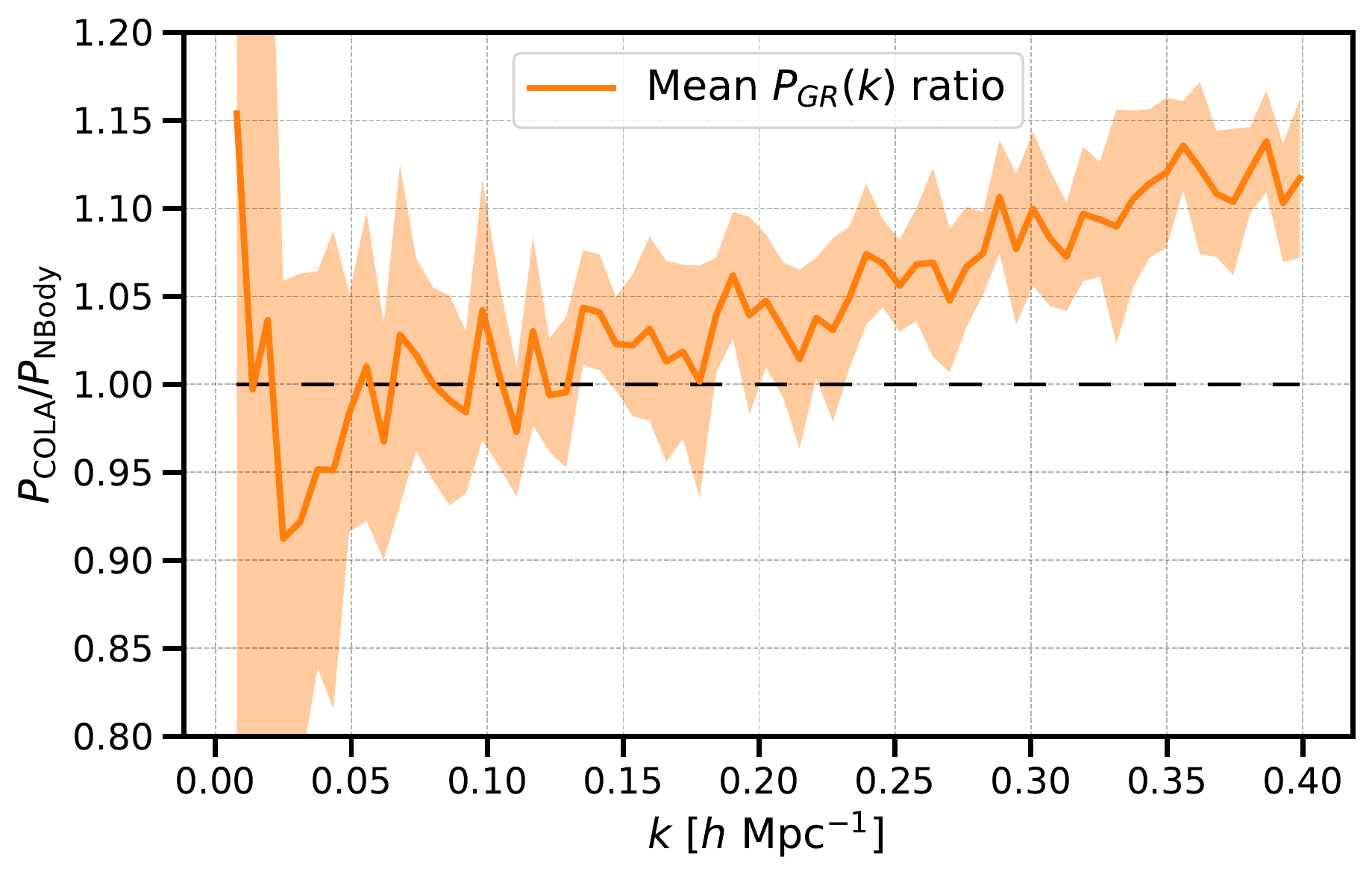}
        }
    \caption{\textcode{rockstar} power spectrum comparison between the power spectra of COLA and {\it N}-body in GR. The discrepancy, which is already clear in the left panel, is shown to reach the $10\%$ with a scale-dependent behaviour in the right panel.}
    \label{fig:halo_Pk_Rock}
\end{figure}

We first produce \textcode{rockstar} halo catalogues in COLA and {\it N}-body and compare the cumulative halo mass function (hmf) and the halo power spectrum. 
We use the setting to remove unbound particles and keep only parent halos\footnote{We found that, if we do not remove unbound particles, the agreement between {\it N}-body and COLA is substantially worse for the halo power spectrum}.
The hmf is simply defined as the number density of halos with mass greater than $M$. Before measuring the halo power spectrum, we draw a halo sample defined by an abundance matching procedure, in which the target density is given by a cut at $10^{13} \Msun$ in the {\it N}-body catalogues for GR.
Figures~\ref{fig:hmf_Rock} and~\ref{fig:halo_Pk_Rock} show the comparisons of the hmf and the halo power spectrum for COLA and {\it N}-body \textcode{rockstar} catalogues respectively, both in GR. The hmf is $25\%$ off for masses greater than $10^{13} \Msun$, and the halo power spectrum ratio is well predicted in the large-scale limit but shows a scale-dependent deviation, up to $10\%$ at $k\sim 0.4 \hompc$. These differences are likely due to the lower force resolution and accuracy in recovering the velocity field in COLA simulations.

Given this result, in the rest of the paper, we will use only the FoF finder to generate halo catalogues from COLA and {\it N}-body simulations. However, we still use \textcode{rockstar} in the {\it N}-body simulations to quantify the effects of MG on the concentration and velocity dispersion of halos in F5 simulations (see Section~\ref{sec:F5_NFWtweaks}), which we then incorporate in the prescriptions that we use to populate the COLA simulations with galaxies.

\subsection{Friends-of-Friends halos}
The FoF algorithm links together particles that are separated by a distance smaller than a certain linking length $b$, expressed in units of the mean inter-particle distance. We adopt the standard value $b=0.2$. This value, which is commonly used in {\it N}-body simulations, has been shown to also be valid in COLA simulations (see \cite{Koda:2015mca,Howlett:2014opa,Izard:2015dja}), although, in some other approximate methods that use very low-resolution density fields, $b$ has been tweaked, see \cite{Manera:2012sc}.  
At this point, we are interested in converting FoF masses, $M_{\rm FoF}$, to  spherical over-density (SO) masses, $M_{\rm200c}$, which we adopt in the rest of the paper when expressing halo masses.
$M_{\rm200c}$ denotes the mass enclosed in a spherical region with density 200 times the critical density. We use the fit in \cite{Lukic:2008ds} to convert mass definitions with an accuracy of $5\%$ for most halos in the mass range $10^{12.5}-10^{15.5} \Msun$:
\begin{equation}
    \frac{M_{\rm FoF}}{M_{200 \mathrm{c}}}=\frac{a_{1}}{c_{200}^{2}}+\frac{a_{2}}{c_{200}}+a_{3}
    \label{eq:mfof_m200_conversion} \, ,
\end{equation}
where the coefficients $a_1$, $a_2$ and $a_3$ depend on the number of particles (see their Table 1), and $c_{200}$ is the concentration parameter, which we compute using the empirical formula from \cite{Dutton:2014xda}:

\begin{align}
   \log _{10} c_{200} &=a+b \log _{10}\left( \frac{M_{\rm 200 c}}{10^{12} \Msun}\right), \label{eq:concentration}\\
   a& =0.520+(0.905-0.520) \exp \left(-0.617 z^{1.21}\right), \nonumber \\ 
   b& =-0.101+0.026 z. \nonumber
\end{align}

\begin{figure}
        \centering 
        \subfloat[][Mass conversion factor]{
        \includegraphics[width=.48\textwidth,clip]{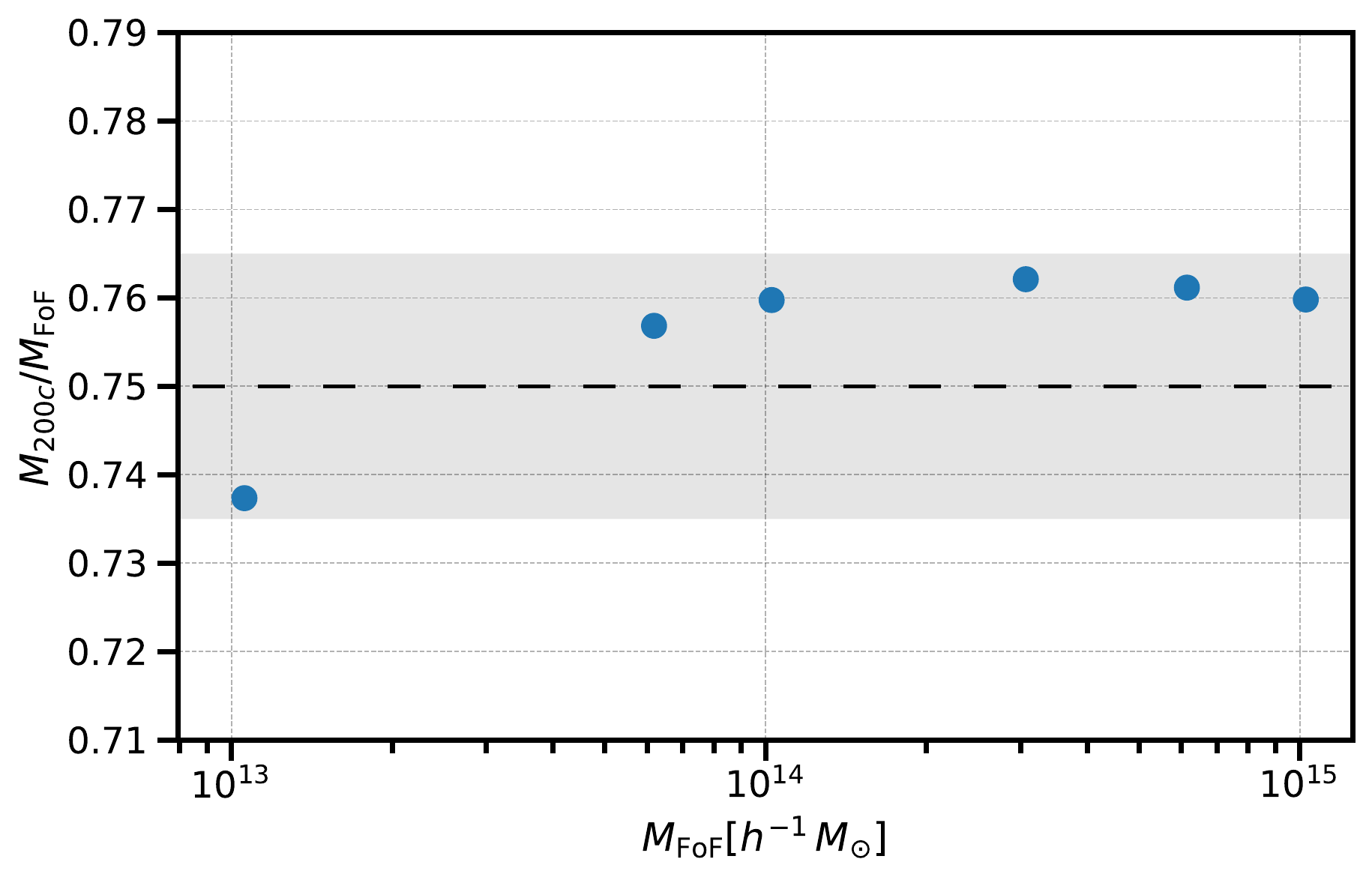}
        }
        \hfill
        \subfloat[][\textcode{rockstar} comparison]{
        \includegraphics[width=.48\textwidth]{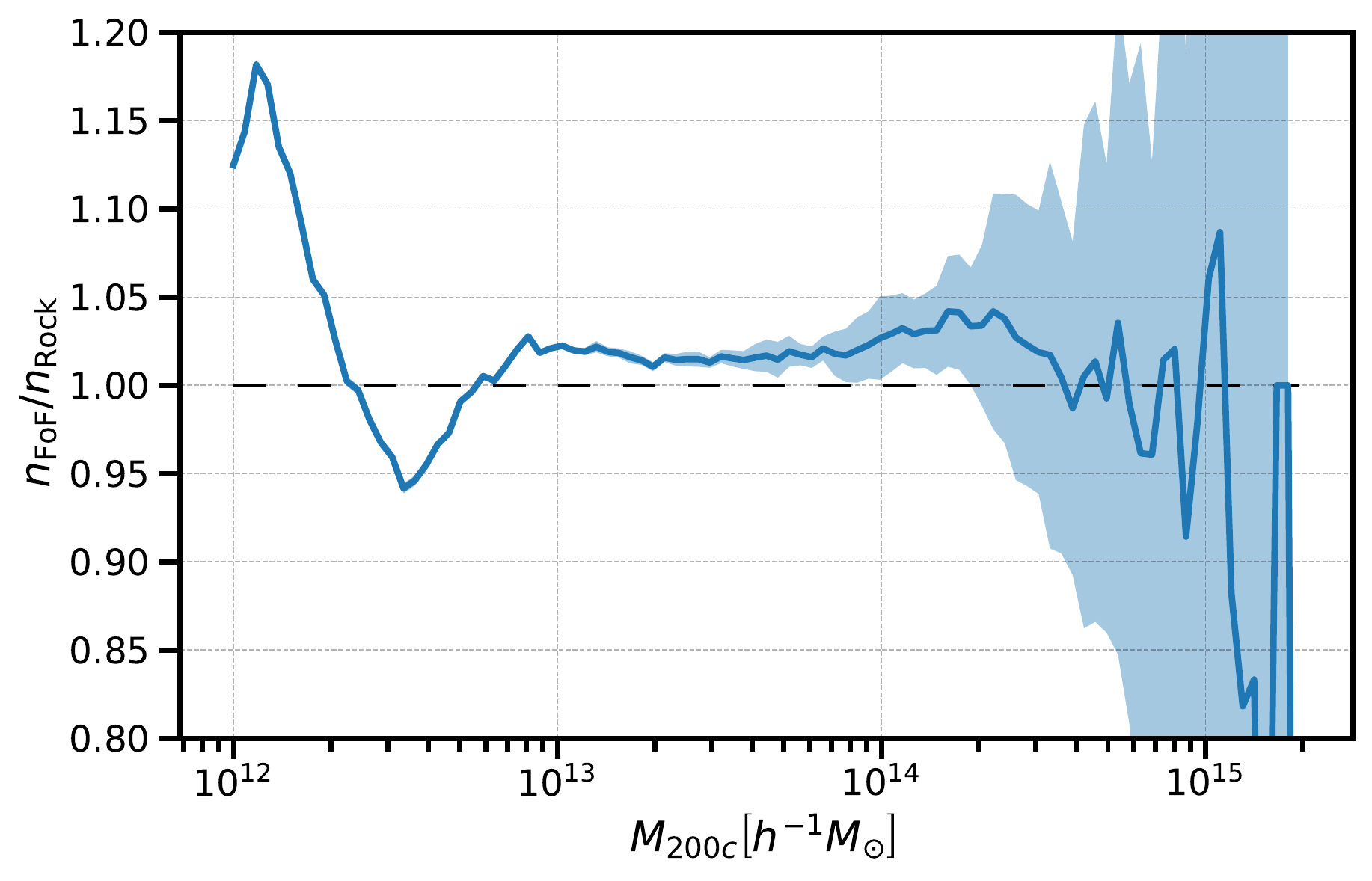}
        }
    \caption{Mass conversion between FoF and $M_{200c}$ spherical over-density masses. The conversion factor is evaluated for 6 mass values (left panel), corresponding to the SO mass of halos composed of 100, 600, 1000, 3000, 6000 and 10000 DM particles. All the six points lie within $2\%$ from the value $0.75$ which we select as a constant to convert the masses of the FoF halos to SO masses. In the right panel, we use the $N$-body simulations in GR to compare the FoF hmf resulting after this conversion with the hmf of \textcode{rockstar} catalogues. The ratio between the two (right panel) shows a better than $3\%$ agreement for masses above $10^{13} \Msun$.}
    \label{fig:mass_conversion}
\end{figure}

The left panel in Figure~\ref{fig:mass_conversion} shows the conversion factor of Eq.~\eqref{eq:mfof_m200_conversion} (for a set of values corresponding to SO masses of 100, 600, 1000, 3000, 6000 and 10000 particles). There is very little dependence on halo mass, with all the values within $2\%$ of 0.75. Hence we use the constant value 0.75 to convert the FoF masses to SO masses. 
To validate this mass conversion, we compare the mass function for the FoF halo catalogues $n_{\rm FoF}$ with rescaled $M_{200 \mathrm{c}}=0.75 M_{\rm FoF}$ to the mass function as measured by \textcode{rockstar} $n_{\rm Rock}$ in terms of $M_{\rm 200c}$. The right panel of Figure~\ref{fig:mass_conversion} shows this comparison for {\it N}-body GR halo catalogues, and the agreement of these two mass functions is within $3\%$ for masses grater than $10^{13} \Msun$. 
We thus produce halo catalogues for both the {\it N}-body and COLA simulations using the FoF method and assign the over-density mass $M_{200 \mathrm{c}}=0.75 M_{\rm FoF}$.

\begin{figure}
\centering 
\subfloat[][GR]{
\includegraphics[width=.48\textwidth,clip]{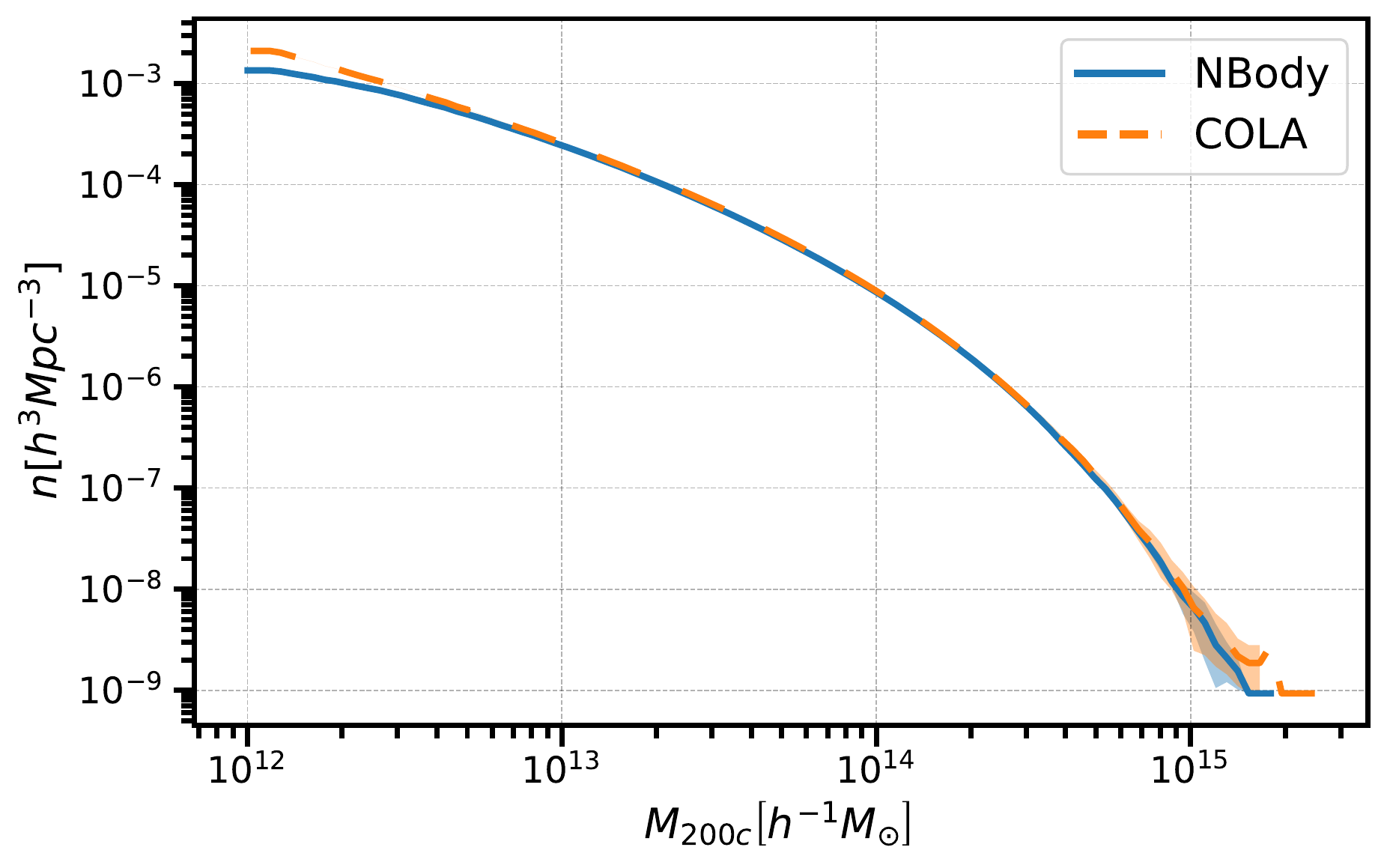}
}
\hfill
\subfloat[][GR ratio]{
\includegraphics[width=.48\textwidth]{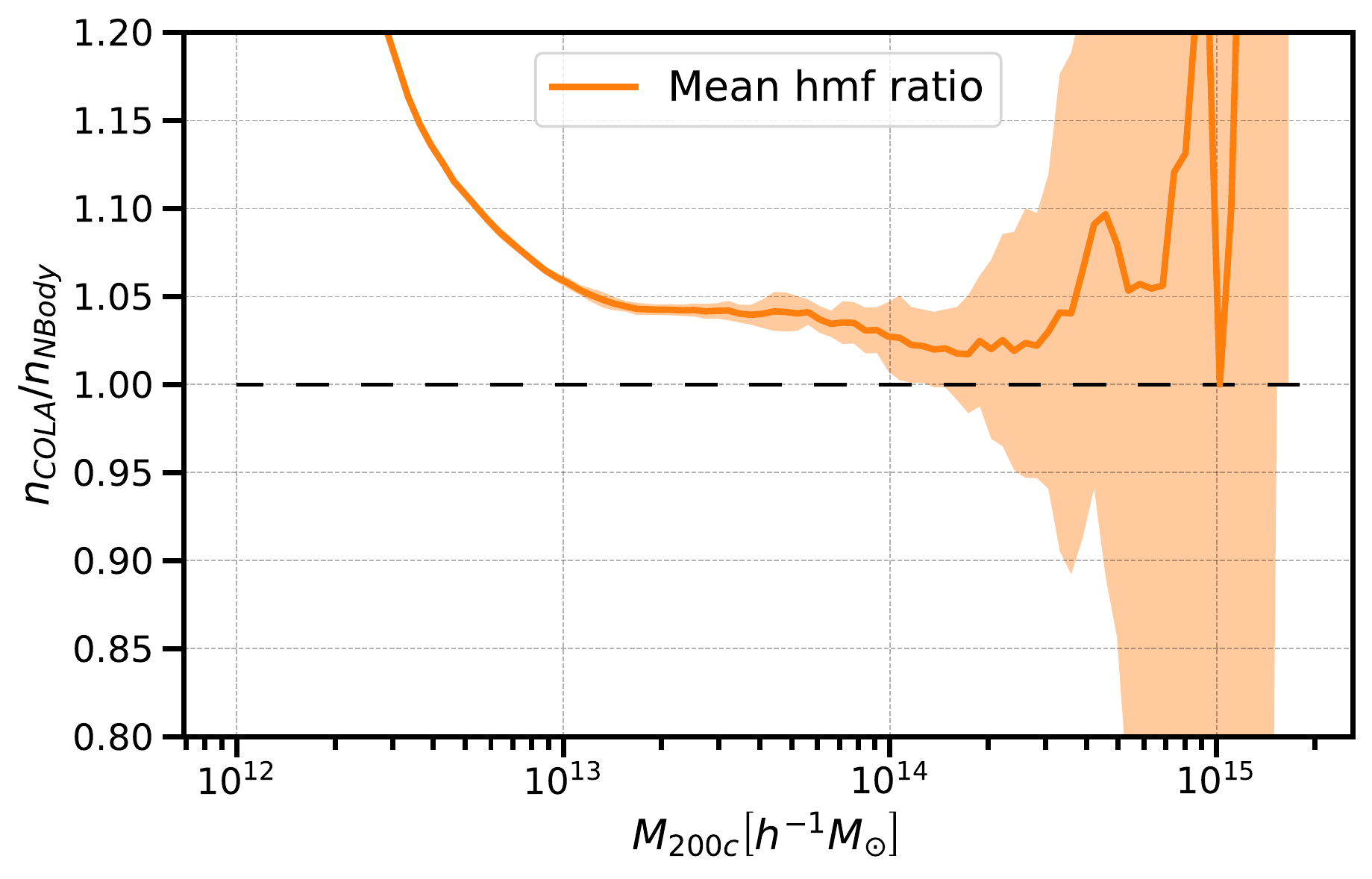}
}
\vfill
\subfloat[][F5 boost factor]{
\includegraphics[width=.48\textwidth]{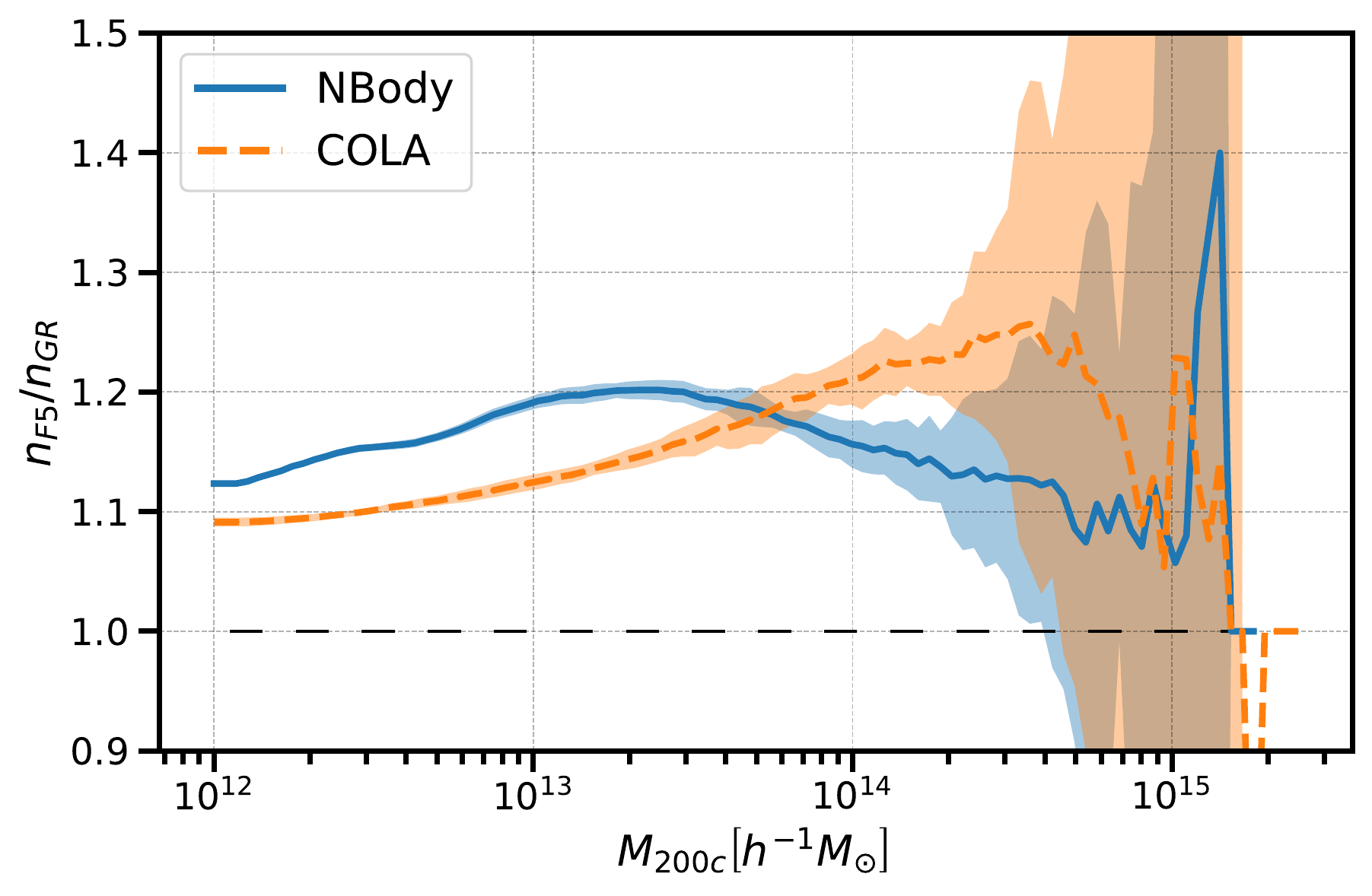}
}
\hfill
\subfloat[][N1 boost factor]{
\includegraphics[width=.48\textwidth]{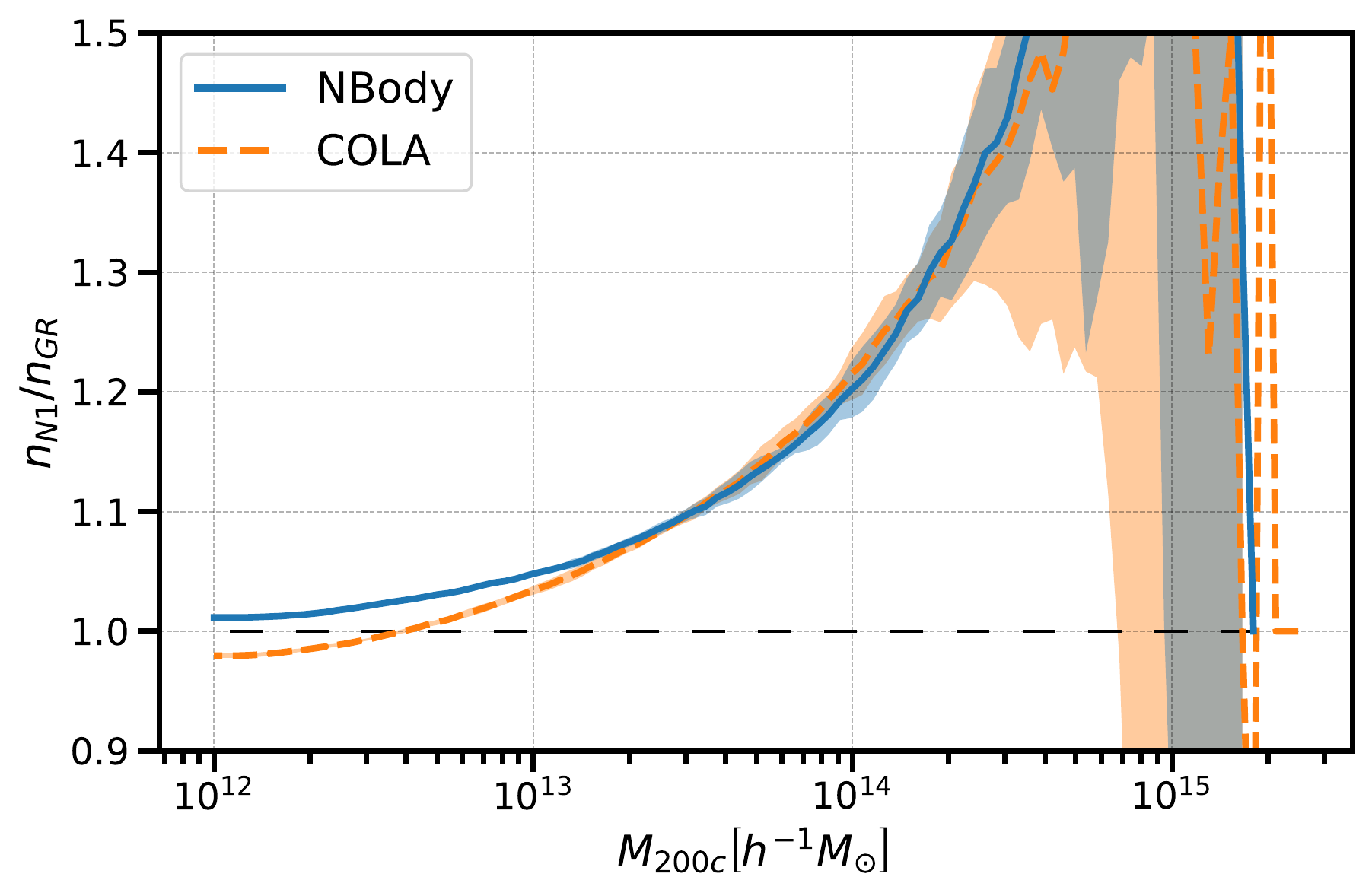}
}
\caption{Halo mass function comparison between COLA (orange dashed lines) and {\it N}-body (blue solid lines) obtained by taking the average over 5 realisations. The shaded regions represent the standard deviation over the 5 realisations. The hmf in GR (top panels) show a $\sim 5 \%$ accuracy for masses above $10^{13} \Msun$. The boost factor in N1 (bottom right) in COLA agree with that in {\it N}-body within the $5 \%$ at all masses, while the boost factor in F5 (bottom left) features an agreement within the $10 \%$.}
\label{fig:Halo_hmf}
\end{figure}

\begin{figure}
\centering 
\subfloat[][GR]{
\includegraphics[width=.48\textwidth,clip]{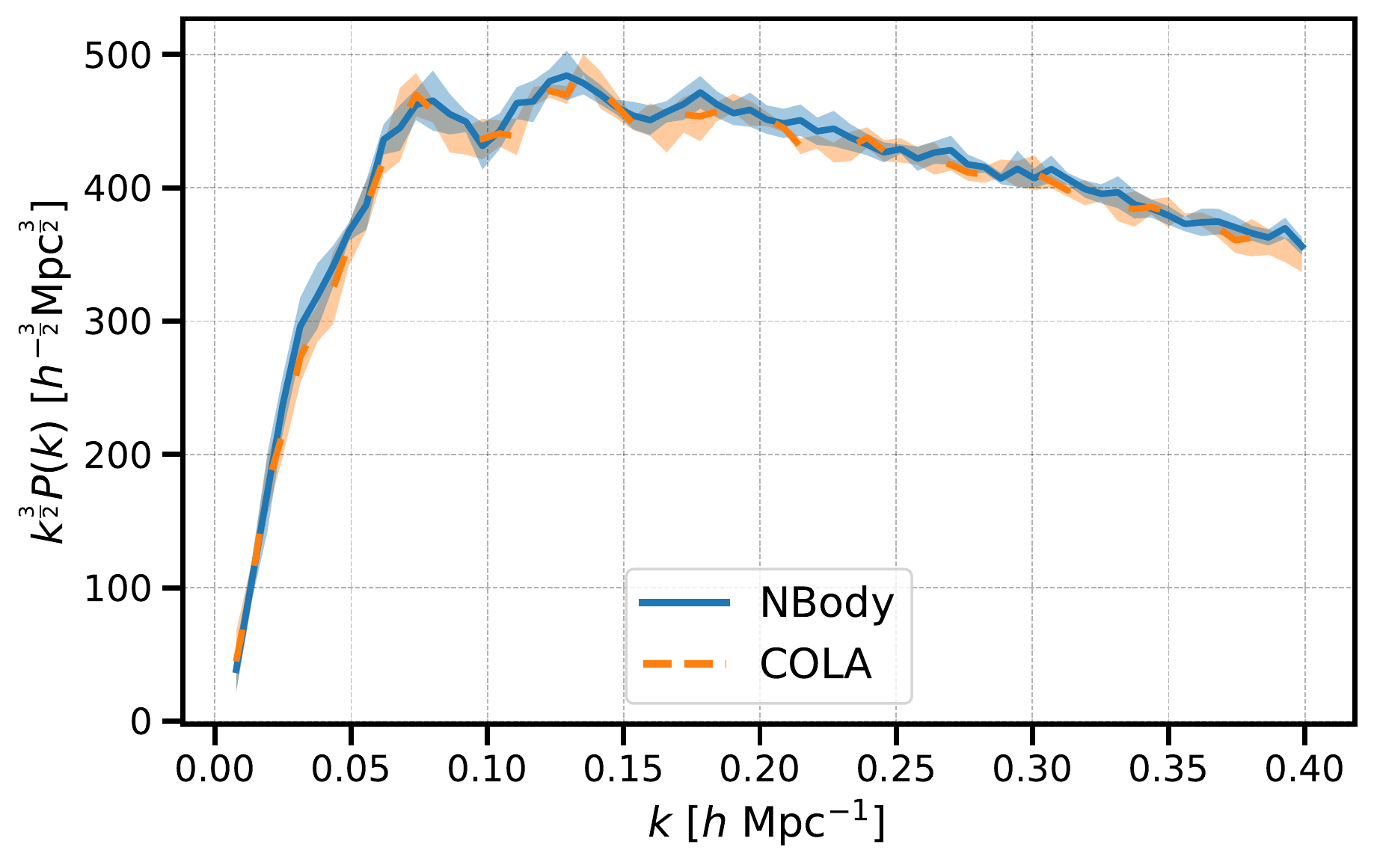}
}
\hfill
\subfloat[][GR ratio]{
\includegraphics[width=.48\textwidth]{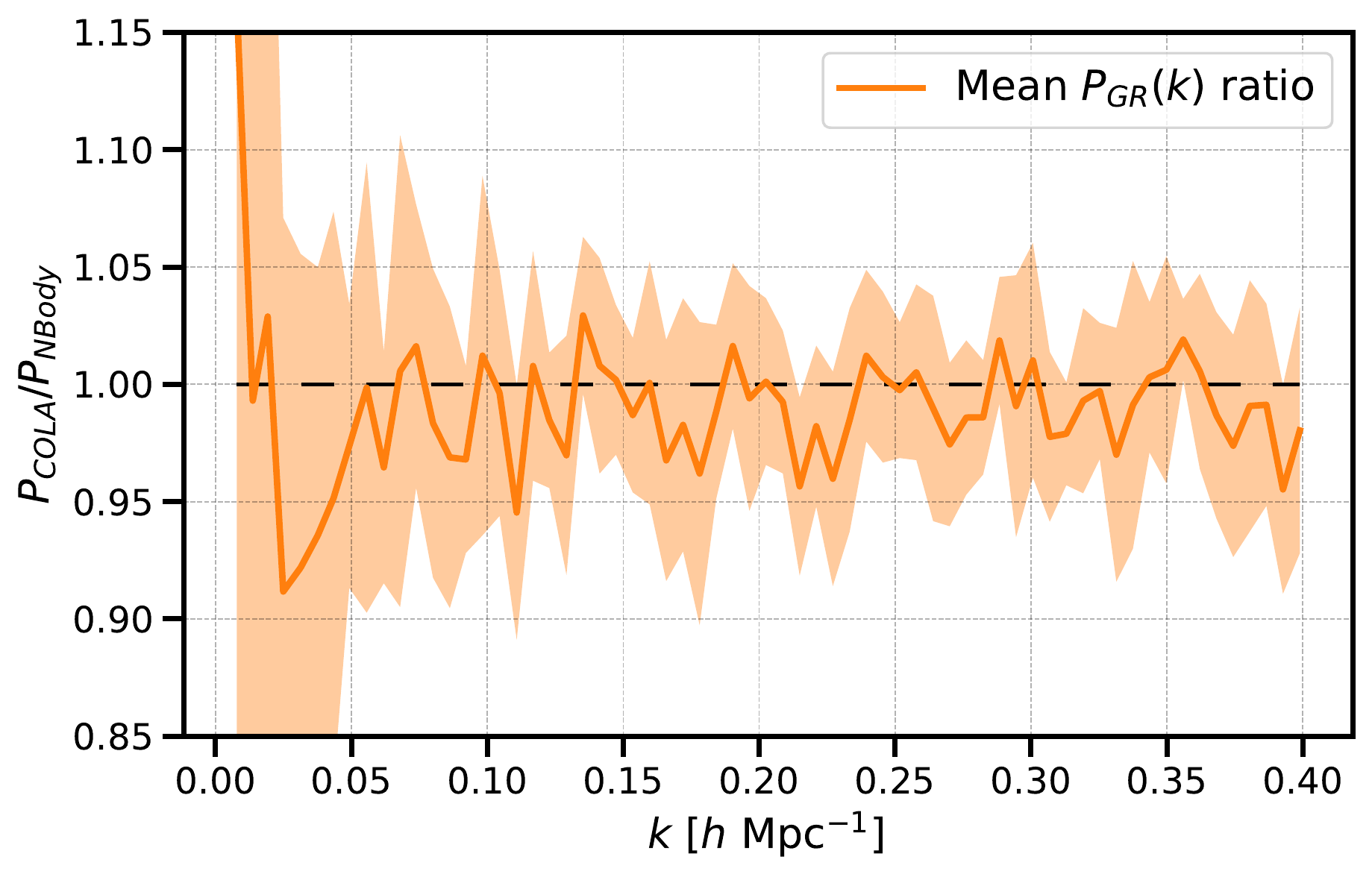}
}
\vfill
\subfloat[][F5 boost factor]{
\includegraphics[width=.48\textwidth]{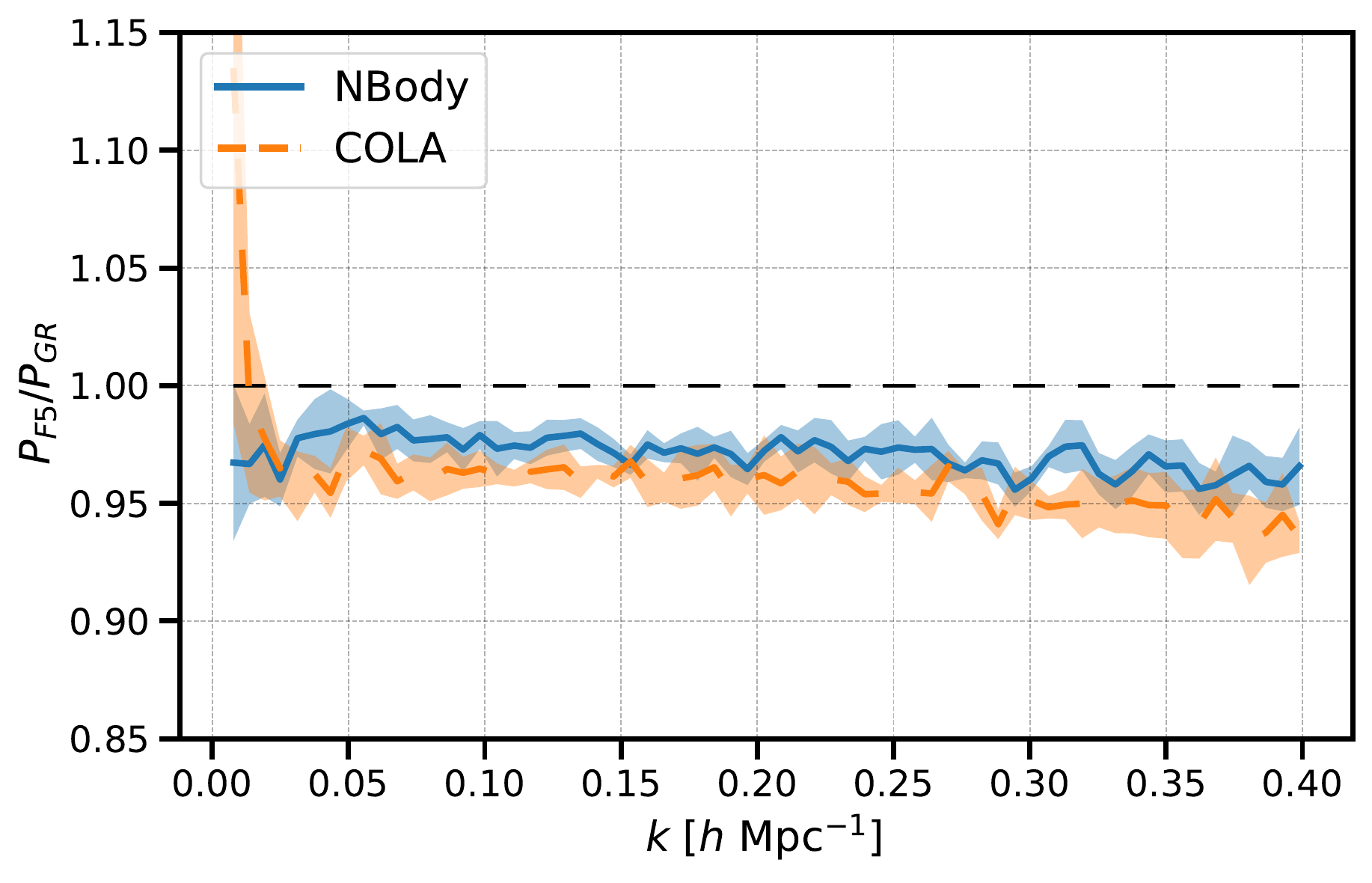}
}
\hfill
\subfloat[][N1 boost factor]{
\includegraphics[width=.48\textwidth]{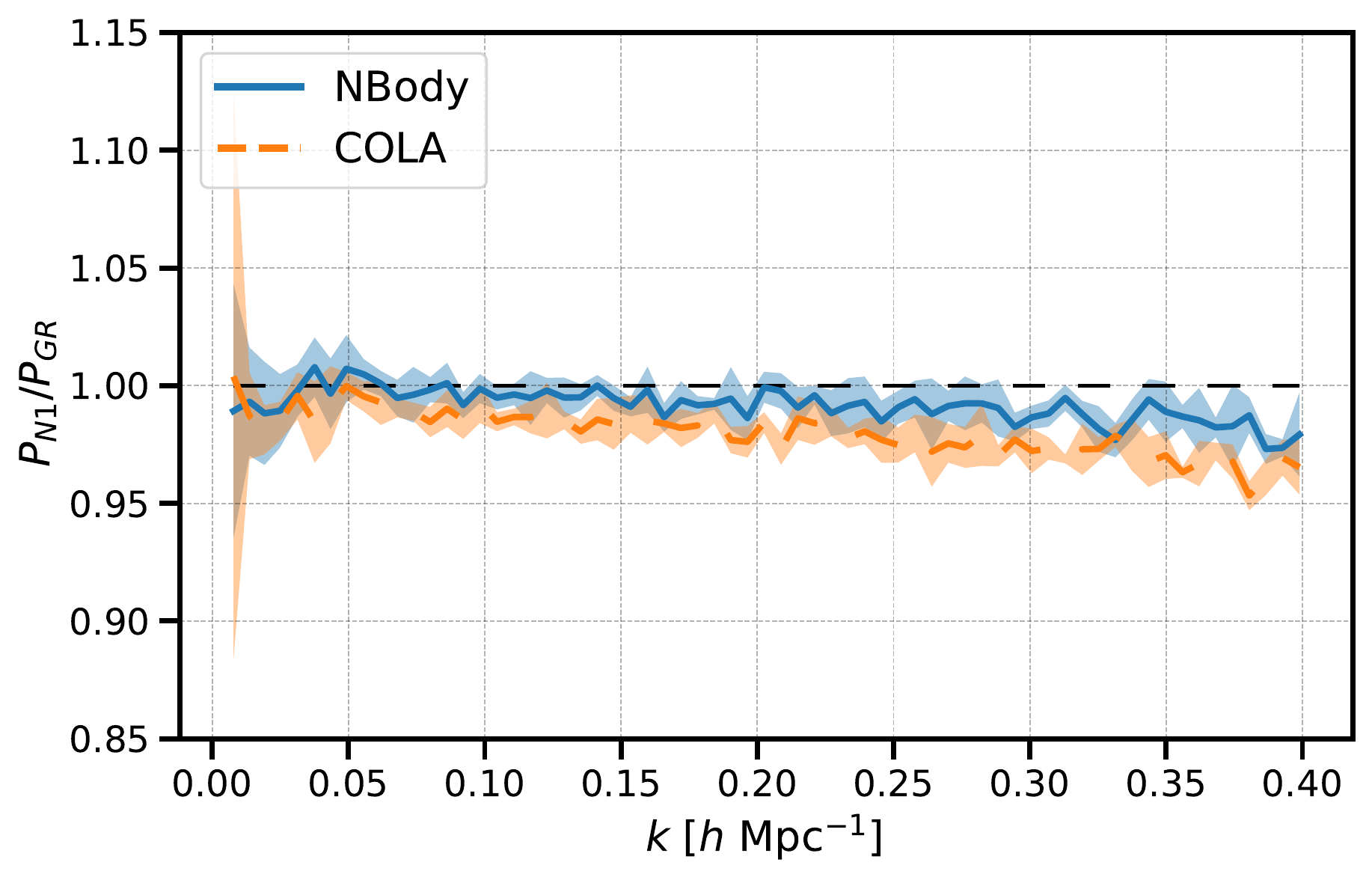}
}
\caption{Halo power spectrum comparison between COLA (orange dashed lines) and {\it N}-body (blue solid lines) obtained by taking the average over 5 realisations. The shaded regions represent the standard deviation over the 5 realisations. The halo power spectrum in GR (top panels) and the boost factors in F5 and N1 (bottom panels) show agreement within the variance up to $k \sim 0.4 \hompc$.}
\label{fig:Halo_Pk}
\end{figure}

To validate the COLA halo catalogues we compare against the {\it N}-body halo catalogues for the following two summary statistics: the halo mass function and the halo power spectrum, which are shown in Figure~\ref{fig:Halo_hmf} and Figure~\ref{fig:Halo_Pk} respectively. 
For the halo power spectrum, we draw a halo sample defined by an abundance matching procedure, in which the target density is given by a cut at $10^{13} \Msun$ in the {\it N}-body catalogues for GR. We see that COLA reproduces the {\it N}-body mass function with an accuracy better than $5\%$ for halos with more than 130 particles (or $10^{13}\Msun$) in GR, while the hmf boost-factors agree to better than $10\%$ for all the halo masses. In regards to the halo power spectrum, Figure~\ref{fig:Halo_Pk} shows that COLA and {\it N}-body are in agreement within the variance both for GR and for the boost-factors. 

\subsection{Internal structure of halos}
\label{sec:F5_NFWtweaks}
Certain halo properties are needed to create galaxy mock catalogues with the HOD model discussed in the next section; the phase space distribution of satellite galaxies is based on the NFW profile, which in turn is completely determined by the concentration parameter and the mass of the halo. 

COLA is purposely designed to accurately determine basic halo quantities (positions, velocities and masses) but not the internal halo structure, which would require simulations with a much higher computational cost. Because of that, in the pipeline developed for this paper, we use several prescriptions calibrated from high resolution {\it N}-body simulations in $\Lambda$CDM to assign certain halo properties in COLA (see Eq.~\eqref{eq:mfof_m200_conversion} and Eq.~\eqref{eq:concentration}). Furthermore, MG may slightly modify some of these relationships (e.g. the concentration parameter and the velocity dispersion). We use the \textcode{elephant} simulations to estimate these effects, which we then introduce in the prescription to create galaxy mock catalogues in the MG models.

In F5 the fifth force is unscreened for halos of mass lower than $10^{14} \Msun$, affecting the radial density profile and the virial equilibrium for such halos. We account for this by implementing corrections to the concentration parameter and the velocity dispersion. 
Detailed studies on the effect of the fifth force on these quantities were performed in \cite{Mitchell:2018qrg} and \cite{Mitchell:2019qke} using higher-resolution simulations, which derived the fitting formulae for them as a function of halo mass. However, these studies used a different halo mass definition, thus we use simple hyperbolic functions to fit concentration parameter and velocity dispersion boost-factors measured with the \textcode{rockstar} halos in \textcode{elephant} simulations. Although the concentration parameter measurement for halos with less than $500-1000$ particles, which corresponds to $(4-7) \times 10^{13} \Msun$ in \textcode{elephant}, needs to be treated with care, the study using higher-resolution simulations found a similar increase of the concentration parameter for unscreened halos \cite{Mitchell:2019qke}.
In the case of the concentration parameter we use the fitting formula:
\begin{equation}
    \frac{c_{\rm F5}}{c_{\rm GR}} = 1+ a_c \, (1- \tanh{ ( b_c \, \log_{10}(M_{\rm 200 c}) - M_c^* } ) \, .
    \label{c_fit}
\end{equation}
The velocity dispersion is obtained by solving the equations for the virial equilibrium between kinetic energy and gravitational energy of DM in halos: the solution depends linearly on the square root of the effective gravitational constant, $G_{\rm eff}$. To fit the transition of the velocity dispersion between screened and unscreened halos we use the fitting function:
\begin{equation}
    \frac{V_{\rm rms, F5}}{V_{\rm rms, GR}} = 1+ a_V \, (1- \tanh{ ( b_V \, \log_{10}(M_{\rm 200 c}) - M_V^* } ) \, .
    \label{V_rms_fit}
\end{equation}

\begin{figure}
        \centering 
        \subfloat[][Concentration]{
        \includegraphics[width=.48\textwidth,clip]{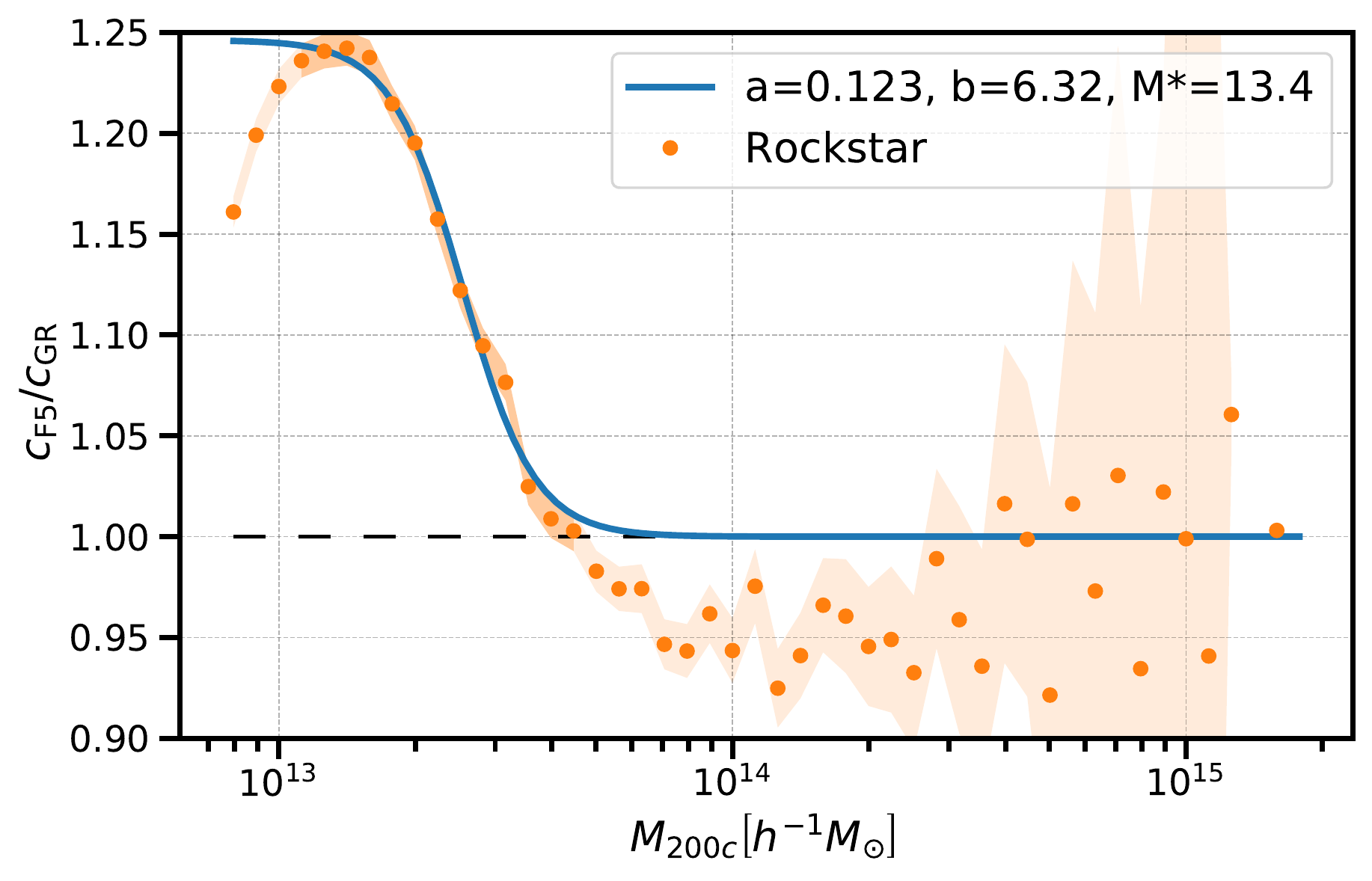}
        }
        \hfill
        \subfloat[][Velocity dispersion]{
        \includegraphics[width=.48\textwidth]{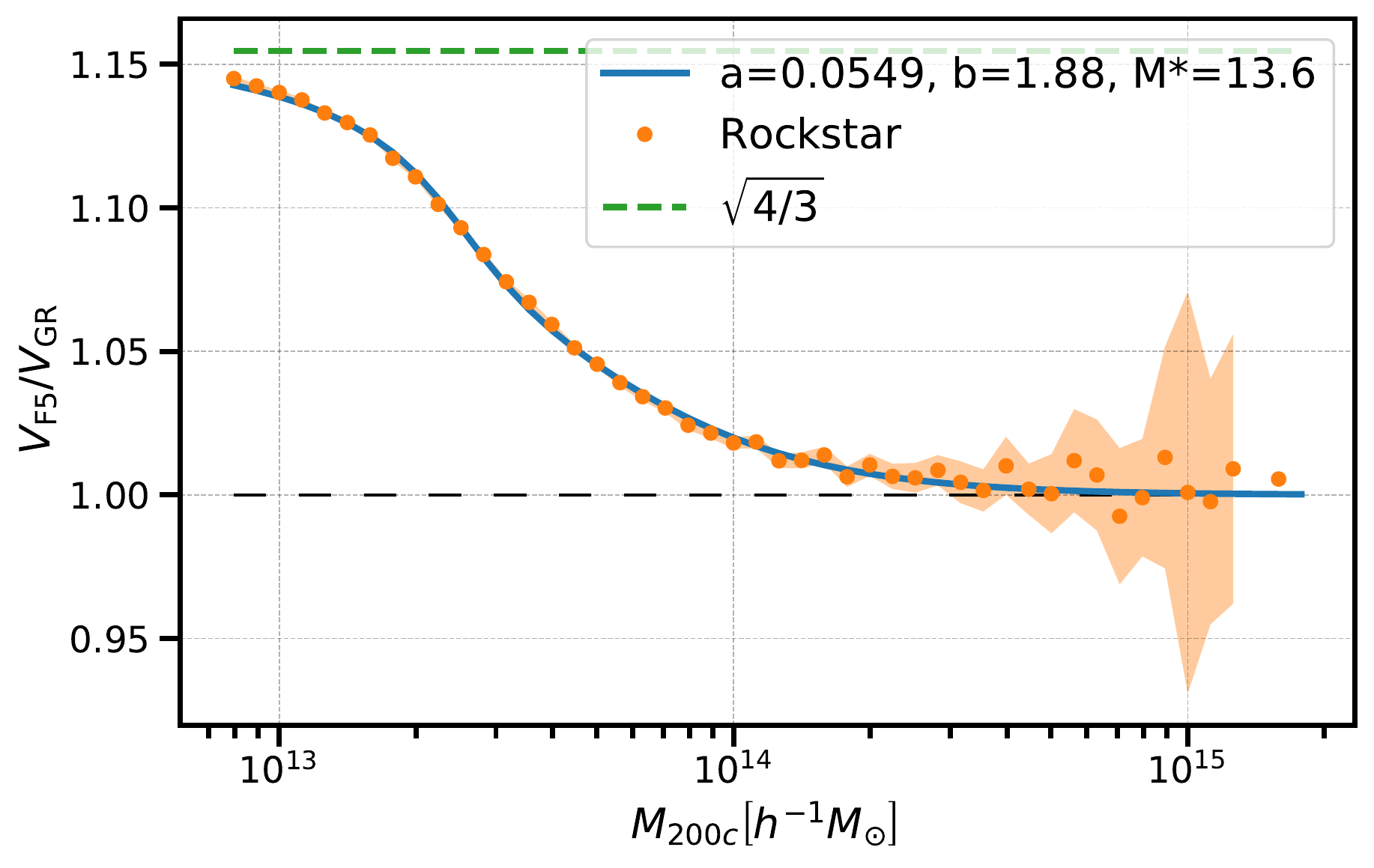}
        }
    \caption{Fit of F5 boost factors for concentration parameter (left panel) and velocity dispersion (right panel). The values measured from \textcode{rockstar} catalogues (orange dots) are used to fit the functions in Eq.~\eqref{c_fit} and Eq.~\eqref{V_rms_fit}. Only halos with masses above $10^{13} \Msun$ are used to perform the fits.}
    \label{fig:NFWfitF5}
\end{figure}

The results of the fit compared with the data from \textcode{rockstar} are shown in Figure~\ref{fig:NFWfitF5}: the hyperbolic functions correctly represents the transition from unscreened to screened halos and provides an effective description of the fifth force effect on these halo properties. The fitted parameters, $a$, $b$ and $M_*$, are shown in Figure~\ref{fig:NFWfitF5}. Note that these fitted parameters are valid only for $|f_{R0}|=10^{-5}$ and they need to be re-fitted for different values of $f_{R0}$.

In the case of N1, the Vainshtein mechanism operates efficiently irrespectively of halo mass and we confirmed that the concentration parameter and the velocity dispersion in N1 are not modified from GR.

\section{Galaxy catalogues}
\label{sec:Galaxies}

The Halo Occupation Distribution (HOD) technique is an empirical model to connect galaxy and dark matter properties. It allows us to produce galaxy mock catalogues from low-resolution dark matter simulations that do not resolve the internal structure of halos. 

\subsection{HOD model}
In this paper, due to the limited resolution of \textcode{elephant} simulations, we create mock galaxies similar to those found in the Baryon Oscillation Spectroscopic Survey CMASS galaxy sample. These galaxies can be modelled by the HOD model proposed in \cite{Zheng:2007zg}\footnote{
We used \texttt{Halotools} \cite{Hearin:2016uxs} (\url{https://halotools.readthedocs.io}) to implement the HOD model in our simulations.}. In this model, galaxies are split between centrals and satellites and the probabilities of a halo of mass $M$ to host either galaxy type are given by:  
\begin{equation}
   \left\langle N_{\mathrm{cen}}(M)\right\rangle =\frac{1}{2}\left[1+\operatorname{erf}\left(\frac{\log M-\log M_{\min }}{\sigma_{\log M}}\right)\right], \quad
   \left\langle N_{\mathrm{sat}}(M)\right\rangle = \left\langle N_{\mathrm{cen}}(M)\right\rangle\left(\frac{M-M_{0}}{M_{1}}\right)^{\alpha} \,,
\end{equation}
where each halo can have only one central galaxy but multiple satellite galaxies. The parameters $M_{\min }$ and $\sigma_{\log M}$ control respectively the scale and the width of the transition between probability 0 and 1 to host a central galaxy. The number of satellites is instead modelled as a power law with exponent $\alpha$, scale $M_{1}$ and cut-off $M_{0}$, multiplied by the number of central galaxies. 

Once the halo occupation has been determined, we need to assign the position and velocity for each galaxy. We place central galaxies at the halo centre of mass position and assign them the velocity of the host halo. On the other hand, to place satellites we sample a NFW halo density profile \cite{Navarro:1995iw} centred at the halo centre of mass and with a concentration parameter determined by Eq.~(\ref{eq:concentration}). We set the velocities to the halo velocity plus a dispersion term that corresponds to the halo virial velocity, as calculated in Eq.~(24) in \cite{More:2008yy}. In the case of F5, we correct the concentration parameter and the velocity dispersion as discussed in Section~\ref{sec:F5_NFWtweaks} employing the fitting formulae in Eq.~\eqref{c_fit} and Eq.~\eqref{V_rms_fit}.   

\subsection{HOD parameters fitting}
The 5 parameters of this HOD model are to be determined by requiring that the galaxy clustering reproduces a reference clustering signal, typically coming from real observations. In this paper, instead, we use as reference the clustering of the galaxy catalogues measured in the {\it N}-body halo catalogues in GR with a fiducial set of parameters that were constrained in \cite{Manera:2012sc} to fit the clustering of the CMASS galaxies in BOSS.
Since we can measure our reference signal directly from simulations, we can choose the statistics most convenient for our purpose. The results of \cite{Hernandez-Aguayo:2018oxg} show the importance of RSD in breaking the degeneracy between galaxy bias and MG parameters. Furthermore, it has been shown that the multipoles of the power spectrum can break the degeneracy between different HOD parameters better than the projected correlation function \cite{Hikage:2014bza}. In light of these works, we employ the multipoles of the galaxy power spectrum in redshift space to fit the HOD parameters.

\begin{table}[]
    \centering
    \begin{tabular}{lrclrcl}
    \toprule
    \toprule
    {} & \multicolumn{3}{c}{COLA} & \multicolumn{3}{c}{{\it N}-body} \\
    \toprule
    {} &      F5 &      GR &      N1 &      F5 &      \textbf{GR} &      N1 \\
    \midrule
    $\log_{10} \left(M_{\min }\right)$    &  13.205 &  13.155 &  13.170 &  13.174 &  \textbf{13.090} &  13.116 \\
    $\sigma_{\log_{10} M}$ &   0.579 &   0.589 &   0.590 &   0.578 &   \textbf{0.596} &   0.600 \\
    $\log_{10} \left(M_{0}\right)$      &  13.070 &  13.093 &  12.809 &  12.832 &  \textbf{13.077} &  12.677 \\
    $\log_{10} \left(M_{1}\right)$      &  13.969 &  13.955 &  14.056 &  14.030 &  \textbf{14.000} &  14.107 \\
    $\alpha$      &   1.059 &   1.044 &   1.043 &   1.040 &   \textbf{1.013} &   1.057 \\
    \midrule
    $n_s [10^{-4}(\hompc)^3]$      &   3.210 &   3.218 &   3.217 &   3.231 &   \textbf{3.226} &   3.230 \\
    \bottomrule
    \bottomrule
    \end{tabular}
    \caption{HOD parameters and number density of galaxies obtained by minimising the objective function in Eq.~\eqref{RSD_ObjFun} with the simplex search algorithm, except for the {\it N}-body GR column where the HOD parameters are the fiducial parameters taken from \cite{Manera:2012sc}.}
    \label{tab:RSD_HOD_params}
\end{table}

The galaxy power spectrum in redshift space can be decomposed in angular moments using the projection
\begin{equation}
    P_{\ell}(k)=(2 \ell+1) \int_{0}^{1} d \mu P(k, \mu) \mathcal{L}_{\ell}(\mu) \, ,
\end{equation}
where $\mathcal{L}_{\ell}(\mu)$ is the Legendre polynomial of order $\ell$. In the Kaiser approximation \cite{Kaiser:1987qv}, one can see how the quadrupole and the hexadecapole moments carry information on the redshift space distortion, which is key to break the degeneracy between cosmological and modified gravity parameters. 
The hexadecapole, however, is strongly affected by noise if estimated in the cosmological volume covered by our simulations. Thus, the objective function used to fit the HOD parameters includes only the monopole and the quadrupole. We use a Gaussian model \cite{Taruya:2010mx} to estimate the covariance matrix $\text{Cov}_{\ell, \ell^\prime}$ in terms of the volume spanned by the 5 realisations ($\sim5 (h^{-1} \, {\rm Gpc})^3$), and the galaxy number density and bias measured from {\it N}-body simulations in GR.  We add the tuning of the number density of galaxies to define the objective function as:
\begin{equation}
    \chi^2 = \sum_{\ell, \ell^\prime \in 0,2} \sum_{i \in k_{\rm bins}} \left(P_{\ell, i} - P^{\text{ref}}_{\ell, i}\right)\text{Cov}_{\ell, \ell^\prime, i}^{-1} \left(P_{\ell^\prime, i} - P^{\text{ref}}_{\ell^\prime, i} \right) + W_{n_{s}} \left( \frac{n_s - n_{s,\text{ref}}}{n_{s,\text{ref}}} \right)^2	\, .
\label{RSD_ObjFun}
\end{equation}
Here $W_{n_{s}}$  is the weight to control the importance of the number density tuning. We will set $W_{n_{s}}=10^4$ to enforce roughly a one per cent agreement in the number density. The multipoles of the power spectrum and the covariance matrix are evaluated in 25 linearly distributed bins in the range $k=[0.05,0.3] \hompc$.
We use the simplex search algorithm \cite{Nelder:1965zz} to find the set of HOD parameters that minimise the objective function \eqref{RSD_ObjFun}. The reference power spectra are computed from {\it N}-body GR simulations by populating galaxies using the fiducial HOD parameters of \cite{Manera:2012sc}, marked in bold in Table~\ref{tab:RSD_HOD_params}. We average over the 5 realisations of halo catalogues in each model to produce the multipole moments that are used in the objective function. The best fit HOD parameters resulting from the simplex search are summarised in Table~\ref{tab:RSD_HOD_params} together with the number density of galaxies $n_s$.

Using the best-fit parameters, we produce 5 HOD realisations from each halo catalogue and we average over them to compute galaxy statistics. We add redshift space distortions along the three different axes, producing 3 redshift space galaxy catalogues for each real space galaxy catalogue, and use the average over these when computing redshift space clustering statistics. 

\begin{figure}
        \centering 
        \subfloat[][Mass Distribution Centrals]{
        \includegraphics[width=.48\textwidth,clip]{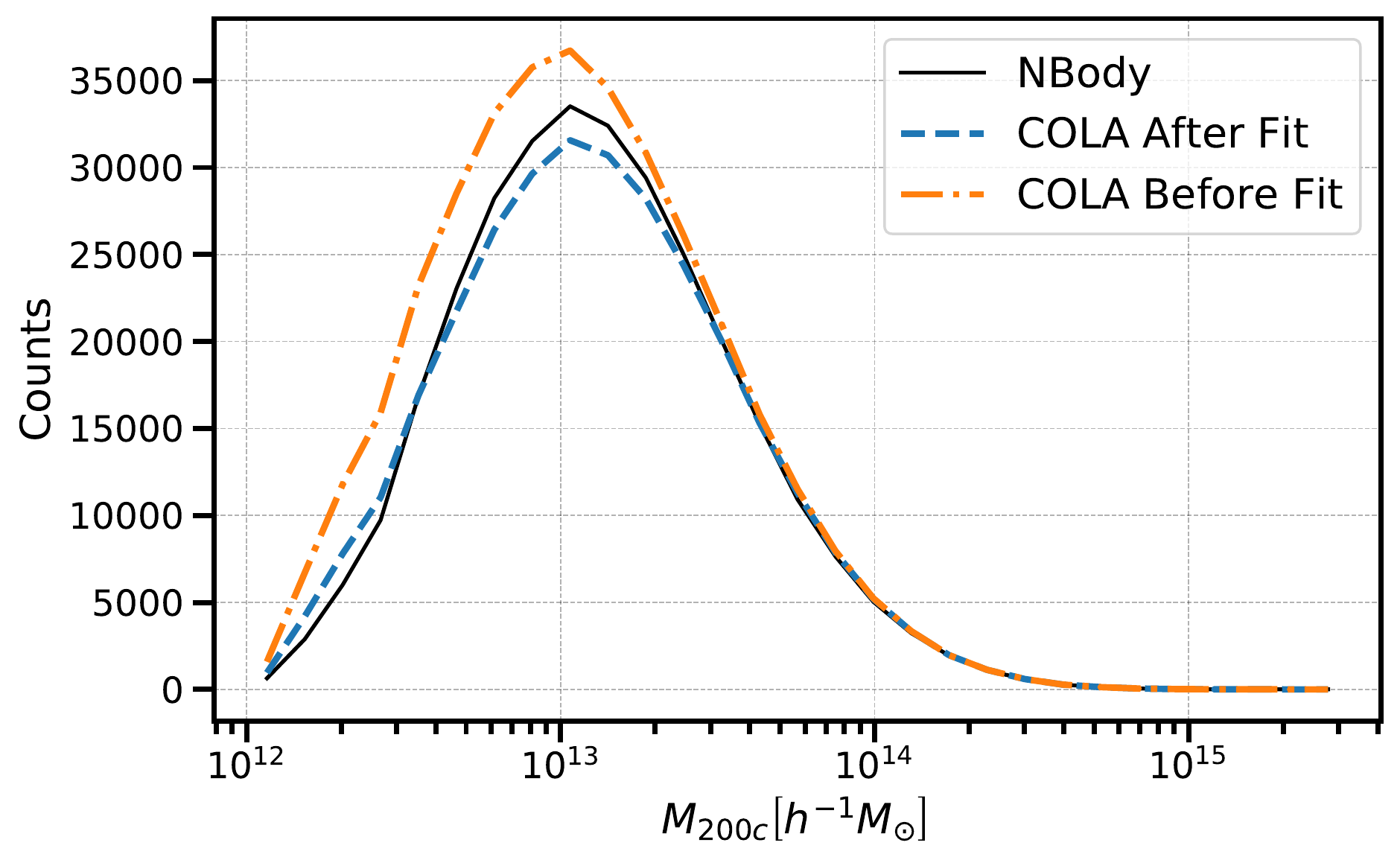}
        }
        \hfill
        \subfloat[][Mass Distribution Satellites]{
        \includegraphics[width=.48\textwidth]{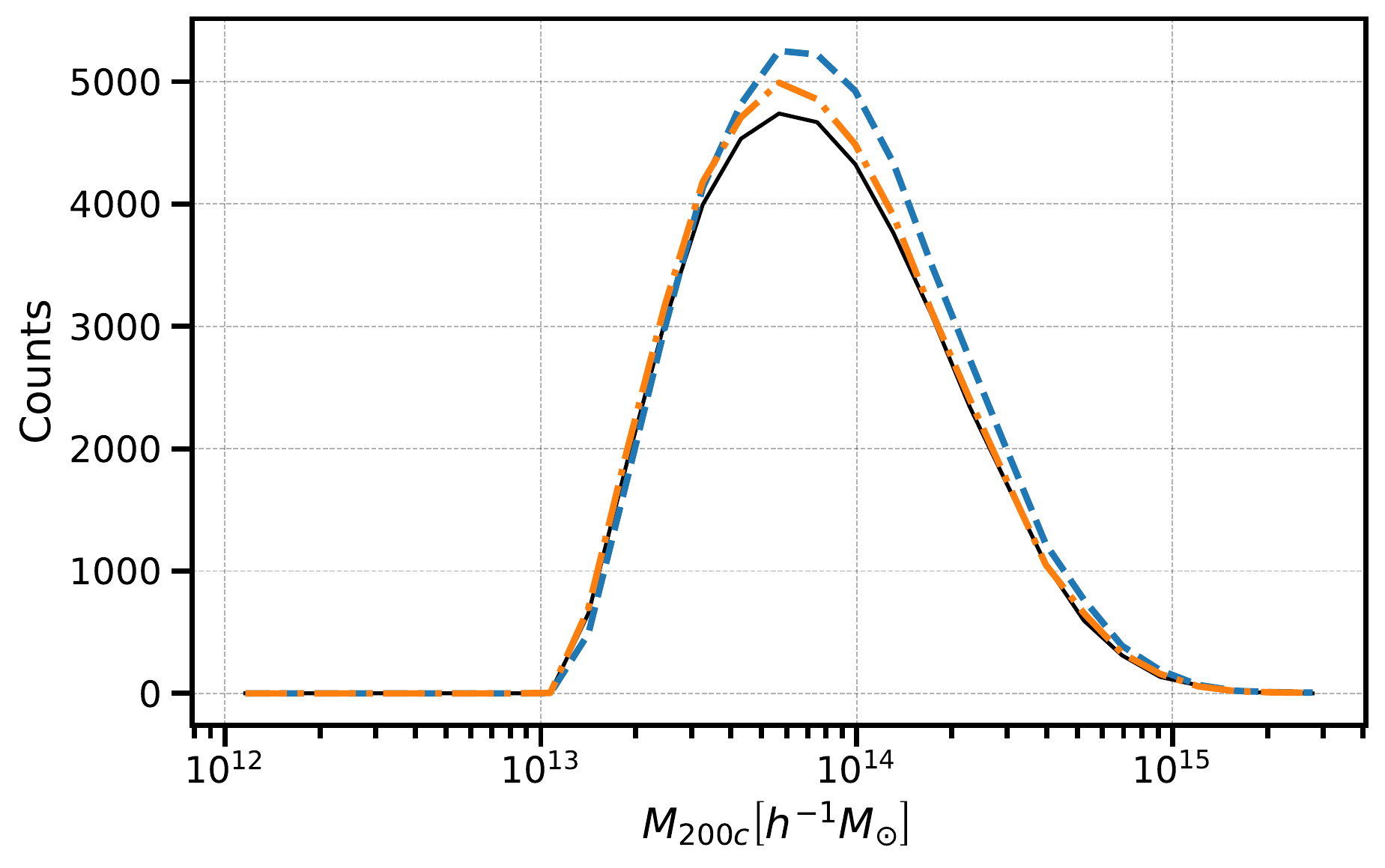}
        }
        \vfill
        \subfloat[][Velocity Distribution Centrals]{
        \includegraphics[width=.48\textwidth,clip]{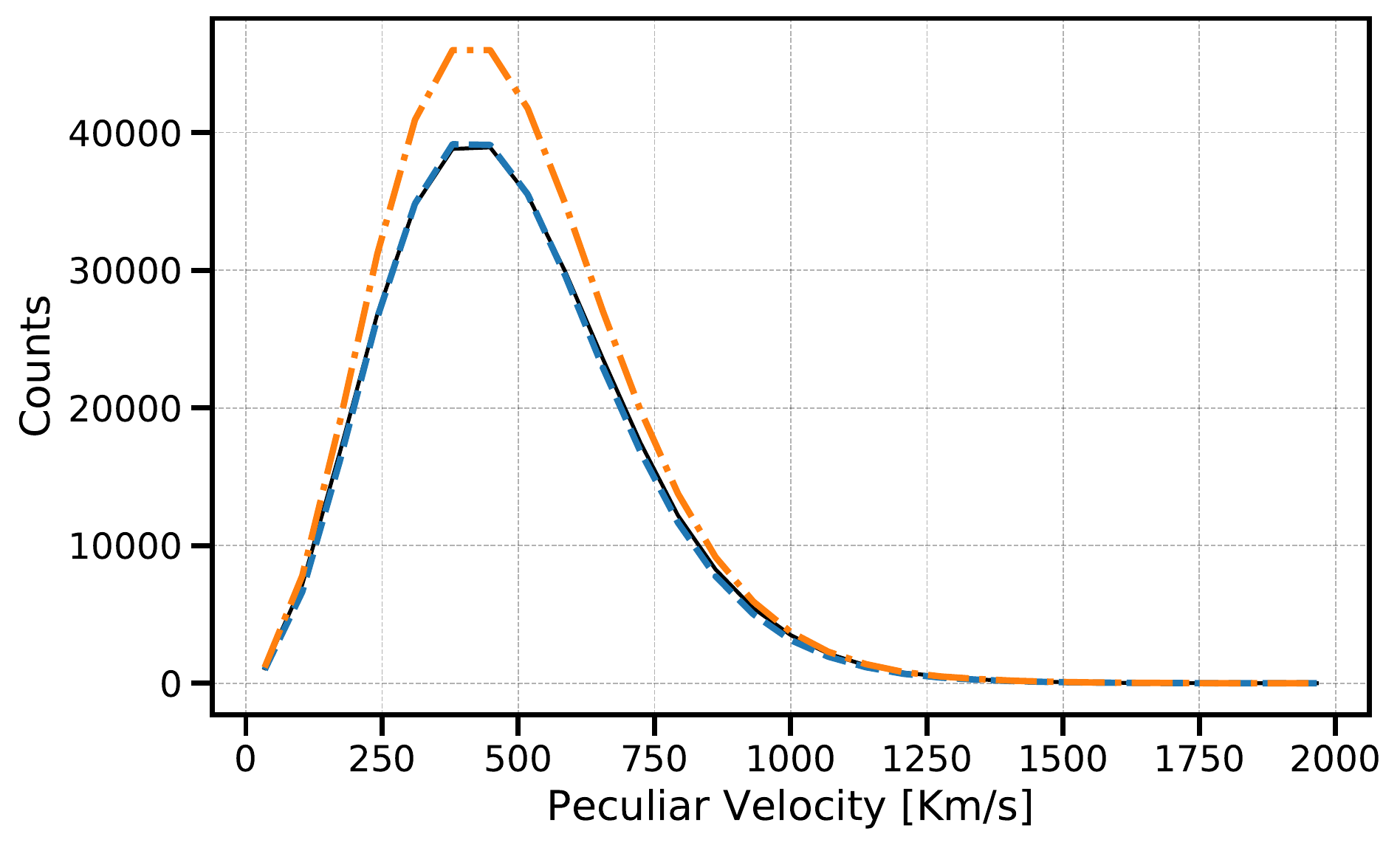}
        }
        \hfill
        \subfloat[][Velocity Distribution Satellites]{
        \includegraphics[width=.48\textwidth]{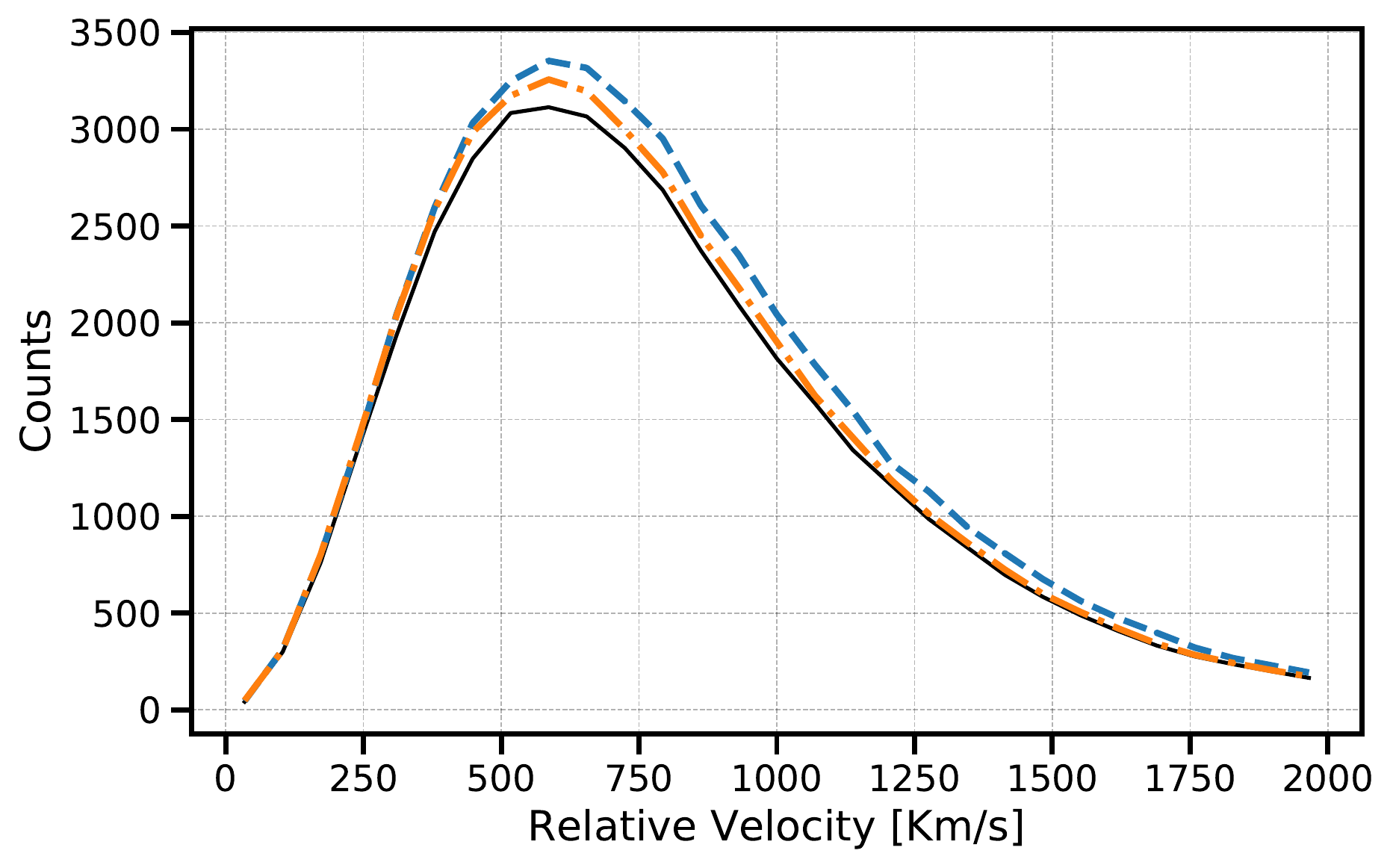}
        }
\caption{\label{fig:GalDistr_NoFit}Halo mass (top panels) and velocity distributions (bottom panels) of central (left panels) and satellite (right panels) galaxies in GR. The velocity distribution shows the number of galaxies in each velocity bin, which corresponds to peculiar velocities for centrals and velocities with respect to the host halo rest frame for satellites. The distributions of galaxies in COLA before (orange dot-dashed lines) and after (blue dashed lines) the fit compared with the distributions of {\it N}-body in GR (black solid lines) show that the fit is mostly driven by central galaxies.}
\end{figure}

\begin{figure}
        \centering 
        \subfloat[][Mass Distribution Centrals]{
        \includegraphics[width=.48\textwidth,clip]{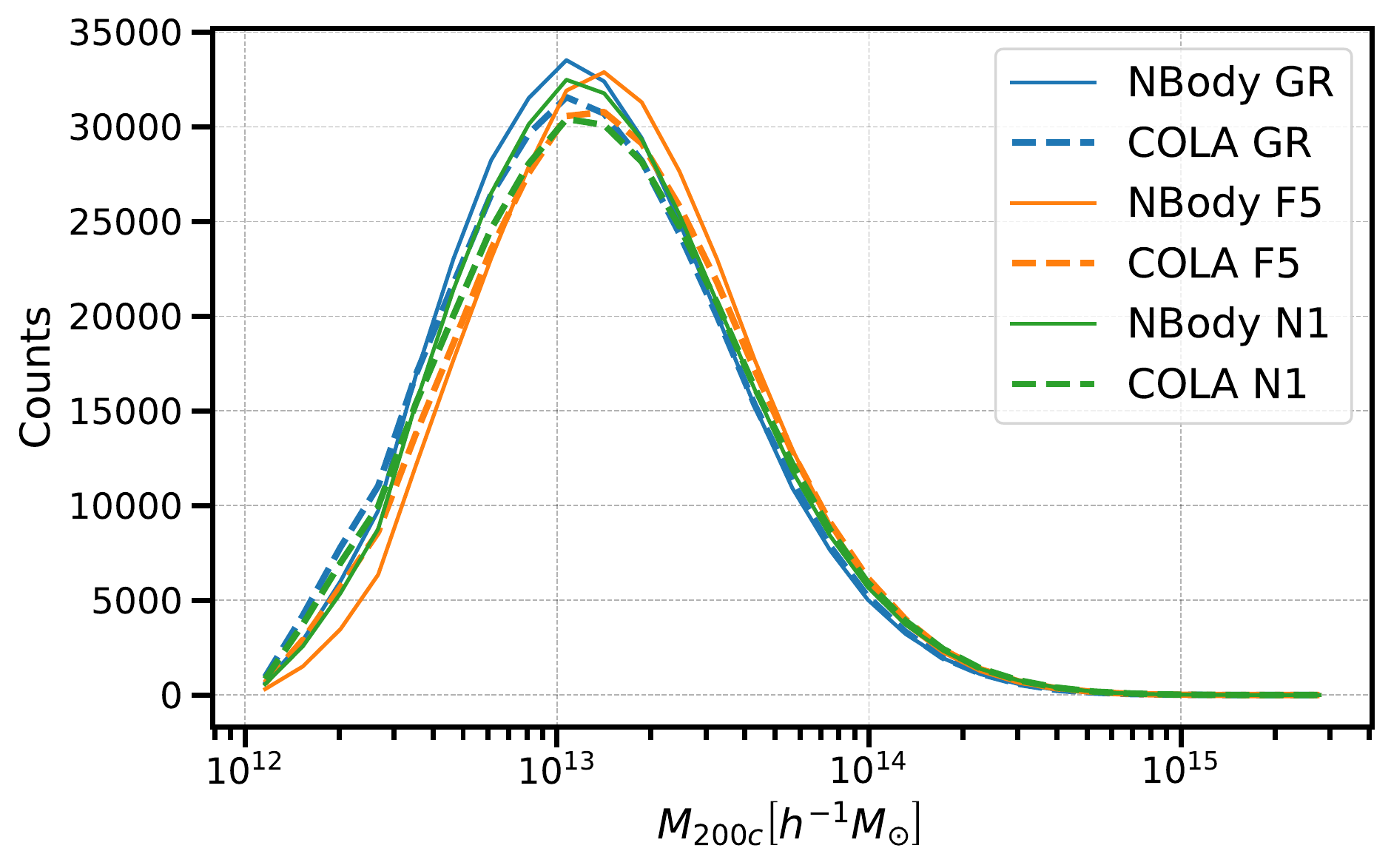}
        }
        \hfill
        \subfloat[][Mass Distribution Satellites]{
        \includegraphics[width=.48\textwidth]{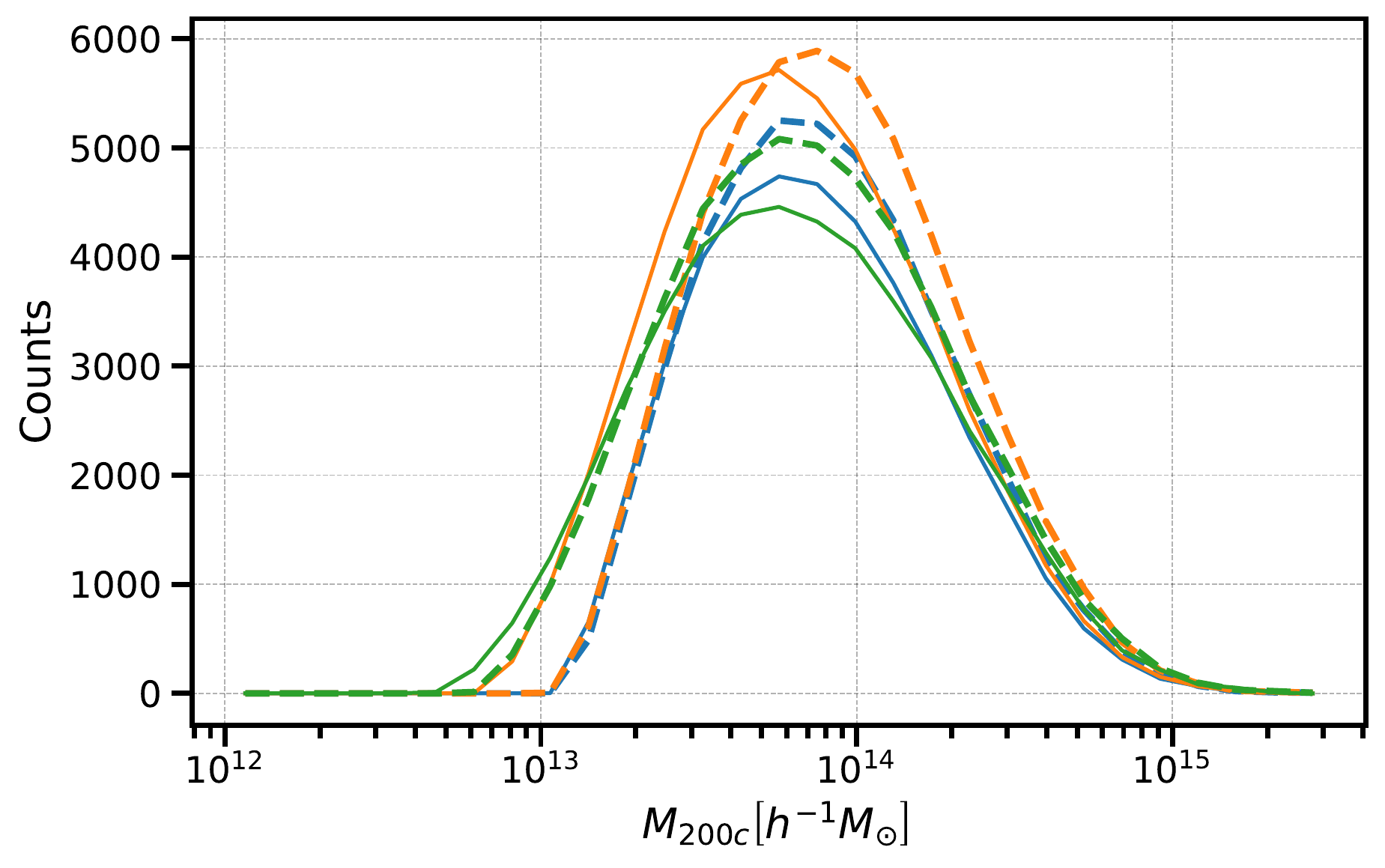}
        }
        \vfill
        \subfloat[][Velocity Distribution Centrals]{
        \includegraphics[width=.48\textwidth,clip]{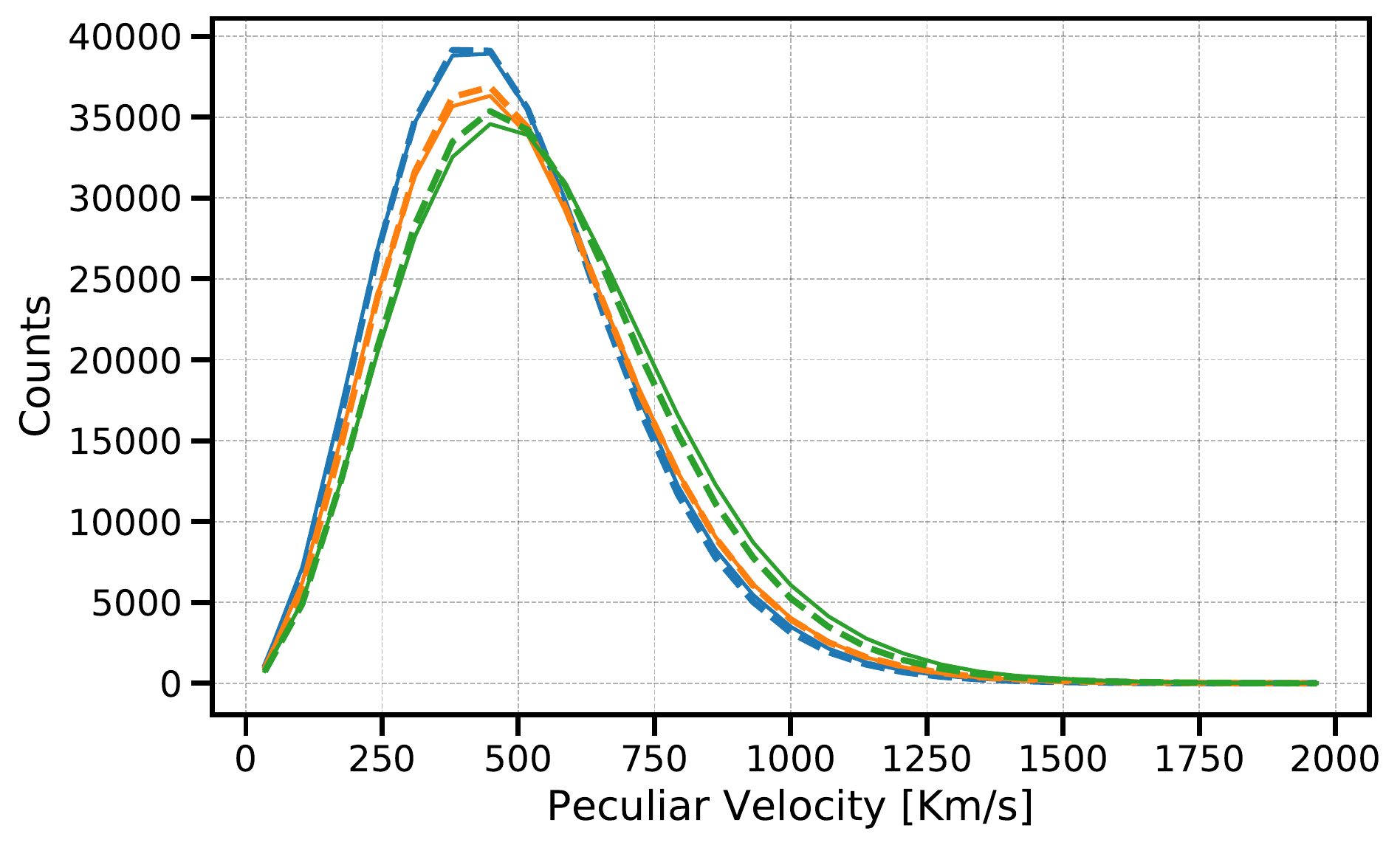}
        }
        \hfill
        \subfloat[][Velocity Distribution Satellites]{
        \includegraphics[width=.48\textwidth]{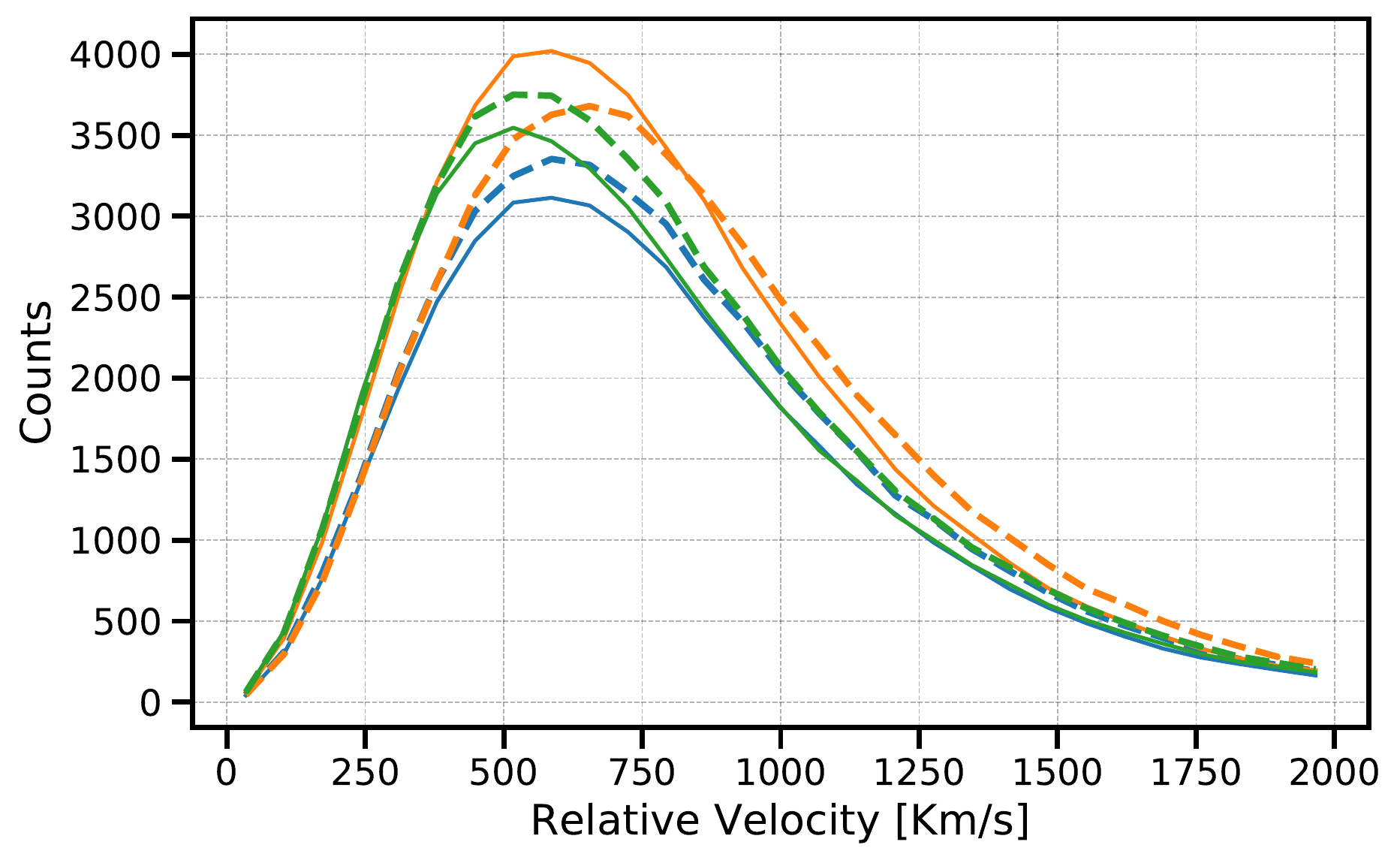}
        }
\caption{\label{fig:GalDistr_MultipolesFit}
Same as Figure~\ref{fig:GalDistr_NoFit} but for all the gravity models (GR, F5 and N1 in blue, orange and green respectively) after fitting the HOD parameters, which are shown in Table~\ref{tab:RSD_HOD_params}.
}
\end{figure}

To interpret the results of the HOD parameters tuning, we measure in the GR mocks the galaxy number count as a function of the halo mass and the peculiar velocity (which for satellites we take as the velocity with respect to the reference frame of the hosting halo).
Figure~\ref{fig:GalDistr_NoFit} compares the counts before\footnote{The HOD parameters before the fit are the same as the default values for the GR case in $N$-body, highlighted in bold in Table~\ref{tab:RSD_HOD_params}.} and after fitting the HOD parameters in COLA and $N$-body in GR.
We can see how the fit is driven by central galaxies, due to their higher abundance and the stronger signal-to-noise ratio in the monopole. The larger value of $M_{\mathrm{min}}$ in COLA places central galaxies in more massive halos, which are less abundant, compensating for the excess of halos below $~10^{13} \Msun$. The smaller values of $M_1$ compensates for the increase of $M_{\mathrm{min}}$, while the larger value of $\alpha$ is responsible for the heavier right tail of the satellite distributions.
Figure~\ref{fig:GalDistr_MultipolesFit} also shows the host halo mass and velocity distribution but for all the gravity models (GR, F5 and N1) after the fit of the HOD parameters, which are summarised in Table~\ref{tab:RSD_HOD_params}.
We can see that the differences in the velocity distribution of central galaxies in different models are well captured in COLA. 

\begin{figure}
\centering 
\subfloat[][GR]{
\includegraphics[width=.48\textwidth,clip]{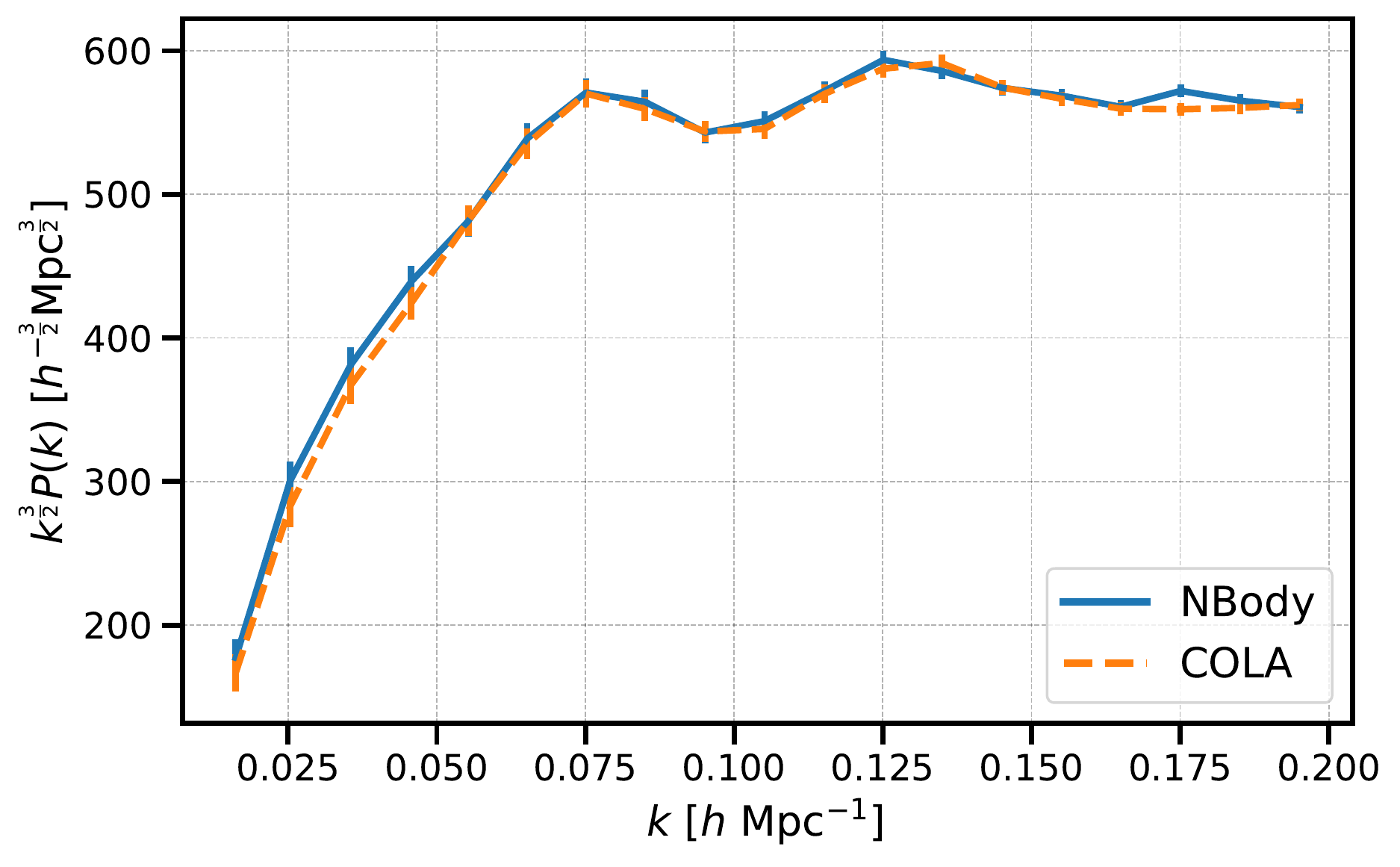}
}
\hfill
\subfloat[][GR ratio]{
\includegraphics[width=.48\textwidth]{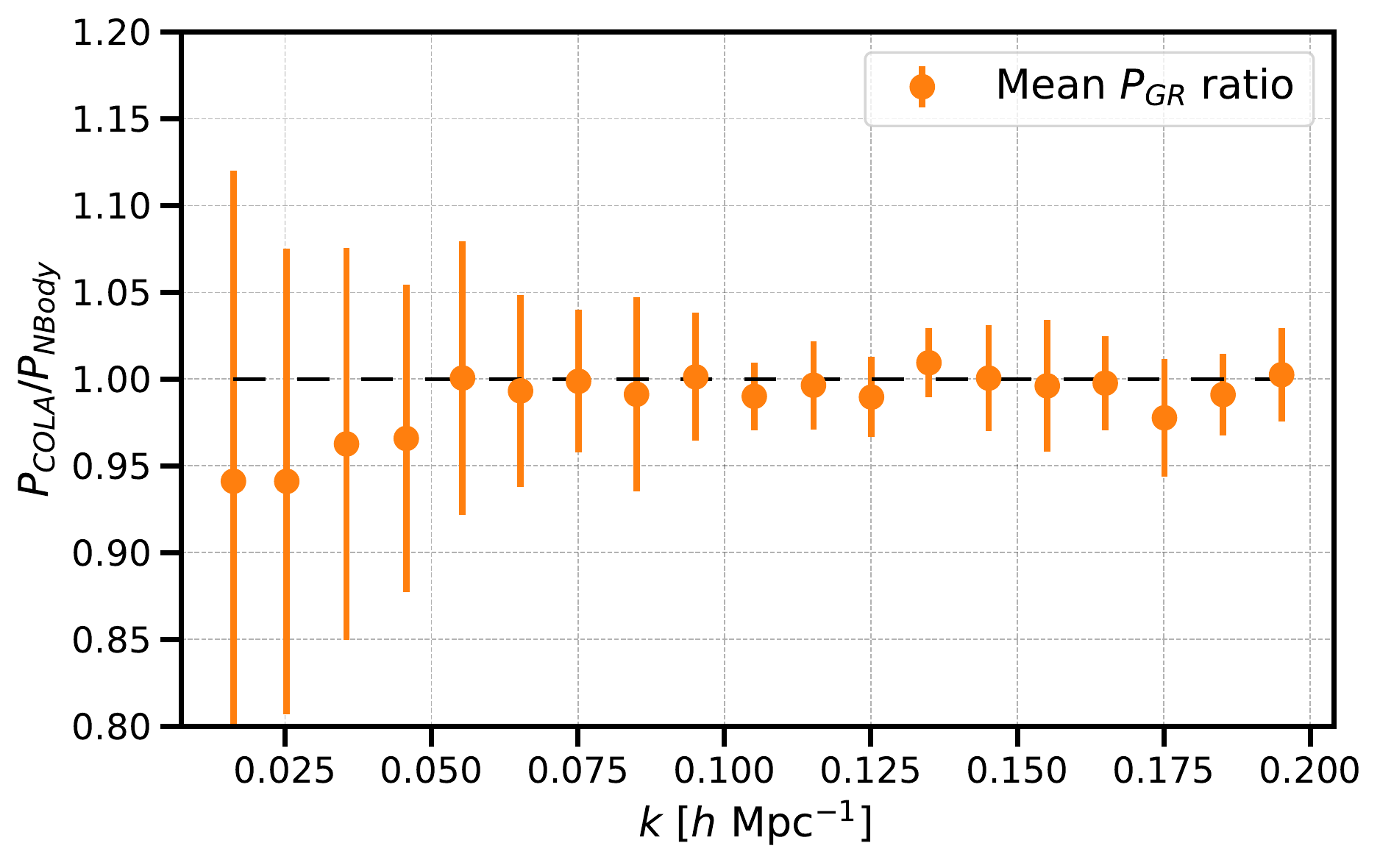}
}
\vfill
\subfloat[][F5 boost factor]{
\includegraphics[width=.48\textwidth]{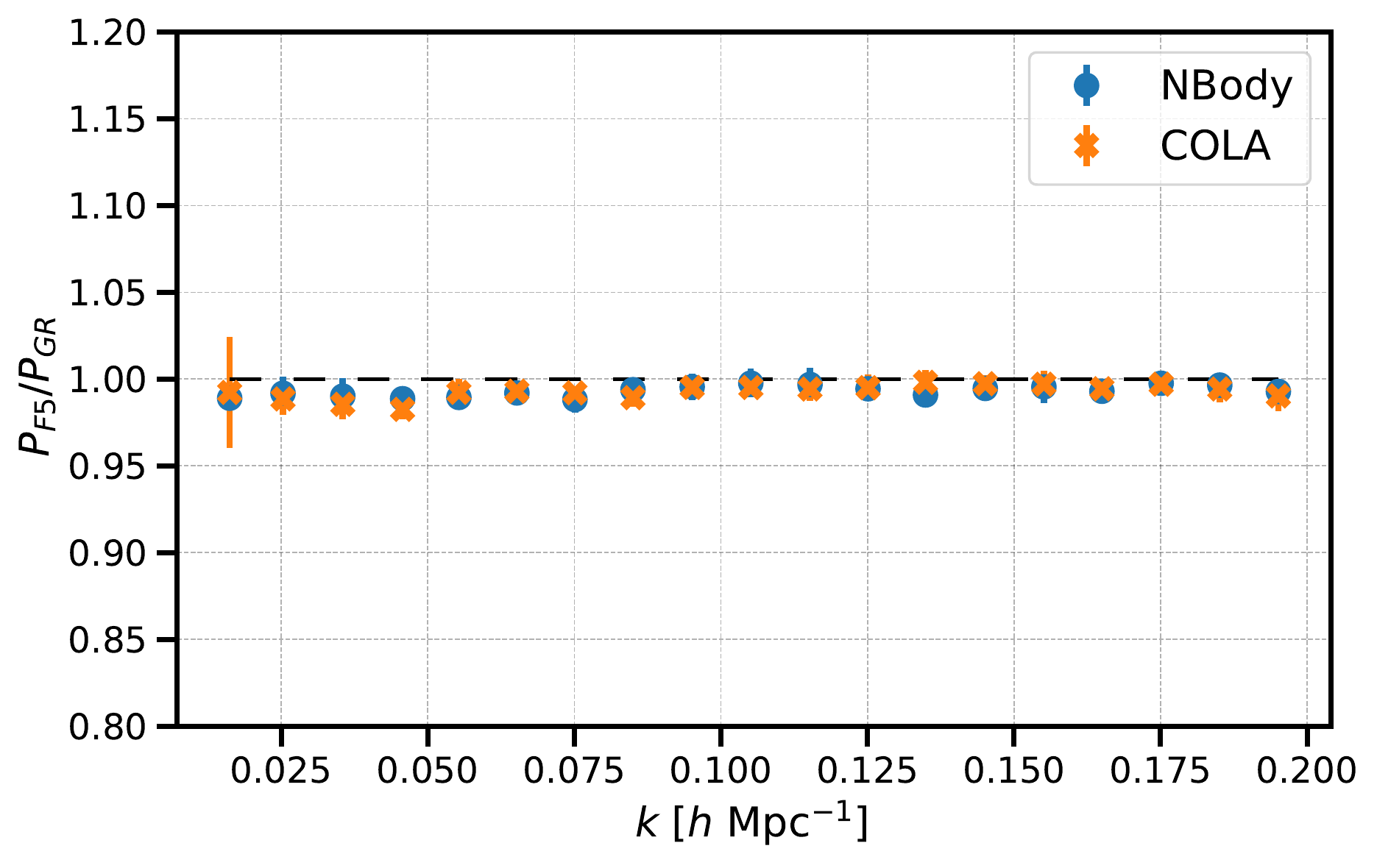}
}
\hfill
\subfloat[][N1 boost factor]{
\includegraphics[width=.48\textwidth]{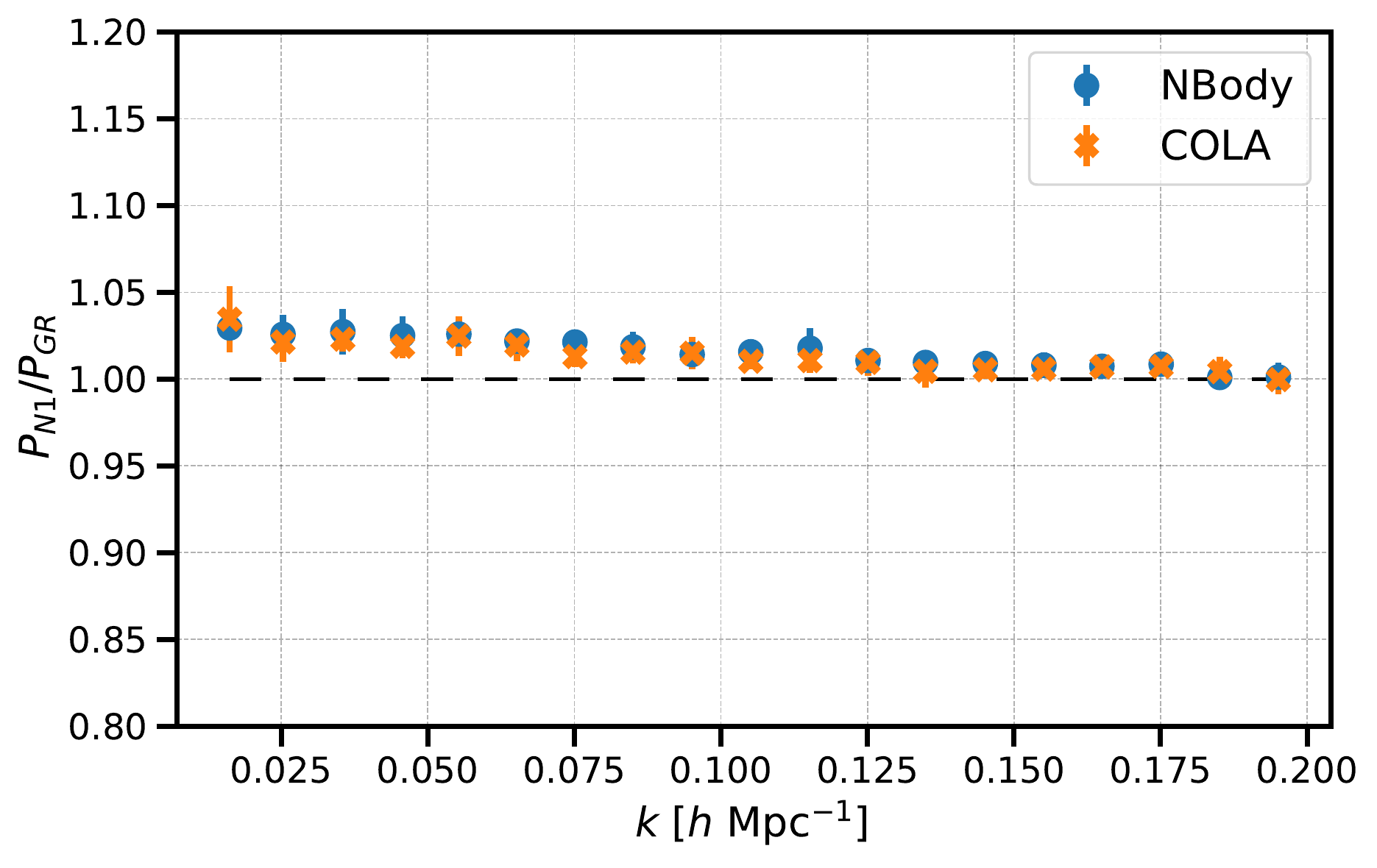}
}
\caption{Comparison of the monopole of the galaxy power spectrum in redshift space between the COLA (in orange) and {\it N}-body (in blue) catalogues. The power spectrum in GR (top panels) and the boost factors in F5 and N1 (bottom panels) show an agreement within the variance between COLA and {\it N}-body.}
\label{fig:RSD_P0}
\end{figure}

\begin{figure}
\centering 
\subfloat[][GR]{
\includegraphics[width=.48\textwidth,clip]{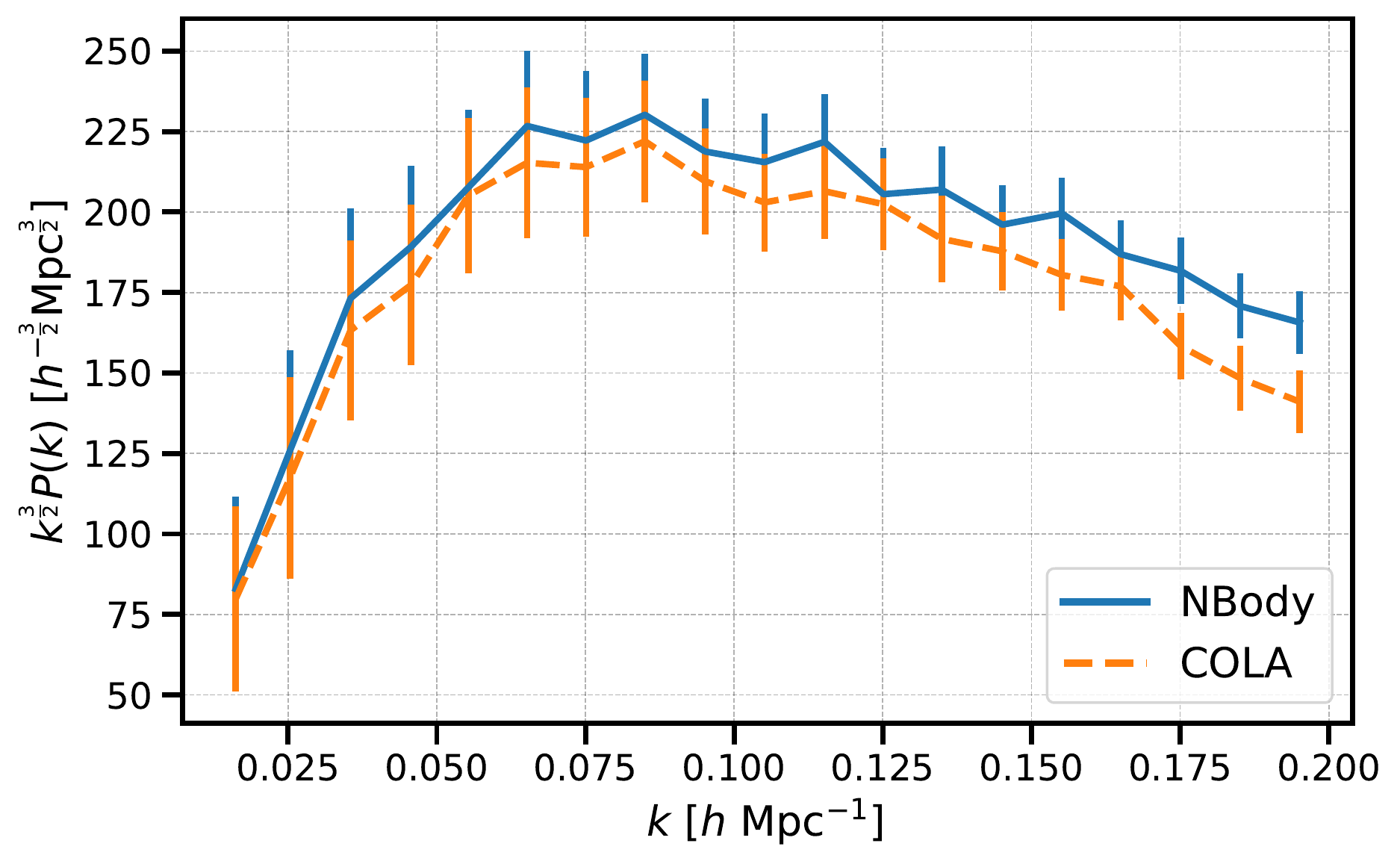}}
\hfill
\subfloat[][GR ratio]{
\includegraphics[width=.48\textwidth]{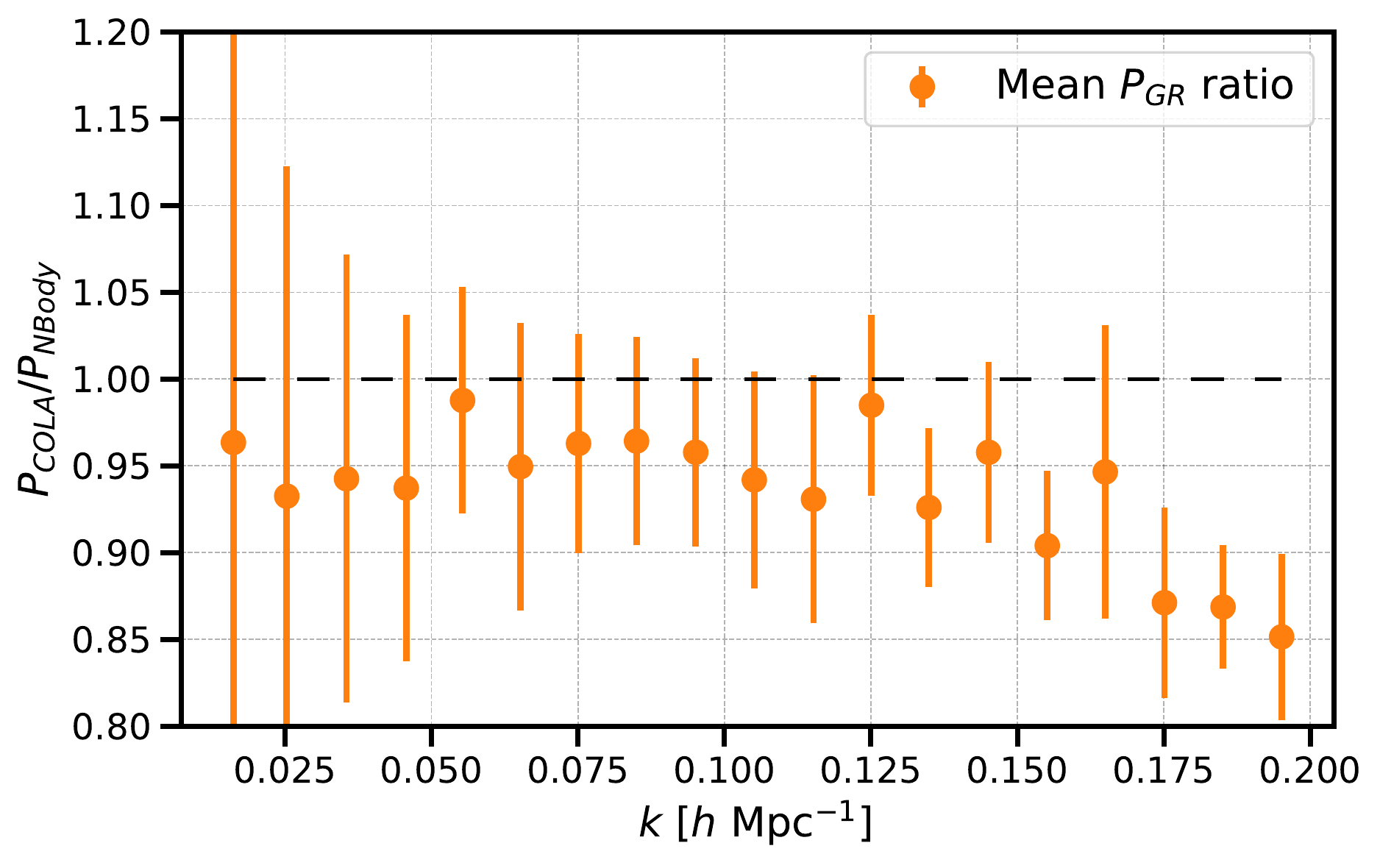}}
\vfill
\subfloat[][F5 boost factor]{
\includegraphics[width=.48\textwidth]{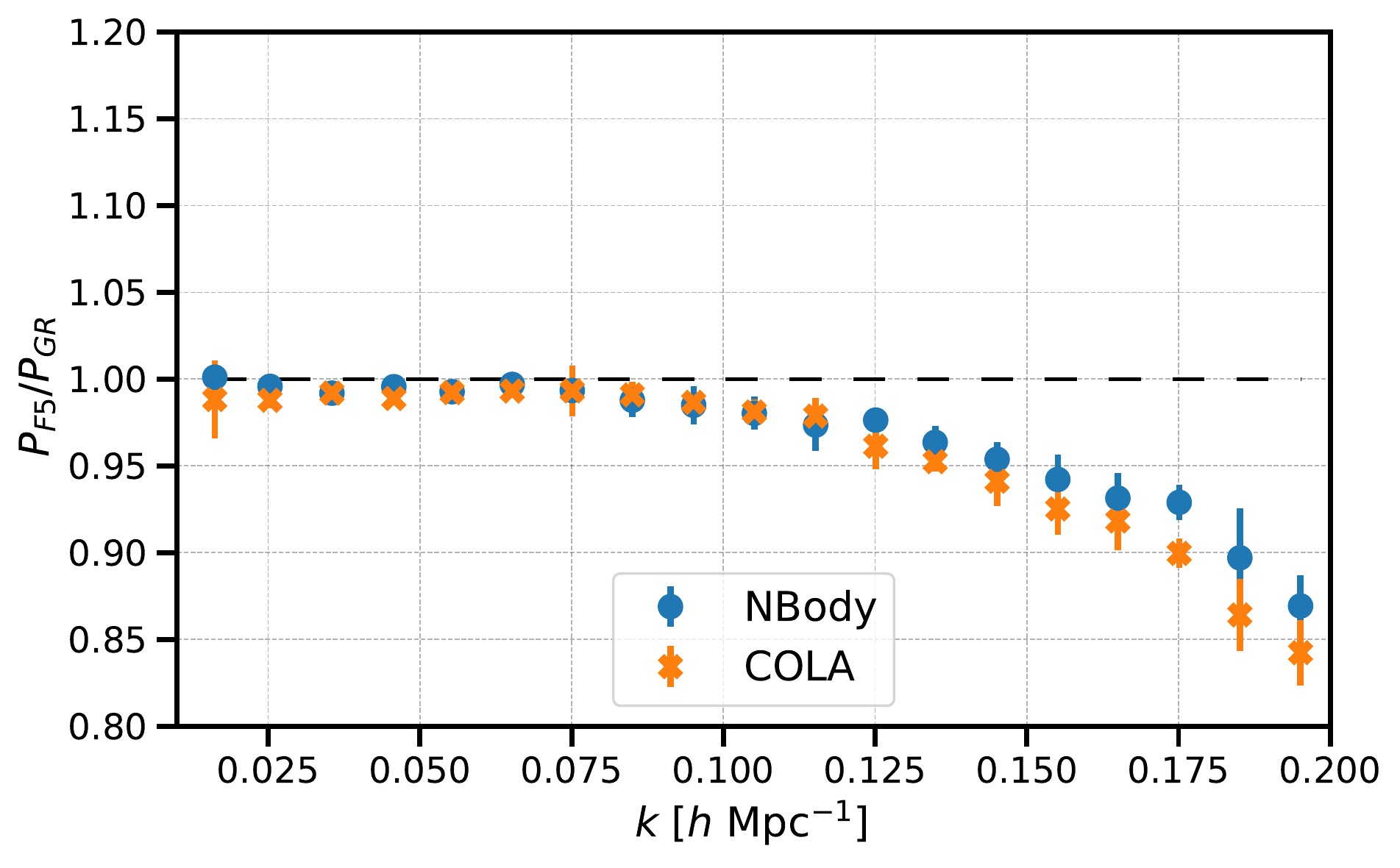}}
\hfill
\subfloat[][N1 boost factor]{
\includegraphics[width=.48\textwidth]{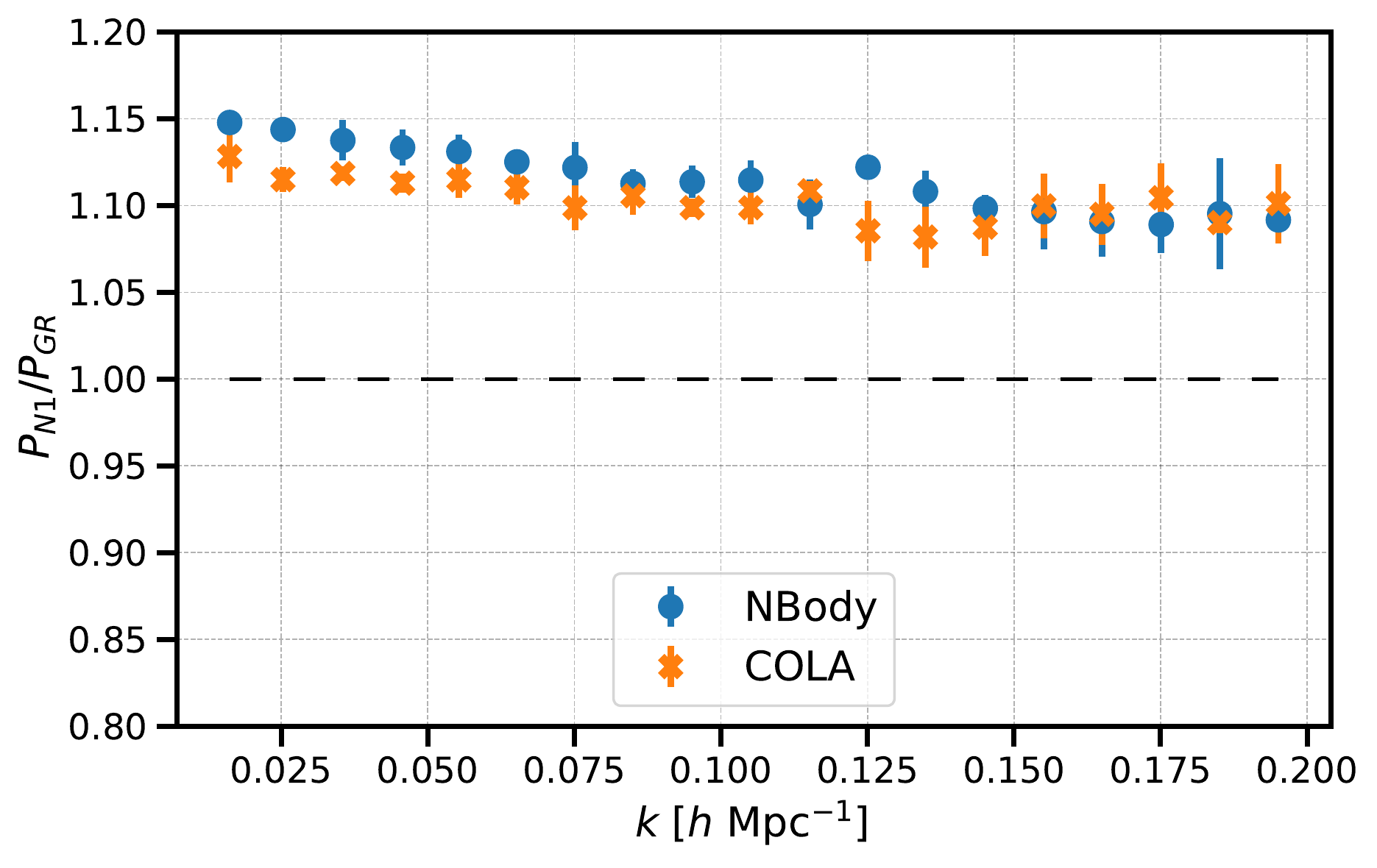}}
\caption{Comparison of the quadrupole of galaxy power spectrum in redshift space between COLA (in orange) and {\it N}-body (in blue) catalogues. The power spectrum in GR (top panels) show a $5\%$ agreement up to $k \sim 0.17 \hompc$. The boost factors in F5 and N1 (bottom panels) show show a better than $5\%$ agreement up to $k \sim 0.2 \hompc$, despite $\sim 10\%$ deviations appear above $k \sim 0.17 \hompc$ in GR.}
\label{fig:RSD_P2}
\end{figure}

We compare the monopole and the quadrupole of the galaxy power spectrum in Figures~\ref{fig:RSD_P0} and \ref{fig:RSD_P2} respectively. The agreement between COLA and {\it N}-body for the monopole is within the variance for all the gravity models up to $k \sim 0.2 \hompc$. The GR quadrupole in COLA agrees with that in {\it N}-body within $5\%$ up to $k \sim 0.17 \hompc$. The agreement for the quadrupole boost factors is within $5\%$ up to $k \sim 0.2 \hompc$ for both F5 and N1. Additionally, the quadrupole boost factor in F5 shows a strong scale dependence, with a small departure from GR on linear scales that increases on non-linear scales. The quadrupole boost factor in N1, instead, shows a large departure from GR already on linear scales.
This different behaviour of the quadrupole boost factors is due to the difference in the linear growth rates of these two models, as well as the different screening mechanisms. This shows that even after the HOD tuning to obtain a similar monopole, the difference between models shows up in the quadrupole. Thus RSD provides a powerful way to distinguish between different MG models \cite{Hernandez-Aguayo:2018oxg}.

We note that even though we fit multipole moments up to $k_{\rm max}=0.3 \hompc$, the agreement between {\it N}-body and COLA degrades substantially at $k>0.2 \hompc$ for quadrupole. This is because the slight difference in satellite distributions gives a large effect on the higher multipoles at large $k$ due to the Finger-of-God effect \cite{Hikage:2014bza}, which is highly sensitive to velocity dispersion of galaxies on small scales. This is not necessarily an issue for the modelling of cosmological and MG effects in mock galaxies as the velocity dispersion is usually treated as a nuisance parameter in cosmological analyses.

\begin{figure}
\centering 
\subfloat[][F5]{
\includegraphics[width=.48\textwidth]{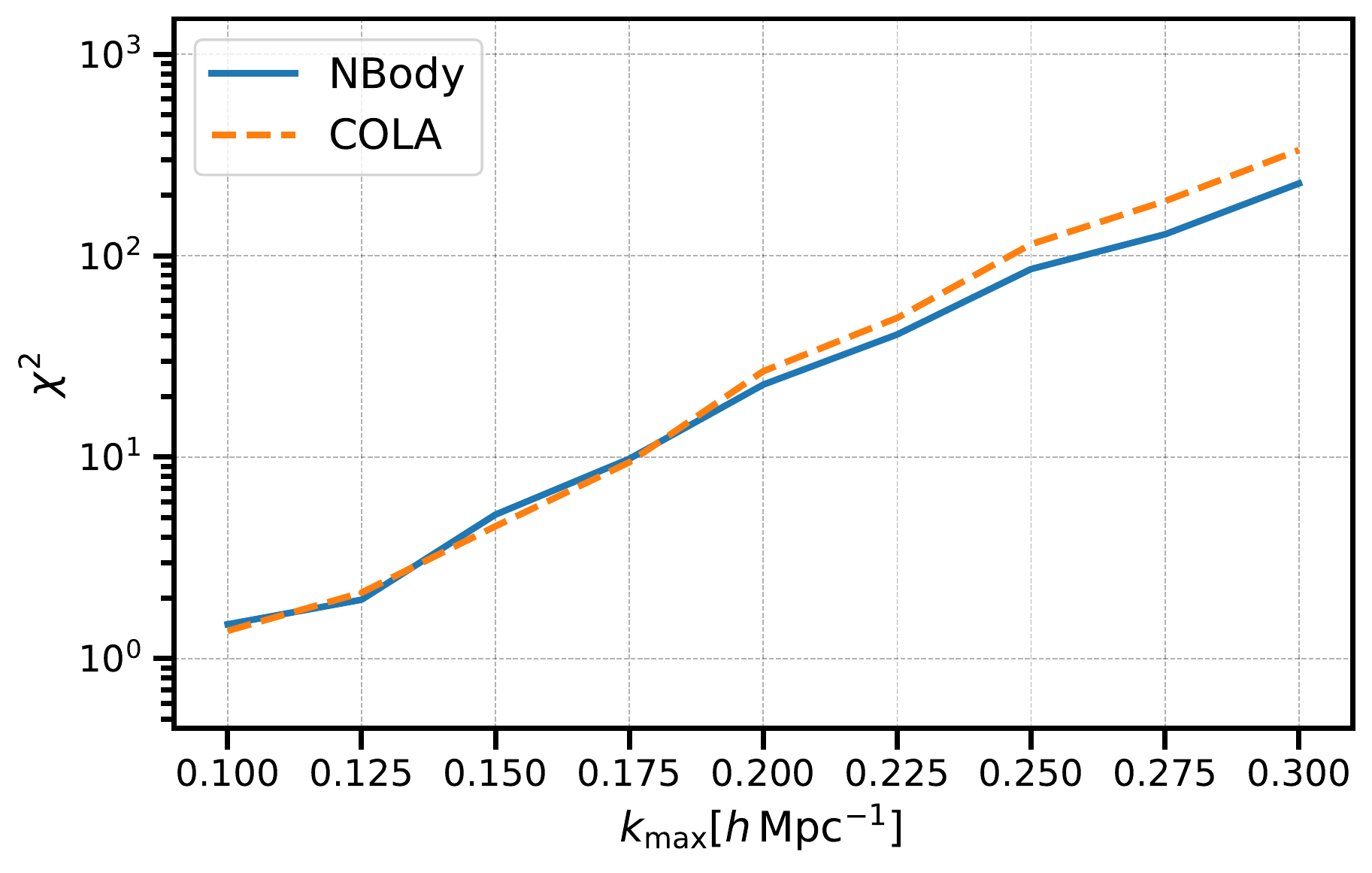}}
\hfill
\subfloat[][N1]{
\includegraphics[width=.48\textwidth]{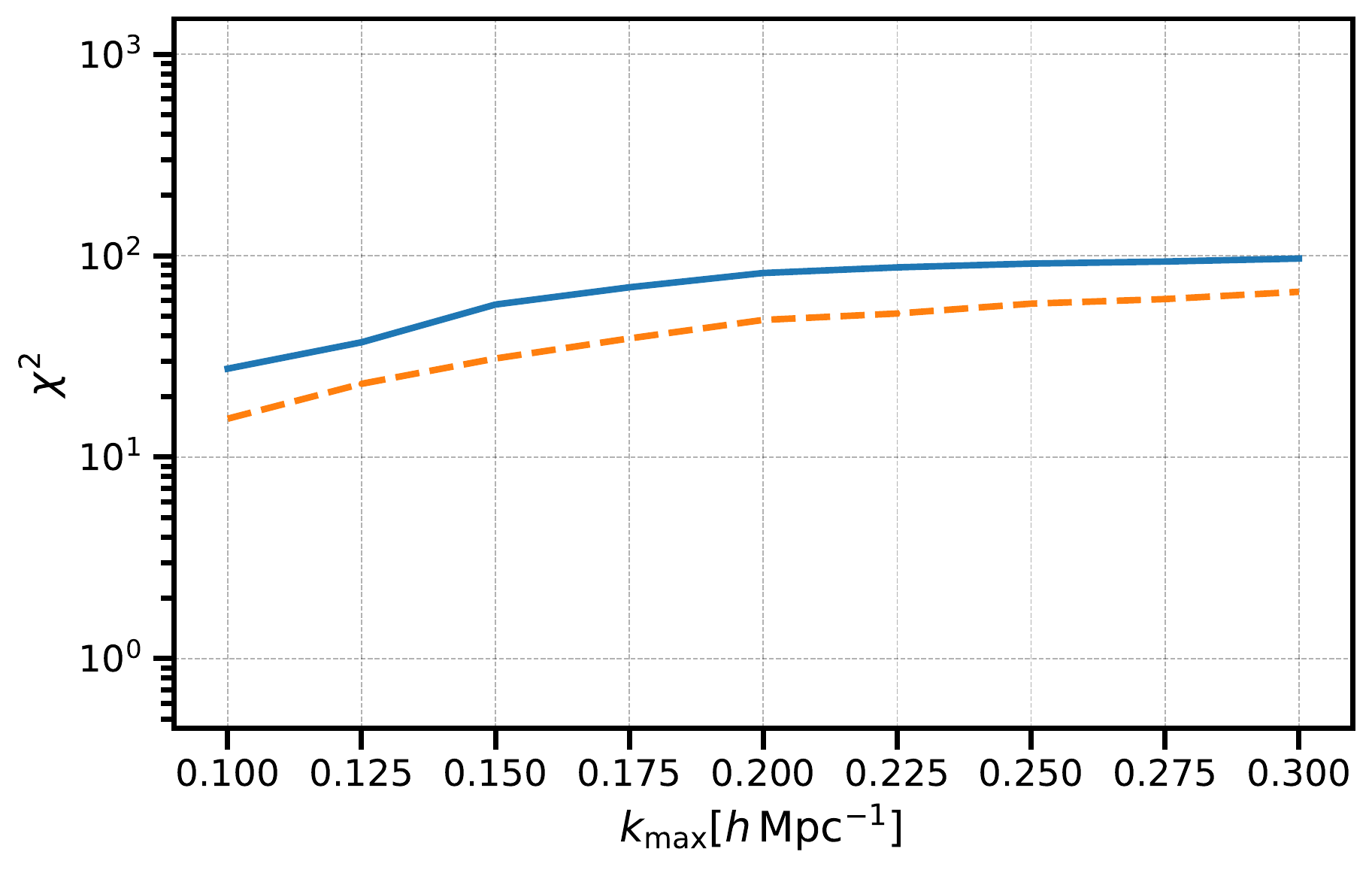}}
\caption{Value of $\chi^2$ as defined in Eq.~\eqref{RSD_ObjFun} in COLA (orange dashed lines) and {\it N}-body (blue solid lines) evaluated using the signal in GR for the corresponding simulation technique as the target function, as a function of the small scales cut-off $k_{\rm max}$. The $\chi^2$ in F5 (left panel) shows a strong scale dependence, with small values on linear scales and larger values on mildly non-linear scales, highlighting the importance of mildly non-linear scales in constraining MG parameters. The $\chi^2$ in N1 (right panel) shows little scale dependence, reflecting the features of Vainshtein screening. The results in COLA and {\it N}-body are in good agreement.}
\label{fig:RSD_chi2}
\end{figure}

To determine the detectability of MG with summary statistics of galaxy catalogues, we compute $\chi^2$ from Eq.~\eqref{RSD_ObjFun} in F5 and N1 for both the COLA and {\it N}-body catalogues while using the clustering signal in GR as a reference.
To investigate the role of the different scales, we evaluate the $\chi^2$ for different small scale cut-off $k_{\rm max}$ in the range between $0.1$ and $0.3 \hompc$.
The results in Figure~\ref{fig:RSD_chi2} highlight the intrinsic difference between F5 and N1, with the former being significantly different from GR only on non-linear scales while the latter showing a strong departure from GR already on linear scales. COLA reproduces the $\chi^2(k_{\rm max})$ behaviour of the {\it N}-body very well, demonstrating that it captures the MG effects accurately. 

\section{Conclusions}
\label{sec:Conclusions}
We have extended, applied and validated the HOD formalism to COLA simulations in MG, enabling the fast generation of mock galaxy catalogues that are required to test gravity on cosmological scales in Stage IV surveys. 

Using a full {\it N}-body simulation suite as a benchmark, we produced an equivalent simulation suite with \textcode{mg-picola}, focusing on the following gravity models: GR, braneworld nDGP model (N1) and $n=1$ Hu-Sawicki $f(R)$ gravity (F5). Looking at the power spectra of the density and velocity divergence fields, we confirmed the good accuracy of COLA in reproducing the density field (agreement within the variance up to $k \sim 1 \hompc$) and have given the first measure of its accuracy in the velocity field (within $4 \%$ up to $k \sim 0.4 \hompc$).

To apply the HOD formalism to COLA simulations, we first need to employ a halo finding algorithm to find dark matter halos. We found that the \textcode{rockstar} halo-finder, which uses the six-dimensional phase-space information to find halos, did not perform well with our COLA simulations with the default setting, resulting in a disagreement between COLA and {\it N}-body halo statistics ($\sim 25 \%$ in the hmf and $10\%$ in the power spectrum at $k \sim 0.4 \hompc$). On the other hand, finding halos with a simple FoF finder and converting the halo mass with the results of \cite{Lukic:2008ds} and \cite{Dutton:2014xda}, we have produced halo catalogues in COLA and {\it N}-body which are in good agreement, as shown by the hmf (better than $10\%$ agreement) and the halo power spectrum (agreement within the variance).

Since COLA simulations do not resolve the internal halo structure, we have assumed a NFW profile and the concentration-mass relation of \cite{Dutton:2014xda} in GR. To take into account the effect of the fifth force on the internal structure of halos, we have measured the boost factors of the concentration and velocity dispersion mass relations in MG with \textcode{rockstar} halo catalogues in {\it N}-body. We have found that no significant difference is present in the N1 boost factors, which allows us to use the standard NFW profile in N1. In F5, instead, there is a clear transition from screened to un-screened halos both in the concentration and the velocity dispersion boost factors. This is in line with the findings of \cite{Mitchell:2018qrg, Mitchell:2019qke}. Using simple fitting formulae that we calibrated on these results, we have been able to assign realistic concentration and velocity dispersion to halos in F5 depending on their mass.

With the HOD model proposed in \cite{Zheng:2007zg}, we have fitted the HOD parameters of the different models to match the galaxy clustering of {\it N}-body catalogues in GR. The fit performed with a simplex search algorithm is mostly driven by the central galaxies, whose final distributions in velocity are different in each model but similar between COLA and {\it N}-body. Despite that the objective function for the fit is designed to minimise the difference of the monopole and the quadrupole of the galaxy power spectrum with the target signals, we have shown that MGs leave a characteristic imprint in RSD statistics, especially in the quadrupole. To estimate the constraining power of galaxy statistics, we have computed the $\chi^2$ of the difference between MG and GR multipoles. 
We confirmed that non-linear scales are important in constraining F5, while N1 can be constrained on large scales due to the scale-independent modification of the growth rate. We also found a good agreement between COLA and {\it N}-body galaxy statistics. 

The accuracy of COLA galaxy catalogues may be further improved by tuning the COLA halo catalogues to reproduce an analytic halo mass function. 
This can be achieved by introducing a mass-dependent conversion factor when converting $M_{\rm FoF}$ to $M_{\rm200c}$. However, this requires an accurate measurement of the mass function from {\it N}-body simulations in MG, which undermines the predictability of COLA, one of the major advantages that COLA has over other approximate methods. In this paper, we focused on the power spectrum in redshift space. It is an open question whether \textcode{mg-picola} can also produce higher-order statistics in MG such as the bispectrum (because of the approximations in the screening, although \cite{Colavincenzo_2018} showed that COLA can reproduce the bispectrum in GR) as well as other non-linear structures such as voids. These studies are left for future work.

Using the results of this work, we plan to produce a large number of mock galaxy catalogues in MG with \textcode{mg-picola}. This will allow us to study the degeneracy between MG and cosmological parameters by looking at different probes and formulate the best estimators to use in Stage IV experiments to constrain gravity on cosmological scales.

\acknowledgments
KK and AI are supported by the European Research Council under the European Union's Horizon 2020 programme (grant agreement No.646702 ``CosTesGrav"). KK is also supported by the UK STFC grant ST/S000550/1. BSW is supported by the Royal Society grant number RGF\textbackslash{}EA\textbackslash{}181023. BL is supported by the European Research Council under the European Union's Horizon 2020 programme (grant agreement No.~716532 ``PUNCA"), and UK STFC Consolidated Grants ST/L00075X/1, ST/P000451/1. Numerical computations were done on the Sciama High Performance Compute (HPC) cluster which is supported by 
the ICG, SEPNet and the University of Portsmouth. Some of the simulations described in this work used the DiRAC Data Centric system at Durham University, operated by the Institute for Computational Cosmology on behalf of the STFC DiRAC HPC Facility (www.dirac.ac.uk). This equipment was funded by BIS National E-infrastructure capital grant ST/K00042X/1, STFC capital grants ST/H008519/1 and ST/K00087X/1, STFC DiRAC Operations grant ST/K003267/1 and Durham University. DiRAC is part of the National E-Infrastructure. Part of the simulations were run by Wojciech Hellwing at the Interdisciplinary Center for Mathematical and Computational Modelling (ICM) at University of Warsaw, under Grants No.~GA67-17 and No.~GA65-30.

\bibliographystyle{JHEP}
\bibliography{References}

\end{document}